\documentclass[manuscript,screen]{acmart}

\usepackage{lineno,hyperref}
\usepackage{booktabs} 
\usepackage{upgreek} 
\usepackage{calc}
\usepackage{tikz-cd}
\usepackage{tikz}
\usetikzlibrary{shapes,arrows}
\usepackage{graphicx,verbatimbox,booktabs}
\usepackage{listings}
\usepackage{subcaption}
\usepackage{mathtools}
\usepackage{color}
\usepackage{tabularx,ragged2e}
\usepackage{multirow}
\usepackage{mathtools}
\usepackage{enumitem}
\usepackage[british]{babel}
\usepackage{todonotes}
\usepackage{comment}
\usepackage{color}
\usepackage{mathtools}
\usepackage{afterpage}
\DeclarePairedDelimiter{\ceil}{\lceil}{\rceil}

\newcommand*\circled[1]{\tikz[baseline=(char.base)]{
  \node[shape=circle,fill,inner sep=0.6pt] (char) {\bf\small\textcolor{white}{#1}};}}
\usepackage{soul}

\usepackage{footnote}
\makesavenoteenv{tabular}
\makesavenoteenv{table}

\newcommand\Vtextvisiblespace[1][.3em]{%
  \mbox{\kern.06em\vrule height.3ex}%
  \vbox{\hrule width#1}%
  \hbox{\vrule height.3ex}}
\modulolinenumbers[5]

\AtBeginDocument{%
  \providecommand\BibTeX{{%
    \normalfont B\kern-0.5em{\scshape i\kern-0.25em b}\kern-0.8em\TeX}}}

\setcopyright{acmcopyright}
\copyrightyear{2018}
\acmYear{2018}
\acmDOI{10.1145/1122445.1122456}

\begin{document}

\title{Pseudo Relevance Feedback with Deep Language Models and Dense Retrievers: Successes and Pitfalls}

\author{Hang Li}
\email{hang.li@uq.edu.au}
\orcid{0000-0002-5317-7227}
\author{Ahmed Mourad}
\email{a.mourad@uq.edu.au}
\orcid{0000-0002-9423-9404}
\author{Shengyao Zhuang}
\email{s.zhuang@uq.edu.au}
\orcid{0000-0002-6711-0955}
\affiliation{%
  \institution{IElab, The University of Queensland}
  \city{St. Lucia}
  \state{Queensland}
  \country{Australia}
}

\author{Bevan Koopman}
\email{bevan.koopman@csiro.au}
\orcid{0000-0001-5577-3391}
\affiliation{%
  \institution{Australian E-Health Research Centre, CSIRO}
  \city{Herston}
  \state{Queensland}
  \country{Australia}}

\author{Guido Zuccon}
\email{g.zuccon@uq.edu.au}
\orcid{0000-0003-0271-5563}
\affiliation{%
  \institution{IElab, The University of Queensland}
  \city{St. Lucia}
  \state{Queensland}
  \country{Australia}
}

\renewcommand{\shortauthors}{Li and Mourad, et al.}

\begin{abstract}
  Pseudo Relevance Feedback (PRF) is known to improve the effectiveness of bag-of-words retrievers. At the same time, deep language models have been shown to outperform traditional bag-of-words rerankers. However, it is unclear how to integrate PRF directly with emergent deep language models. This article addresses this gap by investigating methods for integrating PRF signals with rerankers and dense retrievers based on deep language models. We consider text-based, vector-based and hybrid PRF approaches and investigate different ways of combining and scoring relevance signals. An extensive empirical evaluation was conducted across four different datasets and two task settings (retrieval and ranking).

\textit{Text-based PRF} results show that the use of PRF had a mixed effect on deep rerankers across different datasets. We found that the best effectiveness was achieved when (i) directly concatenating each PRF passage with the query, searching with the new set of queries, and then aggregating the scores; (ii) using Borda to aggregate scores from PRF runs. 

\textit{Vector-based PRF} results show that the use of PRF enhanced the effectiveness of deep rerankers and dense retrievers over several evaluation metrics. We found that higher effectiveness was achieved when (i) the query retains either the majority or the same weight within the PRF mechanism, and (ii) a shallower PRF signal (i.e., a smaller number of top-ranked passages) was employed, rather than a deeper signal. Our vector-based PRF method is computationally efficient; thus, this represents a general PRF method others can use with deep rerankers and dense retrievers.

\end{abstract}



\maketitle

\section{Introduction}

Pseudo Relevance Feedback (PRF) assumes the top-ranked passages from any phase of retrieval contain relevant signals and thus modifies the query by exploiting these signals in a bid to reduce the effect of query-passage vocabulary mismatch and improve search effectiveness~\cite{azad2019query}. Previous research has considered PRF in the context of traditional bag-of-words retrieval models such as probabilistic~\cite{robertson2009probabilistic}, vector space~\cite{rocchio1971rocchio}, and language models~\cite{lv2014revisiting,zhai2001model,lavrenko2017relevance}. PRF methods such as Rocchio~\cite{rocchio1971rocchio}, relevance models~\cite{lavrenko2017relevance}, RM3~\cite{lv2009comparative}, and KL expansion models~\cite{zhai2001model} analyse the top-ranked passages to expand the query or to modify the query weights. The query and the passage are represented as either text or vectors, hence the categorisation of \textit{text-based} and \textit{vector-based} PRF approaches hereafter. Empirically, these approaches improve the initial retrieval effectiveness~\cite{azad2019query}.

Recently, Transformer~\cite{vaswani2017attention} based deep language models~\cite{devlin2018bert,yang2019xlnet,dai2019transformer,radford2019language, raffel2019exploring} have been adopted with promising results in information retrieval~\cite{lin2020pretrained,guo2020deep}. Seminal in this context is the work of \citet{nogueira2019passage} who fine tuned BERT~\cite{devlin2018bert} as a reranker on top of BM25. In this article, we investigate how to integrate PRF signals, effective for bag-of-words models, with deep language model rerankers, e.g. BERT (other models such as RoBERTa~\cite{liu2019roberta}, query likelihood models~\cite{zhuang2021deep,zhuang2021tilde,dos2020beyond} can be applied as well), and dense retrievers, (specifically ANCE~\cite{xiong2020approximate}, RepBERT~\cite{zhan2020repbert}, TCT-ColBERT V1~\cite{lin2020distilling}, TCT-ColBERT V2 HN+~\cite{lin2021batch}, DistilBERT KD~\cite{hofstatter2020improving}, DistilBERT Balanced~\cite{hofstatter2021efficiently}, and SBERT~\cite{reimers2019sentence}); extensive evaluations are done towards the proposed PRF methods along the side.

Our experiments investigate two alternative paths to integrate PRF signals with deep language models: text-based and vector-based. The \emph{text-based PRF} approach is an obvious direction as the concatenation of the query text and the PRF passages text is used as the new formulated query to feed into the deep language model (e.g., BERT). However, this approach has two significant impediments: (i) the lengthy concatenated text would often exceed the allowed input size (input vector length) of these deep language models~\cite{yang2019simple,ding2020cogltx} and (ii) it is computationally expensive or infeasible as it requires additional deep language model inferences at query time~\cite{jiao2020tinybert,liu2020fastbert}. To solve the first challenge, we propose three different text handling methods to generate text partitions from the full concatenated text such that each of the partitions is within the length limit of the deep language models. Furthermore, because we split the concatenated text into partitions, we also propose three different score aggregation methods (Average, Borda, and Max) to aggregate the scores from each partition to calculate the final scores for each passage.

To address the computational complexity challenge, we use model pre-generated embeddings to represent text~\cite{khattab2020colbert,zhan2020repbert,xiong2020approximate,ding2020rocketqa}. Query latency is reduced to the time of generating the query embeddings because the passages embeddings are pre-generated. In the context of PRF, we further utilise these pre-generated passages embeddings to efficiently integrate the relevance signals while eliminating the input size limit of deep language models, which we refer to as \emph{vector-based PRF} approach. Each feedback passage is pre-generated as embeddings (vectors) in this approach. We adopt two different vector fusion methods (Average and Rocchio) to integrate the feedback vectors into the query vectors. The Rocchio method has two parameters: the query vector and the feedback passage vector weights. We empirically investigate the influence of query and feedback passages through weighting within the Rocchio PRF approach.

To evaluate these PRF approaches, we use the TREC Deep Learning Track Passage Retrieval Task (TREC DL) 2019~\cite{craswell2020overview} and 2020~\cite{craswell2021overview}, the TREC Conversational Assistance Track 2019 (TREC CAsT)~\cite{dalton2020trec}, the Web Answer Passages (WebAP)~\cite{keikha2014evaluating}, and the DL HARD~\cite{mackie2021dlhard}. TREC CAsT and WebAP are used for the passage retrieval task rather than their original tasks (e.g., for CAsT, we do not consider the multi-turn conversational relationship between queries).

For the text-based PRF approach, we find that our models significantly outperform the baselines across several evaluation metrics on TREC DL 2019 while having mixed results on TREC DL 2020, TREC CAsT and WebAP. For DL HARD, the proposed approach does not have any significant improvements. The results suggest that TREC DL 2019 queries are easier -- the results from the initial ranking contain less noise -- hence, the PRF can add more relevant information to the queries. On the other hand, the queries of TREC DL 2020, TREC CAsT, and WebAP are more challenging ---the results from the initial ranking contain more noise---hence, adding these PRF signals into the query will cause query drift and lead to worse performance. DL HARD is created by selecting queries from TREC DL 2019 and TREC DL 2020 based on the performance systems at TREC had (i.e., select queries for which systems cannot perform well) and the characteristics of the queries~\cite{mackie2021dlhard}. Our results show that text-based PRF did not work on DL HARD, suggesting that the feedback passages do not contain relevant signals or more noise than valuable signals. 

Another challenge for text-based PRF is its computational complexity for the full ranking pipeline. It requires at least two inferences, depending on the text partitioning method. At least, this doubles the total run time compared to that of deep language rerankers without PRF.

For the vector-based PRF approach, we find that our models improve the respective baselines (seven dense retrievers) across all evaluation metrics and all datasets for the retrieval task; the proposed approach also outperforms the strong BM25+BERT ranker across several metrics. This result suggests that encoding the PRF feedback passages into embedding vectors better models the relevance signals exploited by the PRF mechanism. Unlike text-based PRF, the passage vectors are pre-generated and indexed, so the inference steps on passages are not required at retrieval or rerank time. This makes vector-based PRF very efficient: it takes only 1/20th of the time of the BM25+BERT reranker and only about double the time of the simple bag-of-words BM25. In addition, since our proposed approach works directly with the vector embeddings of queries and passages, they can be applied on top of any choice of dense retriever. For the reranking task, we find that our models outperform the BM25 and BM25+RM3 baselines across all metrics and datasets, while they have only mixed improvements over the strong BM25+BERT reranker. Overall, the vector-based PRF approach for retrieval tends to improve deep metrics, while for reranking, they tend to improve shallow metrics.

To summarize in this article we make the following contributions:

\begin{itemize}
	\item We thoroughly investigate the PRF effectiveness under different conditions, in particular how sensitive the effectiveness is to PRF depth, text handling/vector fusion, and score estimation;
	\item We conduct a thorough comparison of text-based and vector-based approaches within the same reranking task;
	\item We conduct a thorough comparison of different vector-based approaches within the same retrieval task;
	\item We study the efficiency of the proposed text-based and vector-based PRF approaches.
\end{itemize}

\section{Related Work}

Pseudo-Relevance Feedback (PRF) is a classic query expansion method that modifies the original query in an attempt to address the mismatch between the query intent and the query representation~\cite{clinchant2013a,wang2020pseudo}. A typical PRF setting uses the top-ranked passages from a retrieval system as the relevant signal to select query terms to add to the original query or to set the weights for the query terms. PRF approaches including Rocchio~\cite{rocchio1971rocchio}, KL expansion~\cite{lv2014revisiting,zhai2001model}, query-regularized mixture model~\cite{tao2006regularized}, RM3~\cite{lv2009comparative}, relevance-feedback matrix factorization~\cite{zamani2016pseudo}, and relevance models~\cite{lavrenko2017relevance} have been well studied. The use of PRF on top of efficient bag-of-words retrieval models is common in information retrieval systems, and it is an effective strategy for improving retrieval effectiveness~\cite{clinchant2013a,lavrenko2017relevance}. Traditional PRF approaches~\cite{rocchio1971rocchio,zhai2001model,lv2009comparative,lavrenko2017relevance} are simple, but more effective, robust and generalisable, in comparison to more complex models~\cite{tao2006regularized,zamani2016pseudo}, which instead achieve marginal gains, may be harder to implement/reproduce or maybe problematic to instantiate across different datasets or domains from those in which they have been originally evaluated. This study employs the most popular PRF method in existing research (RM3) as a baseline. The RM3 considers the original query and the feedback passages when creating a new query by assigning different weights to the original query and feedback terms. RM3 is effective and robust compared to other query expansion methods~\cite{lv2009comparative}, and it is used as a baseline in several pseudo relevance feedback studies~\cite{miao2012proximity,lv2010positional,cao2008selecting}.

Recent research has studied PRF in different settings. \citet{lin2019simplest} considered document ranking as a binary classification problem, combining PRF with text classification by introducing positive and negative pseudo labels. The positive pseudo labels are obtained from the top-$k$ documents, while the negative labels are from the bottom-$n$ documents. The final score is a linear interpolation of the classifier and retriever scores. \citet{li2018nprf} proposed a neural PRF framework, which was further extended by~\citet{wang2020end}, that utilises a feed-forward neural network to determine the target document's relevance score by aggregating the target query and the target feedback relevance scores. However, these proposed models have achieved marginal improvements over the BM25 baseline. Furthermore, the efficiency (run time) of the proposed models have not been reported, and thus it is difficult to establish whether these marginal improvements in effectiveness may be at the cost of efficiency.

Deep language models based on transformers~\cite{vaswani2017attention}, such as BERT~\cite{devlin2018bert}, T5~\cite{raffel2019exploring}, and RoBERTa~\cite{liu2019roberta}, has surpassed the existing state-of-the-art effectiveness in different search tasks. BERT, in specific, has shown to improve over previous state-of-the-art for ad hoc retrieval~\cite{nogueira2019passage}. Recent research has also considered integrating PRF with deep language models. \citet{padaki2020rethinking} integrated RM3 with BERT. The results, however, showed that the selection of highly weighted terms from the feedback passages via RM3 to expand the original query could significantly hurt the ranking quality of a fine-tuned BERT reranker. \citet{yu2021pgt} presented a framework that integrates PRF into a Graph-based Transformer (PGT). It represents each feedback passage as a node, and the PRF signals are captured using sparse attention between graph nodes. While this approach handles the input-size limit of deep language models, it achieves marginal improvements compared to the BERT reranker approach across most evaluation metrics at the cost of efficiency. Specifically, compared to our results, PGT achieves a lower nDCG@10 than our simplest text-based PRF reranking approach; it also achieves lower effectiveness in reranking and similar effectiveness in retrieval than our vector-based PRF with ANCE, but at a much higher computational cost.

\citet{wang2020pseudo} argued that existing PRF research mainly considers relevance matching where terms are used to sort feedback documents. On the contrary, they propose a model that considers both relevance and semantic matching. The relevance score is obtained using BM25. For semantic matching, they split the top-$k$ PRF documents into sentences. For each sentence, they use BERT to estimate the semantic similarity with the query. Scores from the top-$m$ sentences of each document are considered as the semantic score for this document. The final scores of each document are calculated from a linear interpolation of the relevance and semantic scores. The expansion terms are then extracted from the reranked top-$k$ PRF documents and added to the original query for a second retrieval stage. Although the improvements are marginal, they demonstrate that BERT can identify relevance signals from the feedback documents at the sentence level to enhance retrieval effectiveness. However, this marginal improvement is at the expense of efficiency because expansion terms are identified through BERT.

\citet{zheng-etal-2020-bert,zheng2021contextualized} presented a three-phase BERT-based query expansion model: BERT-QE. The first phase is a standard BERT reranking~\cite{nogueira2019passage} step. In the second phase, the top-$k$ passages are selected as feedback passages, further split into overlapping partitions using a sliding window. Together with the original query, these partitions are fed into BERT to get the top-$m$ partitions with the highest scores per passage. The top-$m$ partitions and the candidate passage are fed into BERT in the third phase. The score of a candidate passage in this phase is calculated as a weighted sum, where the weight is the relevance score of each partition in the top-$m$ partitions from phase two, and the score is the relevance score between the top-$m$ partitions and the candidate passage. The final score of a candidate passage is calculated by linear interpolation of the first phase BERT relevance score between the query and the passage, and the third phase weighted sum score between the top-$m$ partitions and the candidate passages. Although BERT-QE achieves significant improvements in effectiveness over BERT reranker, it requires 11.01x more computations than BERT, making it computationally infeasible in many practical applications.

Recently, \citet{yu2021improving} proposed a PRF framework based on ANCE~\cite{xiong2020approximate}, which trains a new query encoder from ANCE that takes in the top-$k$ passages from the first-round ANCE retrieval, then concatenate the passage texts with the original query text to form the new PRF query, without changing ANCE's passage encoder. This newly formed PRF query is passed to the trained query encoder to produce the PRF query representation, and to retrieve the results from the original passage collection index. Although the improvements are significant across several datasets over different metrics, according to a recent reproducibility paper from~\citet{li2021improving}, the proposed model does not generalise well to other dense retriever models and the training process needs to be adjusted accordingly with different dense retriever models, which makes it difficult to achieve the same effectiveness as the one proposed in the original paper. 

Integrating PRF signals with deep language models implies a trade-off between effectiveness and efficiency. While current approaches ignored efficiency, the majority still achieved marginal improvements in effectiveness. In this study, we propose three approaches to integrate PRF signals to improve effectiveness while maintaining efficiency: (i) by concatenating the feedback passages text with the original query to form the new queries that contain the relevant signals, (ii) by pre-generating passage collection embeddings and performing PRF in the vector space, because embeddings promise to capture the semantic similarity between terms~\cite{dalton2019local,diaz2016query,kuzi2016query,naseri2018exploring,roy2016using,zamani2016embedding,zamani2017relevance,ceqe2021shahrzad}, which makes it feasible as a method for first stage retrieval as well, (iii) by combining the previous two approaches into a hybrid approach.

\section{Methodology}
\label{sec:method}

\subsection{Text-Based Pseudo-Relevance Feedback}

BERT is computationally expensive to be applied as a first-stage retriever. Hence, it is commonly employed as a reranker that considers only a subset of the initial retrieval results (usually top 1000). In this approach, we integrate the text-based PRF signal with the BERT reranker. \citet{padaki2020rethinking} demonstrated that the use of RM3~\cite{lv2009comparative} to select highly weighted terms from the feedback passages and construct the new PRF queries significantly hurts the ranking quality of a fine-tuned BERT reranker. Therefore, we use the full passages text to construct the new PRF queries. We address the challenge of the input size limit of BERT by employing three text-based PRF methods:
\begin{enumerate}
    \item \emph{Concatenate and Truncate}: append the query and the top-$k$ \emph{feedback} passages, then truncate to the length of 256 tokens. BERT has an input size limit of 512 tokens; we allocate 256 tokens to the new query and the remaining tokens are left to concatenate the \emph{candidate} passage.
    \item \emph{Concatenate and Aggregate}: append the query to each of the top-$k$ feedback passages to form $k$ new queries. For each new query, use BERT to perform another rerank, resulting in $k$ new ranked lists. The final scores for the candidate passages are generated using different score estimation methods that combine the $k$ ranked lists (but not the ranked list of the original query).
    \item \emph{Sliding Window}: concatenate the top-$k$ passages, then use a sliding window to split the aggregated text into overlapping partitions 
    . Concatenate the query with each partition to create $j$ new queries, then follow the same steps as Concatenate and Aggregate. 
\end{enumerate}

Methods 2 and 3 require the aggregation of multiple ranker lists to estimate the scores and obtain the final ranked list. For this, we aggregate the scores of a candidate passage using several methods:
\begin{enumerate}
    \item \emph{Average}: calculate the average of all the scores per candidate passage.
    \item \emph{Max}: consider only the highest score per candidate passage.
    \item \emph{Borda}: employ the Borda voting rule~\cite{aslam2001models,macdonald2008voting} to calculate the score of each candidate passage.
\end{enumerate}

\begin{figure}[t!]
    \centering
    \includegraphics[width=0.7\columnwidth]{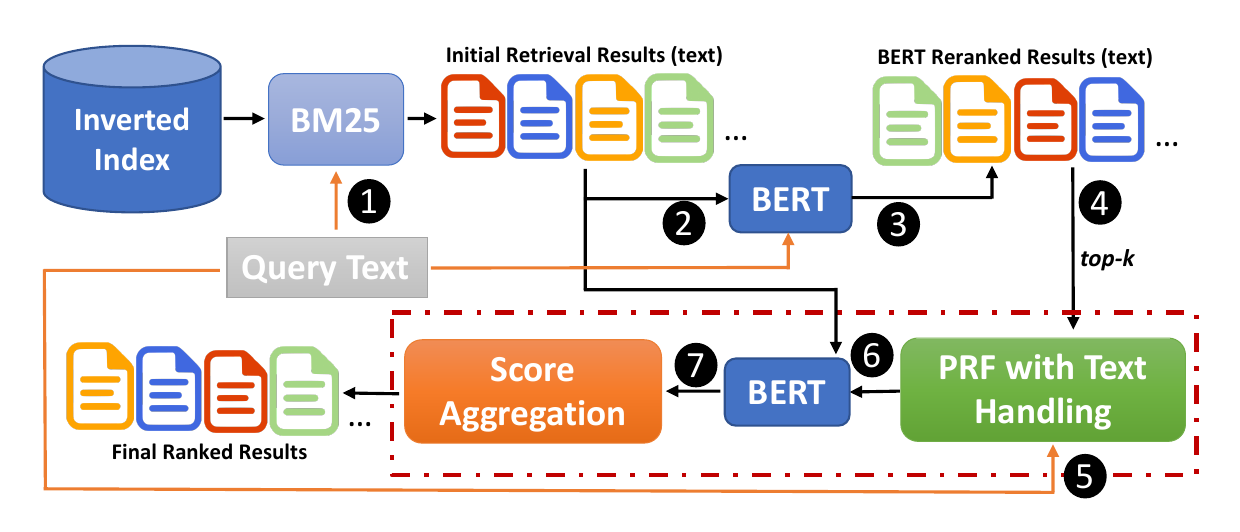}
    \caption{The proposed architecture for integrating Text-based Pseudo-Relevance Feedback with BERT reranker. The initial retriever is a traditional bag-of-words BM25.}
    \label{fig:model}
\end{figure}

Figure~\ref{fig:model} depicts the proposed architecture for integrating text-based PRF signals with BERT reranker. The initial retriever is a traditional bag-of-words BM25 followed by BERT reranker. As shown in step \circled{1}, the query is passed to BM25 to retrieve the initial ranked results from the inverted index. Then the query text and initial retrieval results are passed to BERT for reranking (\circled{2}). The top-$k$ feedback passages from the reranked list are used as PRF relevance signals  (\circled{4}), after mapping them back to their text representation  (\circled{3}). Then, the query and feedback passage texts are combined together to form new query texts (\circled{5}) followed by another BERT-based scoring step (\circled{6}), and finally the individual scores are aggregated per candidate passage to form the final ranking (\circled{7}). The core components of this architecture, which are PRF with Text Handling and Score Estimation, are described in the next two sections.

\subsubsection{Text-Based PRF with Text Handling}

We consider three different approaches to handle the text length that exceeds the BERT input size limit:

\paragraph{Concatenate and Truncate (CT)}

A new query text is generated by concatenating the original query text with the top-$k$ feedback passage texts, separated by a space (\Vtextvisiblespace[0.8em]). If the length of the new query exceeds 256, it will be truncated to the first 256 tokens. Then, we run the new query through BERT reranker. The new query is constructed as follows:
\begin{equation}
    Q_{new,l<=256} = \lceil Q_{original}+\Vtextvisiblespace[0.8em]+p_1+...+\Vtextvisiblespace[0.8em]+p_k \rceil_{256}
    \label{eq:cc}
\end{equation}
where $Q_{new}$ and $Q_{original}$ represent the new query text and the original query text, respectively. $l<=256$ represents the input size limit enforced, which is achieved by truncating the sequence (denoted with $\ceil._{256}$). $p_1,...,p_k$ represent the top-$k$ feedback passages from the BERT reranker. \Vtextvisiblespace[0.8em] is the space in between.

\paragraph{Concatenate and Aggregate (CA)}

This approach generates $k$ new queries by concatenating the original query text with each of the top-$k$ feedback passage texts, separated by a space (\Vtextvisiblespace[0.8em]). Then, each of the new queries is run through another BERT reranking step resulting into $k$ scores per candidate passage, which will be aggregated later to estimate the final score. The new queries are generated as follows:
\begin{equation}\label{eq:ca}
\begin{split}
    Q_{1,new} = Q_{original} + \Vtextvisiblespace[0.8em] + p_1 \\
    ... \\
    Q_{k,new} = Q_{original} + \Vtextvisiblespace[0.8em] + p_k
\end{split}
\end{equation}
where $Q_{1,new},...,Q_{k,new}$ represent the $k$ new queries. $Q_{original}$ represents the original query text. $p_1,..,p_k$ represent the top-$k$ feedback passage texts. \Vtextvisiblespace[0.8em] is the separation token in between.

\paragraph{Sliding Window (SW)}

In this approach, the top-$k$ feedback passage texts are appended together, then a sliding window is applied to split the text into $j$ overlapping partitions with different window size and stride according to different datasets' passage lengths~\cite{dai2019deeper}, as below:
\begin{equation}
    p_1+...+p_k \xrightarrow[]{SW} p_1,...,p_j
    \label{eq:sw-1}
\end{equation}
where $p_1,...,p_k$ represent the top-$k$ feedback passage texts, $SW$ represents the sliding window mechanism, $p_1,...,p_j$ represent the $j$ partitions. Similar to the CA approach, the set of $j$ new queries is generated using Eq.~\ref{eq:ca}.

Note that after generating each new query, the query/passage pair may exceed the BERT input size limit for the CT approach. Under this situation, if the length of the new query exceeds 256, we truncate the new query down to be of length 256. For CA and SW approaches, we also applied the same methodology to guarantee all the new queries are below the length of 256.

\subsubsection{Text-Based with Score Estimation}

CA and SW text-handling approaches generate $k$ and $j$ scores per candidate passage, respectively. To estimate a final score for each candidate passage, we consider the following estimation methods.

\paragraph{Average}

The final score is estimated by calculating the mean of all scores:
\begin{equation}
    S_{final} = Avg(S_1+S_2+...+S_k)
    \label{eq:avg}
\end{equation}
where $S_{final}$ represents the final ranking score for each candidate passage, and $S_1,...,S_k$ represent the $k$ ranking scores for each candidate passage based on each of the $k$ new queries. For the rest of this paper, we refer to this method as Text-Average, represented by T-A for brevity.

\paragraph{Max}

The final score is estimated by taking the highest score per candidate passage:
\begin{equation}
    S_{final} = Max(S_1,S_2,...,S_k)
    \label{eq:max}
\end{equation}
where $S_{final}$ represents the maximum score for each candidate passage, and $S_1,...,S_k$ represent the $k$ ranking scores for each candidate passage based on each of the $k$ new queries. For the rest of this paper, we refer to this method as Max, represented by M for brevity.

\paragraph{Borda}

The final score is estimated by using the Borda voting algorithm. The score of a candidate passage w.r.t a ranked list is the number of candidate passages in the ranked list that are ranked lower. Scores are summed over ranked lists as follows:
\begin{equation}
    S_{final} = \sum_{L_{i}:p \in L_{i}} \frac{n-r_{L_{i}}(p)+1}{n}
\end{equation}
where $L_{i}$ represents the $i$-$th$ ranked list produced using the $i$-$th$ new query, $p$ represents the candidate passage, $r$ is the rank of the candidate passage, and $n$ represents the number of candidate passages in the ranked list. For the rest of this paper, we refer to this method as Borda, represented by B for brevity.

\subsection{Vector-Based Pseudo-Relevance Feedback}
\label{sec:vprf}

Using existing, efficient first stage dense retrievers(RepBERT~\cite{zhan2020repbert}, ANCE~\cite{xiong2020approximate}, TCT-ColBERT V1~\cite{lin2020distilling}, TCT-ColBERT V2 HN+~\cite{lin2021batch}, DistilBERT KD~\cite{hofstatter2020improving}, DistilBERT Balanced~\cite{hofstatter2021efficiently}, and SBERT~\cite{reimers2019sentence}), we employ two vector-based PRF methods for the retrieval task:
\begin{enumerate}
	\item \emph{Average:} the mean of the original query embeddings and the feedback passage embeddings are used to generate the new query representation.
	
	\item \emph{Rocchio:} different weights are assigned to the original query embeddings and the feedback passage embeddings following the intuition provided by the original Relevance Feedback mechanism proposed by Rocchio~\cite{rocchio1971rocchio}.
\end{enumerate}

\begin{figure}[t!]
	\centering
	\includegraphics[width=0.7\columnwidth]{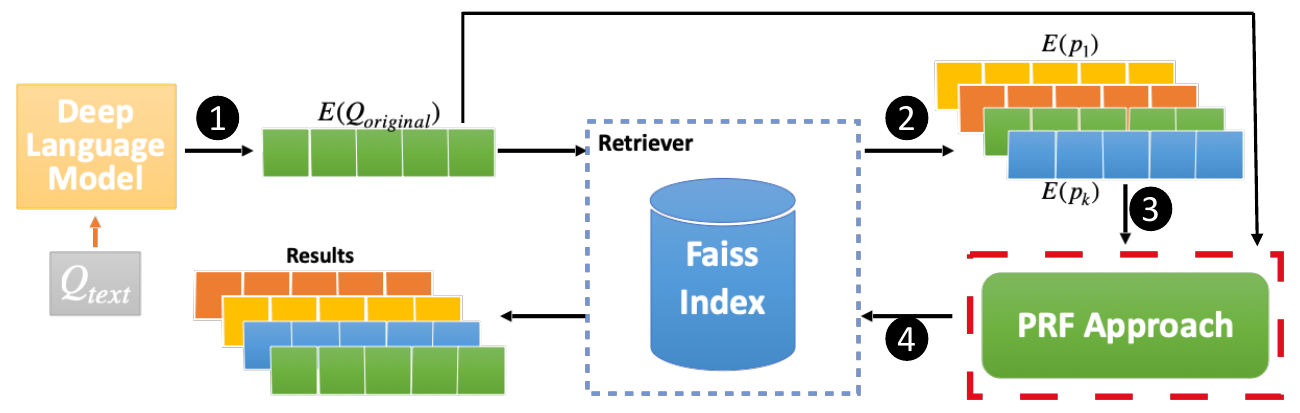}
	\caption{The proposed architecture for integrating Vector-based Pseudo-Relevance Feedback with Deep Language Model dense retrievers for the retrieval task.} 
	\label{fig:model-structure}
\end{figure}

Figure~\ref{fig:model-structure} depicts the proposed architecture for integrating vector-based PRF signals with deep language model dense retrievers. A single deep language model is used to generate offline the embeddings for all passages, which are then stored in a Faiss index~\cite{JDH17}. The deep language model is also used to generate the query embedding at inference time (step \circled{1}). The query embedding is then passed to the dense retriever that exploits the Faiss index to perform the first pass of retrieval to obtain the initial ranked list (\circled{2}).  The top-$k$ feedback passage embeddings from the initial ranked list are used as PRF relevance signals (\circled{3}), using vector operations, and are then used to perform the subsequent retrieval to get the final ranked list (\circled{4}). 

We describe the two proposed vector-based PRF approaches in the next two sections. 

\subsubsection{Vector-Based PRF with Average}

A new query embedding is generated by averaging the original query embedding and the top-$k$ feedback passage embeddings. The intuition is to treat the original query at par of the signal from the top-$k$ feedback passages (i.e., the query weights as much as each passage).
The new query embedding is computed as follows:
\begin{equation}
	E_{Q_{new}}=Avg(E(Q_{original}),E(p_1),...,E(p_k))
	\label{eq:avg}
\end{equation}

where $E$ represents the embeddings of either the query or the feedback passage, $E_{Q_{new}}$ represents the newly formulated query embeddings. We do not generate an actual text query in the vector-based PRF approaches: only the embedding of the new query is generated. $Q_{original}$ represents the original query, $p_1,...,p_k$ represent the top-$k$ passages retrieved by the first stage ranker. In the remainder of the paper we refer to this method as Vector Average, represented by V-A for brevity. 

\subsubsection{Vector-Based PRF with Rocchio}

This method is inspired by the original Rocchio method for relevance feedback~\cite{rocchio1971rocchio} but adapted to deep language models. The intuition  is to transform the original query embedding towards the average of the top-$k$ feedback passage embeddings by assigning different weights to query and (the combination of) feedback passages, thus controlling the contribution of each component toward the final score. Unlike in the original version of Rocchio, in this work we do not model the PRF with non-relevant passages: hence the negative portion of Rocchio is omitted. We note that this could be extended by identifying which passages in the initial ranked list could represent a negative relevance signal (e.g., the bottom passages) -- however we leave this for future consideration.

Thus, our Rocchio PRF approach consists of interpolating the query embedding and the average PRF embedding: 
\begin{equation}
	E_{Q_{new}} = \alpha*E(Q_{original})+\beta*Avg(E(p_1),...,E(p_k))
	\label{eq:rocchio-1}
\end{equation}
where $\alpha$ controls the weight assigned to the original query embedding and $\beta$ the weight assigned to the PRF signal. In the remainder of the paper we refer to this method as Rocchio, represented by $\mathcal{RC}_{\alpha}$ and $\mathcal{RC}_{\alpha,\beta}$ for brevity.

\subsection{Hybrid Pseudo-Relevance Feedback}
\label{sec:hybird}

Text-based PRF is a computationally expensive approach for the reranking task, in our experiments, the BERT inference step is executed twice: one before the PRF, one after the PRF. On the other hand, vector-based PRF is an efficient approach for the retrieval task because of the high efficiency of the dense retriever models. In this section, we investigate a hybrid approach where the architecture of vector-based PRF in Figure~\ref{fig:model-structure} is adapted to the reranking task. The main difference is that the initial ranking of passages is obtained from an inverted-index (Text-based) multi-stage pipeline such as BM25+BERT (as in Figure~\ref{fig:model}). In particular, the initial retrieval results obtained through steps \circled{1} and \circled{2} in Figure~\ref{fig:model-structure} are replaced by steps \circled{1}, \circled{2}, \circled{3}, and \circled{4} in Figure~\ref{fig:model}. The ranked list of passages produced by the BERT reranker is mapped to embeddings using the Faiss index before applying the vector-based PRF methods.


\section{Experimental Setup}

\subsection{Datasets}

Our experiments use the TREC Deep Learning Track Passage Retrieval Task 2019~\cite{craswell2020overview} (DL 2019) and 2020~\cite{craswell2021overview} (DL 2020),  DL HARD~\cite{mackie2021dlhard}, the TREC Conversational Assistance Track  2019~\cite{dalton2020trec} (CAsT 2019), and the Web Answer Passages (WebAP)~\cite{keikha2014evaluating}. The detailed statistics for each dataset are listed in Table~\ref{table:stats}.

\begin{table}[]
\centering
\caption{Statistics of the four datasets considered in our experiments. Where \#Q represents the number of queries, \#P represents the number of passages in the collection, Avg Len represents the average length of passages, Avg \#J/Q represents the average number of judged passages per query, and \#J represents the number of judged passages in total.}
\resizebox{0.6\columnwidth}{!}{%
\begin{tabular}{lrrrrr}
\toprule
               & \#Q & \#P        & Avg Len  & Avg \#J/Q & \#J \\ \midrule
TREC DL 2019   & 43  & 8,841,823  & 64.7  & 215.3    & 9,260 \\
TREC DL 2020   & 54  & 8,841,823  & 64.7  & 210.9    & 11,386 \\
TREC CAsT 2019 & 502 & 38,618,941 & 68.6  & 63.2    & 31,713 \\
WebAP          & 80  & 1,959,777  & 74.5  & 11858.8   & 948,700 \\ 
DL HARD        & 50  & 8,841,823  & 64.7  & 85.1    & 4,256 \\ \bottomrule
\end{tabular}
}
\label{table:stats}
\end{table}

TREC DL 2019 and 2020 contain 200 queries each. However for 2019, only 43 queries have judgements; and thus the remaining 157 queries without judgements are discarded from our evaluation. In 2020, only 54 queries have judgements; and thus the remaining 146 queries are similarly discarded. The relevance judgements for both datasets range from 0 (not relevant) to 3 (highly relevant). The passage collection is the same as the MS MARCO passage ranking dataset~\cite{nguyen2016ms}, which is a benchmark English dataset for ad-hoc retrieval tasks with $\approx$8.8M passages. The difference between TREC DL and MS MARCO is that queries in TREC DL have several judgements per query (215.3/210.9 on average for 2019/2020), instead of an average of one judgement per query for MS MARCO. The very sparse relevance judgements of MS MARCO would not be able to provide detailed, reliable information on the behaviour of the PRF approaches and thus we do not report them in this article. However, we still tried to apply our vector-based PRF for the retrieval task on MS MARCO dev set, which consists of 6,980 queries. We refer the reader to our github page for the full results.\footnote{\url{https://github.com/castorini/pyserini/blob/master/docs/experiments-vector-prf.md}}

DL HARD builds upon the TREC DL 2019/2020 queries: these queries are considered as hard queries on which previous methods do not perform well, and new judgements are provided for the added new queries (originally unjudged in TREC)~\cite{mackie2021dlhard}. While TREC CAsT 2019 is originally constructed for multi-turn conversational search, we treat each turn independently, and we use the manually rewritten topic utterances. WebAP is built from the  TREC 2004 Terabyte Track collection, and it contains 80 queries\footnote{In addition to two queries without relevance judgements, which are excluded in our experiments} and about 2 Million passages (1,1858.8 judged passages per query, on average). The relevance judgements for TREC CAsT 2019 and WebAP ranged from 0 (not relevant) to 4 (highly relevant).

\subsection{Evaluation Metrics}

We employ MAP, nDCG@\{1, 3, 10\}, and Reciprocal Rank (RR)\footnote{If for a query no relevant passage is retrieved up to the considered standard cut-off (1,000), then we assign RR=0.} for the reranking task on both text-based PRF and vector-based PRF. We select these metrics because they are the common measures reported for BERT based models and these datasets -- thus allowing cross-comparison with previous and future work. For the retrieval task on vector-based PRF, we also report Recall@\{1000\}, but it is not considered for text-based PRF approaches because they are built on top of the BERT reranker where the Recall is limited by the initial retriever (BM25) to the top 1,000 passages. We report Recall for its diagnostic ability in informing whether a gain in e.g., MAP is produced because of a higher number of retrieved relevant passages, or because of a better ranking (i.e. ordering of the same number of relevant passages). For the TREC DL 2020 dataset, we follow the instructions from the organisers and consider the label binarized at relevance level 2 for all evaluation metrics. For all results, statistical significance is performed using two-tailed paired t-test.

\subsection{Baselines}

We consider the following baselines:

\begin{itemize}[leftmargin=25pt]
	\item \texttt{BM25}: traditional first stage retriever, implemented using the Anserini toolkit~\cite{yang2018anserini} with its default settings.
	
	\item \texttt{BM25+RM3}: RM3 pseudo relevance feedback method~\cite{abdul2004umass} on top of BM25, as implemented in Anserini. We use this approach as a representative bag-of-words PRF method, since previous research has found alternative bag-of-words PRF approaches achieve similar effectiveness~\cite{miao2012proximity}. We note that BM25+RM3 is a standard baseline for MS MARCO and TREC DL.
	
	\item \texttt{RepBERT (R)}: first stage dense retriever~\cite{zhan2020repbert}. We use the implementation made available by the authors. 
	
	\item \texttt{ANCE (A)}: first stage dense retriever~\cite{xiong2020approximate}. We use the scripts provided by the authors for both data pre-processing and model implementation.
	
	\item \texttt{TCT-ColBERT V1, TCT-ColBERT V2 HN+, DistilBERT KD, DistilBERT Balanced, and SBERT}: first stage dense retrievers employed to evaluate the generalisability of our hypotheses. We use the implementations provided in the pyserini toolkit~\cite{lin2021pyserini}.
	
	\item \texttt{RepBERT+BERT (R+B)}: first stage dense retriever with an additional BERT reranker to rerank the initial results provided by RepBERT.
	
	\item \texttt{ANCE+BERT (A+B)}: first stage dense retriever with an additional BERT reranker to rerank the initial results provided by ANCE.
	
	\item \texttt{BM25+BERT (BB)}: A common two-stage reranker pipeline, first proposed by~\citet{nogueira2019passage}, where the initial stage is BM25, and BERT is used to rerank the results from BM25. BERT is fine-tuned on MS MARCO Passage Retrieval Dataset~\cite{nguyen2016ms}. In all of our experiments, we use the 12 layer uncased BERT-Base provided by \citet{nogueira2019passage}, unless stated otherwise, and we simply refer to it as BERT. In Section~\ref{sec:efficiency} we also use BERT-Large for the efficiency analysis.
\end{itemize}

\subsection{Applying PRF to Rerankers}


\emph{Text-Based Pseudo-Relevance Feedback for Reranking.} We refer to this approach as BB+PRF, where BB represents BM25+BERT.
For the Sliding Window approach, we use the average passage length as the window size, and half of the window size as the stride. Details of the Sliding Window parameters for each dataset are shown in Table~\ref{table:sw-stats}. We experiment by using the top $k=1,3,5,10,15,20$ passages as pseudo relevance feedback.

\begin{table}[]
\centering
\caption{Window size and stride size of the Sliding Window PRF approach for each dataset.}
\resizebox{0.4\columnwidth}{!}{
\begin{tabular}{lcc}
\toprule
               & window size & stride \\ \midrule
TREC DL 2019   & 65  & 32  \\
TREC DL 2020   & 65  & 32  \\
TREC CAsT 2019 & 69  & 34  \\
WebAP          & 75  & 37  \\ 
DL HARD        & 65  & 32  \\ \bottomrule
\end{tabular}
}
\label{table:sw-stats}
\end{table}

\emph{Vector-Based Pseudo-Relevance Feedback for Reranking.} We consider the vector representations (embeddings) generated by RepBERT and ANCE to apply PRF as a second stage ranker, represented as BB+PRF-R and BB+PRF-A, where BB represents BM25+BERT, R represents RepBERT, and A represents ANCE. To achieve this, the top-$k$ passages IDs from BERT are mapped to their vector representations before estimating the final scores. For the Rocchio method, we experiment by assigning weights to the query and the feedback passage within the range of 0.1--1 with a step of 0.1. We experiment by using the top $k=1,3,5,10$ passages as pseudo relevance feedback.

\subsection{Applying PRF to Retrievers}

We choose RepBERT~\cite{zhan2020repbert}, ANCE~\cite{xiong2020approximate}, TCT-ColBERT V1~\cite{lin2020distilling}, TCT-ColBERT V2 HN+~\cite{lin2021batch}, DistilBERT KD~\cite{hofstatter2020improving}, DistilBERT Balanced~\cite{hofstatter2021efficiently}, and SBERT~\cite{reimers2019sentence} as representative first stage dense retrievers because they achieve state-of-the-art effectiveness in previous work on MS MARCO. We note that a host of alternative first stage dense retrievers have been recently proposed, including stronger ones like RocketQA~\cite{ding2020rocketqa} and RocketQAv2~\cite{ren2021rocketqav2}, but most of these retrievers consider more complex training procedures than those selected in this study. We further note that the implementation of the current best first stage dense retriever, RocketQAv2, has only just been made available and is based on PaddlePaddle~\cite{ma2019paddlepaddle}, thus uses a setup that differs from ours and is not selected for simplicity. We expect that findings that apply for the dense retrievers we chose are likely to translate to other dense retrievers, like RocketQA and RocketQAv2.

For the dense retrievers, we utilise the Faiss toolkit~\cite{JDH17} to build the index and perform retrieval. We develop our PRF approaches on top of these dense retrievers. To be consistent with the original dense retriever models, we truncate the query tokens and passage tokens according to the original settings in their papers. For simplicity, we mainly investigate our proposed vector-based PRF models on top of ANCE and RepBERT; the rest of the models are only shown in Table~\ref{tab:other-dense} for validation purposes as well as a demonstration of the generalisability of our proposed models. Therefore, in the following sections, vector-based PRF with RepBERT as base model is represented by R+PRF-R, and with ANCE as base model is represented by A+PRF-A for the retrieval task.

\subsection{Efficiency experiments}

To measure the runtime of each method, we run our experiments on a Unix-based server with the Intel(R) Xeon(R) Gold 6132 CPU @ 2.60GHz for BM25 and BM25 + RM3. For dense retrievers and our approaches, we use a Unix-based server equipped with a single Tesla V100 SMX2 32GB GPU.

\section{Results}

The overarching research question we seek to answer with our experiments is: \emph{What are the successes and pitfalls of integrating PRF into deep language model rerankers and dense retrievers in terms of effectiveness and efficiency?} Each of the following subsections addresses a more specific sub-question.

\subsection{PRF Depth}

\textbf{RQ1: What is the impact of PRF depth on the effectiveness of reranking and retrieval?} To answer this question, we vary the number of top-$k$ passages while displaying the distribution of results over other parameters (text handling and score estimation).

\begin{figure}
	\begin{subfigure}{\columnwidth}
	\centering
	\includegraphics[width=\columnwidth]{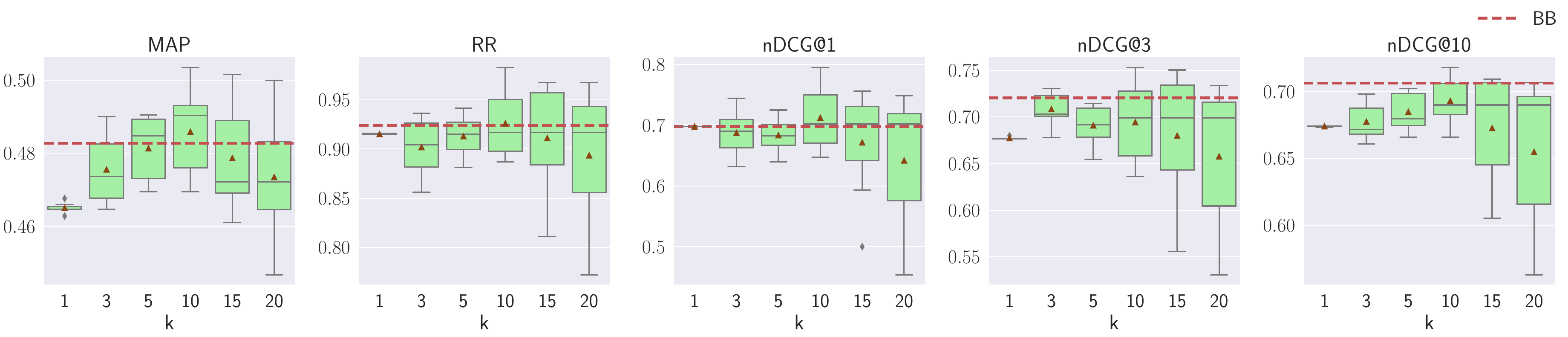}
 	\caption{TREC DL 2019}
	\label{fig:text-prf-depths-trec-2019}
	\end{subfigure}
	\begin{subfigure}{\columnwidth}
	\centering
	\includegraphics[width=\columnwidth]{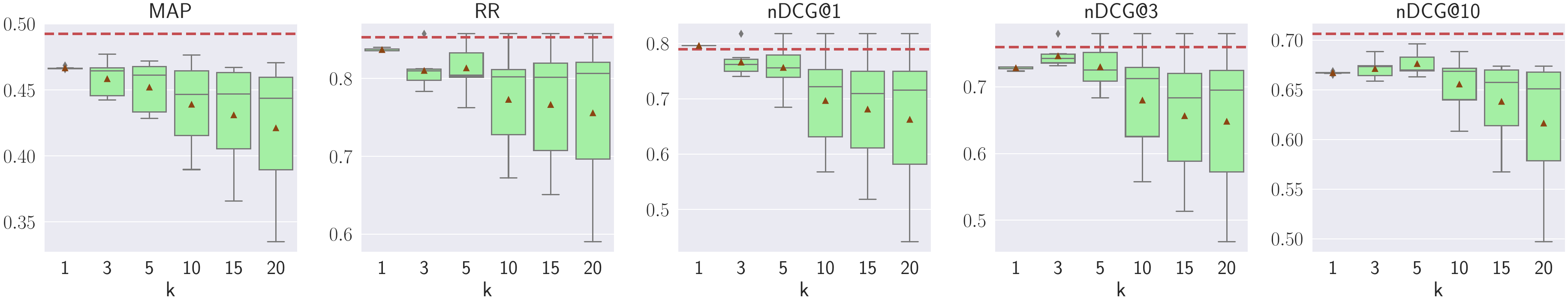}
 	\caption{TREC DL 2020}
	\label{fig:text-prf-depths-trec-2020}
	\end{subfigure}
	\begin{subfigure}{\columnwidth}
	\centering
	\includegraphics[width=\columnwidth]{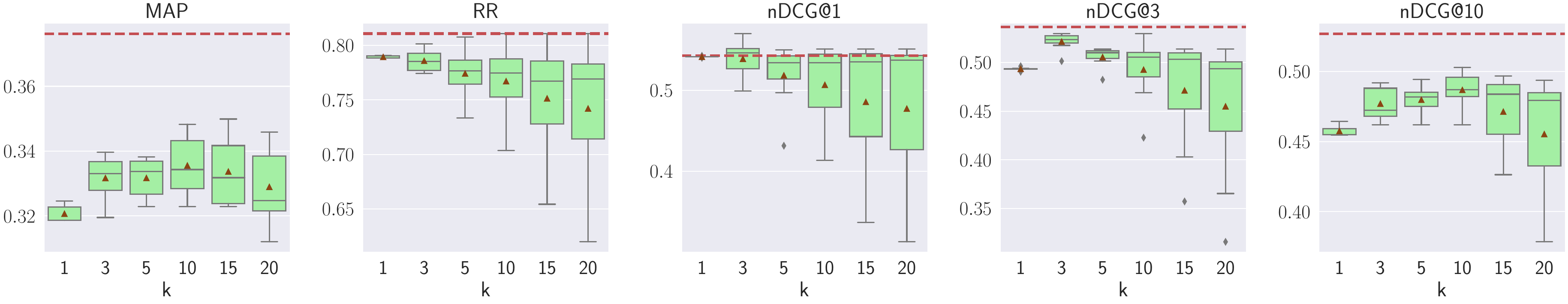}
 	\caption{TREC CAsT}
	\label{fig:text-prf-depths-trec-cast}
	\end{subfigure}
	\begin{subfigure}{\columnwidth}
	\centering
	\includegraphics[width=\columnwidth]{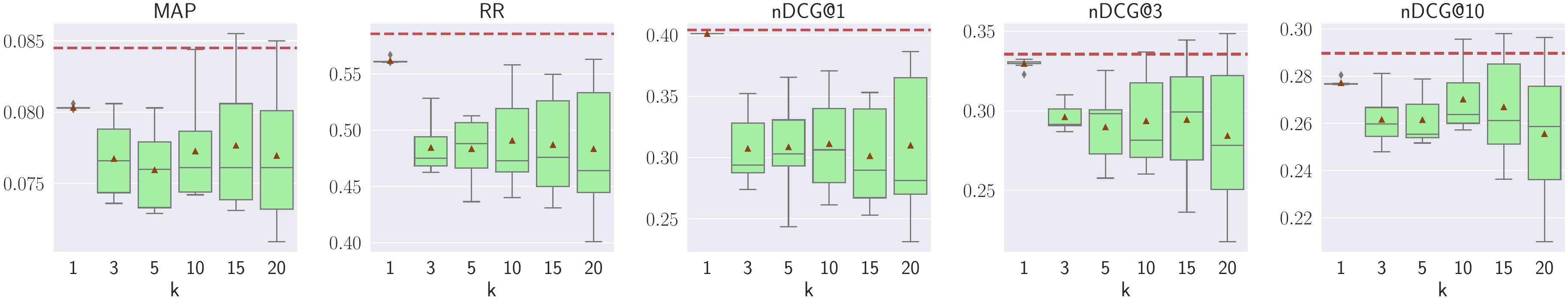}
 	\caption{WebAP}
	\label{fig:text-prf-depths-webap}
	\end{subfigure}
	\begin{subfigure}{\columnwidth}
	\centering
	\includegraphics[width=\columnwidth]{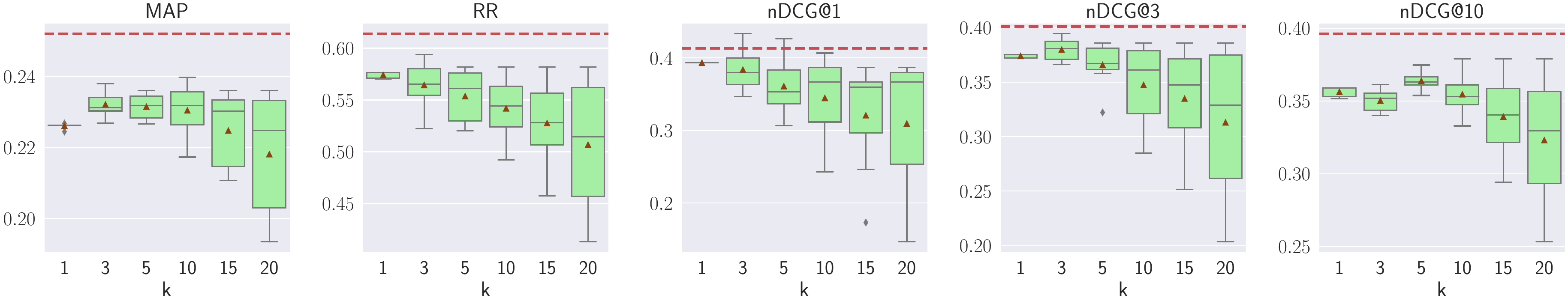}
 	\caption{DL HARD}
	\label{fig:text-prf-depths-dl-hard}
	\end{subfigure}
	\caption{Impact of PRF depth on the effectiveness (y-axis) of BM25+BERT+PRF(BB+PRF) for the task of reranking, $k$ represents different PRF depths. Baseline BM25+BERT(BB) is marked with a dashed red line. PRF depth impacts the effectiveness of text-based reranking models negatively.}
	\label{fig:text-base-prf-depth}
\end{figure}

\begin{figure}
	\begin{subfigure}{\columnwidth}
	\centering
	\includegraphics[width=\linewidth]{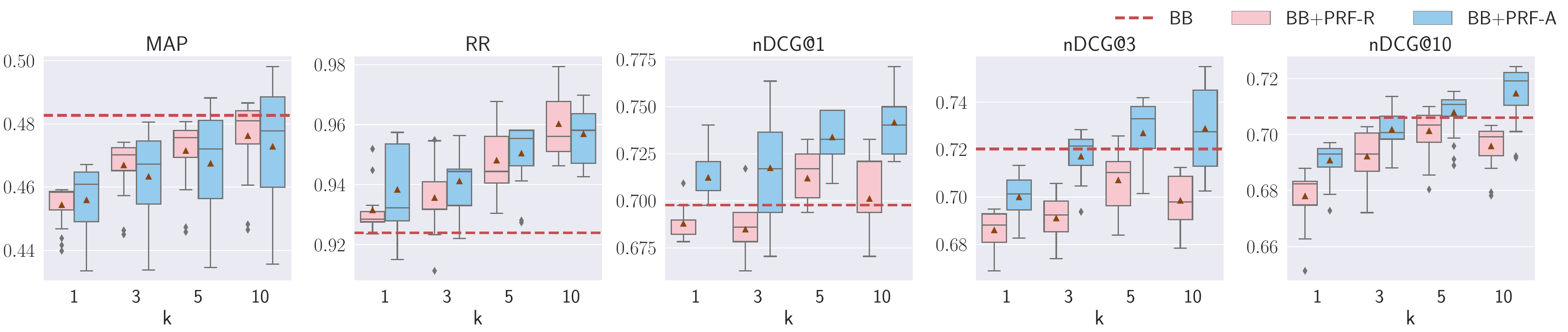}
	\caption{TREC DL 2019}
	\label{fig:repbert-rerank-prf-depth-trec-2019}
	\end{subfigure}
	\begin{subfigure}{\columnwidth}
	\centering
	\includegraphics[width=\linewidth]{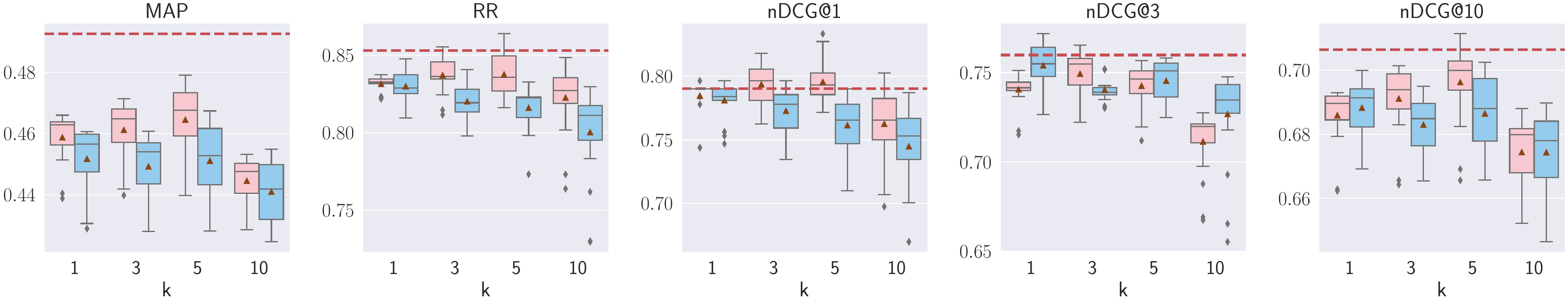}
	\caption{TREC DL 2020}
	\label{fig:repbert-rerank-prf-depth-trec-2020}
	\end{subfigure}
	\begin{subfigure}{\columnwidth}
	\centering
	\includegraphics[width=\linewidth]{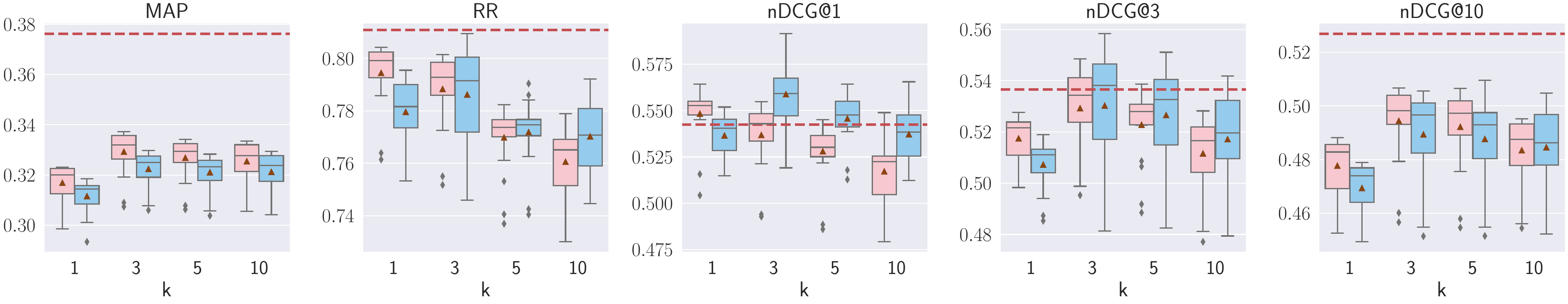}
	\caption{TREC CAsT}
	\label{fig:repbert-rerank-prf-depth-trec-cast}
	\end{subfigure}
	\begin{subfigure}{\columnwidth}
	\centering
	\includegraphics[width=\linewidth]{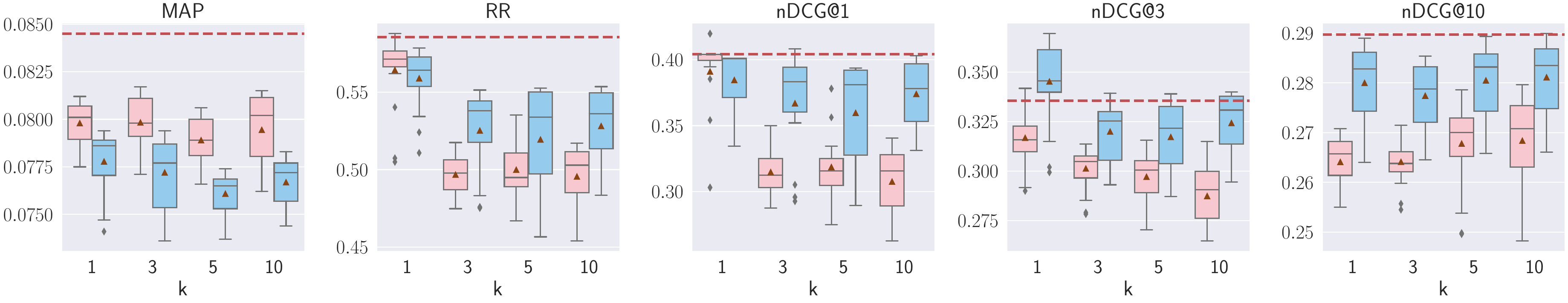}
	\caption{WebAP}
	\label{fig:repbert-rerank-prf-depth-webap}
	\end{subfigure}
	\begin{subfigure}{\columnwidth}
	\centering
	\includegraphics[width=\linewidth]{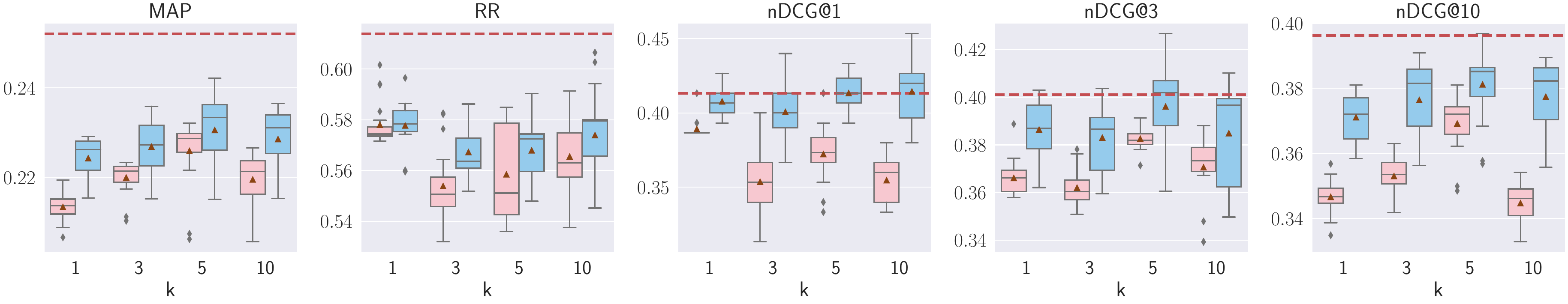}
	\caption{DL HARD}
	\label{fig:repbert-rerank-prf-depth-dl-hard}
	\end{subfigure}
	\caption{Impact of PRF depth on the effectiveness (y-axis) of BM25+BERT+PRF-RepBERT(BB+PRF-R) and BM25+BERT+PRF-ANCE(BB+PRF-A) for the task of reranking, $k$ represents the different PRF depths. Baseline BM25+BERT(BB) is marked with a dashed red line. Increasing PRF depth tends to enhance the effectiveness of hybrid models over shallow metrics (RR, nDCG@\{1,3\}) for reranking.
}
	\label{fig:vector-base-prf-depth-rerank}
\end{figure}

\begin{figure}
	\begin{subfigure}{\columnwidth}
	\centering
	\includegraphics[width=\linewidth]{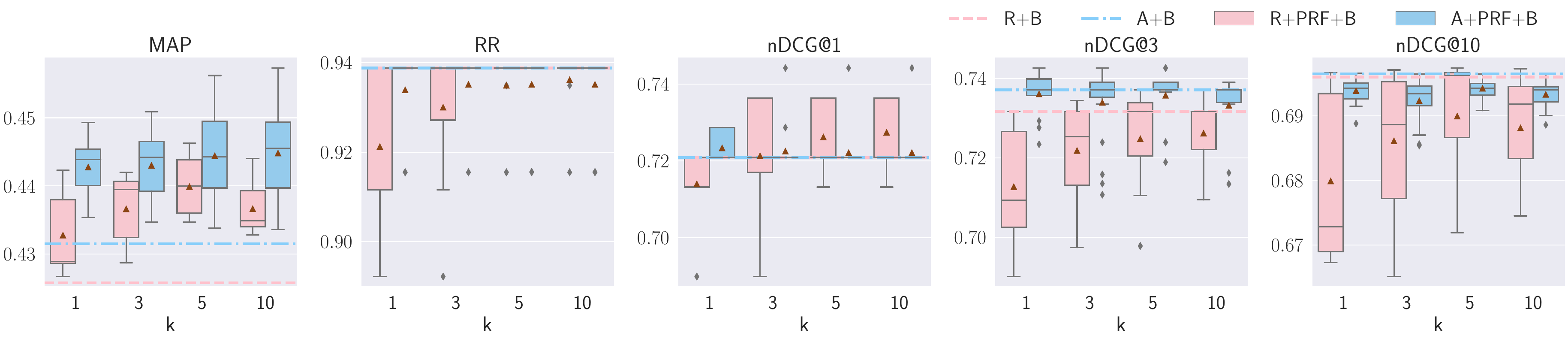}
	\caption{TREC DL 2019}
	\label{fig:vprf_bb-prf-depth-trec-2019}
	\end{subfigure}
	\begin{subfigure}{\columnwidth}
	\centering
	\includegraphics[width=\linewidth]{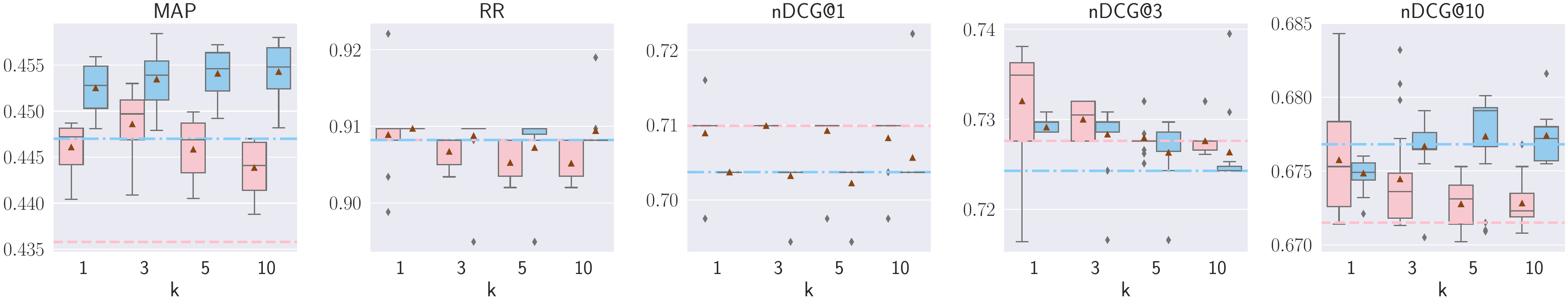}
	\caption{TREC DL 2020}
	\label{fig:vprf_bb-prf-depth-trec-2020}
	\end{subfigure}
	\begin{subfigure}{\columnwidth}
	\centering
	\includegraphics[width=\linewidth]{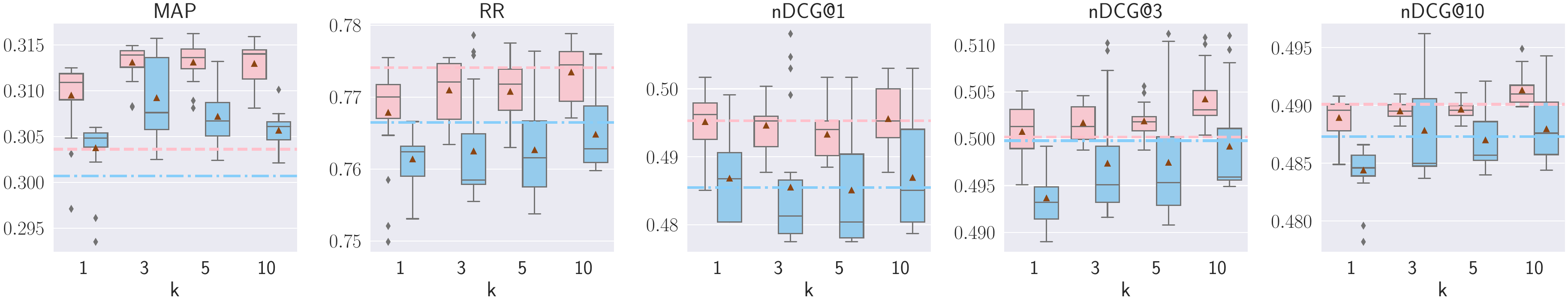}
	\caption{TREC CAsT}
	\label{fig:vprf_bb-prf-depth-trec-cast}
	\end{subfigure}
	\begin{subfigure}{\columnwidth}
	\centering
	\includegraphics[width=\linewidth]{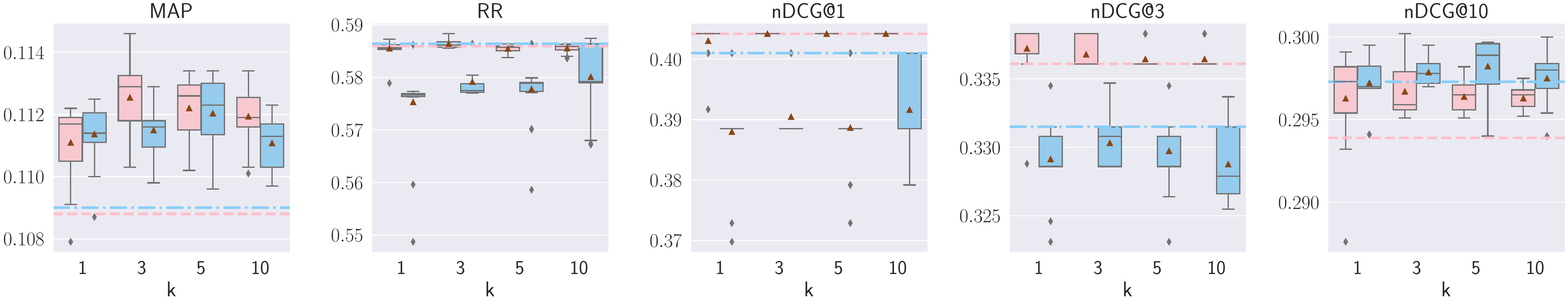}
	\caption{WebAP}
	\label{fig:vprf_bb-prf-depth-webap}
	\end{subfigure}
	\begin{subfigure}{\columnwidth}
	\centering
	\includegraphics[width=\linewidth]{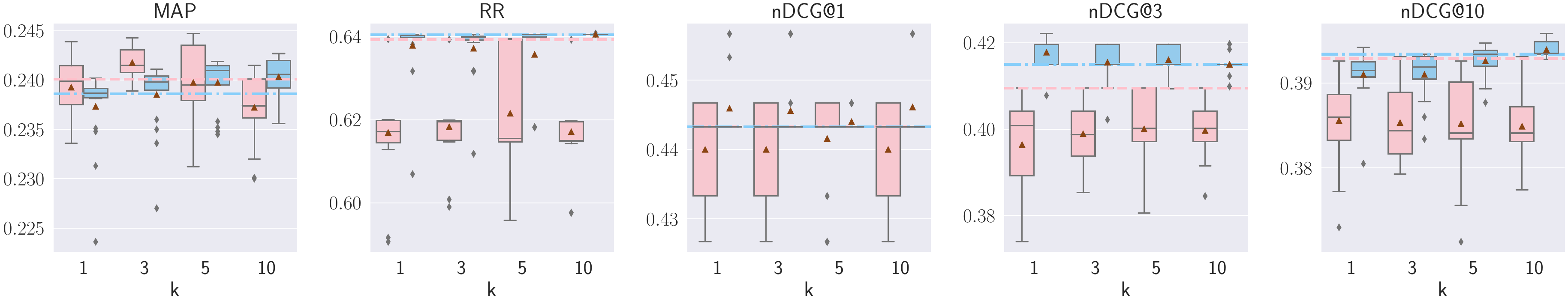}
	\caption{DL HARD}
	\label{fig:vprf_bb-prf-depth-dl-hard}
	\end{subfigure}
	\caption{Impact of PRF depth on the effectiveness (y-axis) of ANCE+PRF+BERT(A+PRF+B) and RepBERT+PRF+BERT(R+PRF+B) for the task of reranking, $k$ represents the different PRF depths. Baseline ANCE+BERT(A+B) and RepBERT+BERT(R+B) are marked with a dash-dot blue line and a dashed red line respectively. Vector-based PRF with BERT reranker does not seem to improve the metrics significantly except MAP, across datasets and PRF depths.
}
	\label{fig:vprf_bb-prf-depth-rerank}
\end{figure}

\subsubsection{Reranking with Different PRF Depths}

Results of \emph{text-based} PRF (BB+PRF) for reranking are shown in Figure~\ref{fig:text-base-prf-depth}. For TREC DL 2019, increased PRF depth is associated with a marginal improvement in effectiveness across most of the evaluation metrics, except for nDCG@10 and, to a minor extent, nDCG@3. On the other hand, increasing PRF depth decreases the effectiveness across the remaining datasets, and none of the PRF configurations is substantially better than the BB baseline. 

Results of \emph{hybrid} PRF models (BB+PRF-R and BB+PRF-A) are shown in Figure~\ref{fig:vector-base-prf-depth-rerank}. For TREC DL 2019, increased PRF depth is associated with substantial improvements in RR and nDCG@1 using both BB+PRF-R and BB+PRF-A, and marginal improvements in nDCG at depths 3--10. For TREC DL 2020 and TREC CAsT, increased PRF depth is associated with marginal improvements in nDCG@\{1,3\}. For the remaining datasets, increased PRF depth shows mixed results, but overall it appears to decrease the effectiveness over all metrics. In addition, we report the results of \emph{hybrid} PRF dense retrievers with a BERT reranker (R+PRF+B and A+PRF+B) in Figure~\ref{fig:vprf_bb-prf-depth-rerank}. We observer that vector-based PRF models with BERT reranker either hurts RR or marginally improves it across all datasets with all PRF depths. On the other hand, all PRF approaches improve MAP across all datasets with PRF depths 3--5, with ANCE-based slightly better than RepBERT-based on TREC DL 2019, TREC DL 2020, RepBERT-based is slightly better than ANCE-based on TREC CAsT. Other datasets show similar effectiveness between these two over different PRF depths. For all other metrics, most of the highest effectivenesses are achieved with PRF depths 3--5, although the improvements are mostly marginal.


\begin{figure}
	\begin{subfigure}{\columnwidth}
	\centering
	\includegraphics[width=\linewidth]{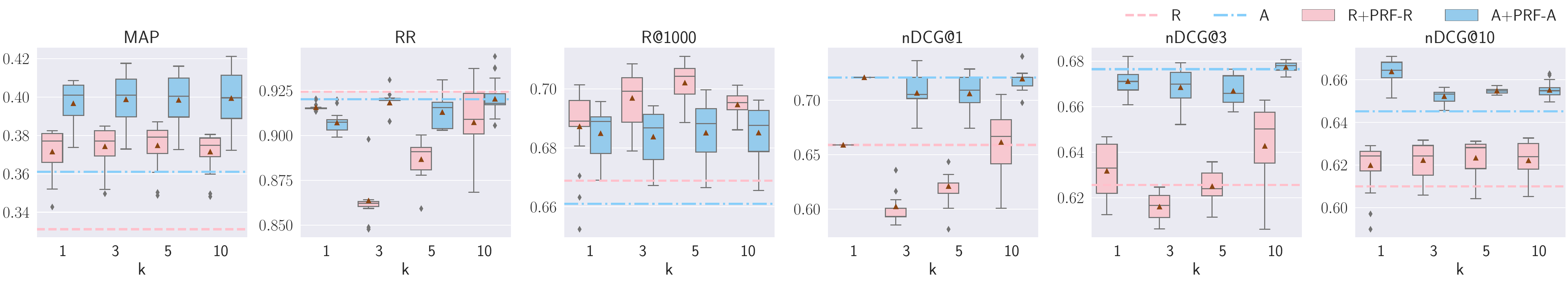}
	\caption{TREC DL 2019}
	\label{fig:repbert-prf-depth-trec-2019}
	\end{subfigure}
	\begin{subfigure}{\columnwidth}
	\centering
	\includegraphics[width=\linewidth]{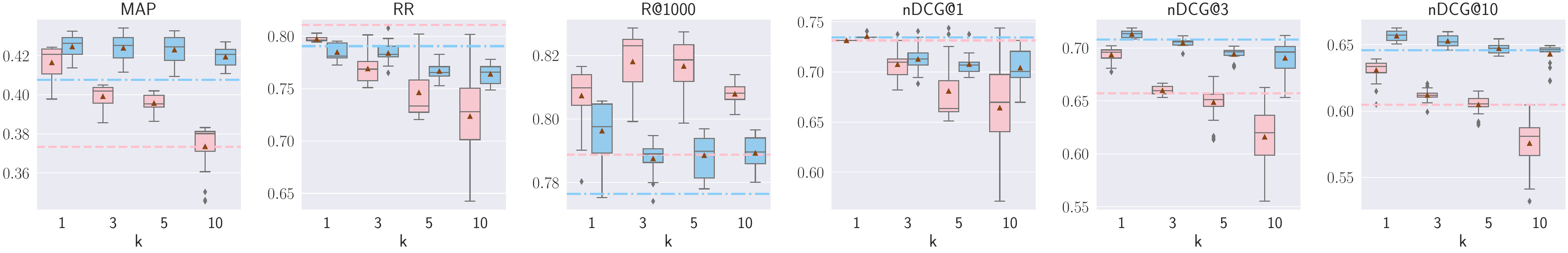}
	\caption{TREC DL 2020}
	\label{fig:repbert-prf-depth-trec-2020}
	\end{subfigure}
	\begin{subfigure}{\columnwidth}
	\centering
	\includegraphics[width=\linewidth]{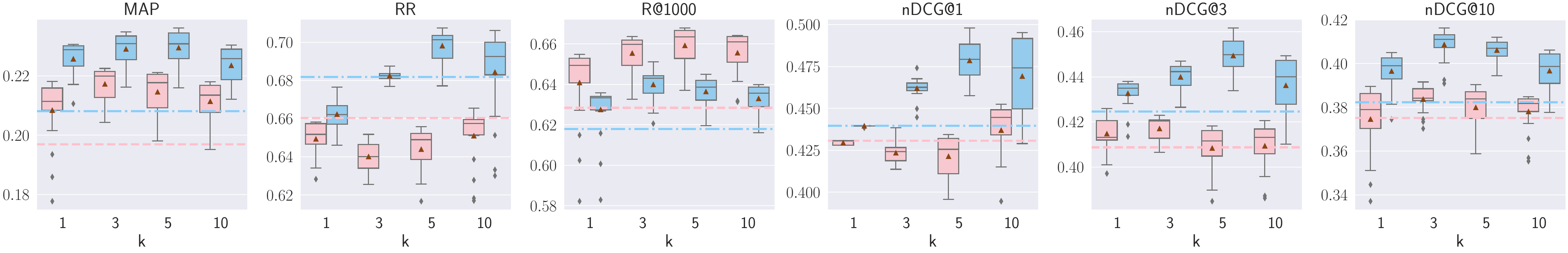}
	\caption{TREC CAsT}
	\label{fig:repbert-prf-depth-trec-cast}
	\end{subfigure}
	\begin{subfigure}{\columnwidth}
	\centering
	\includegraphics[width=\linewidth]{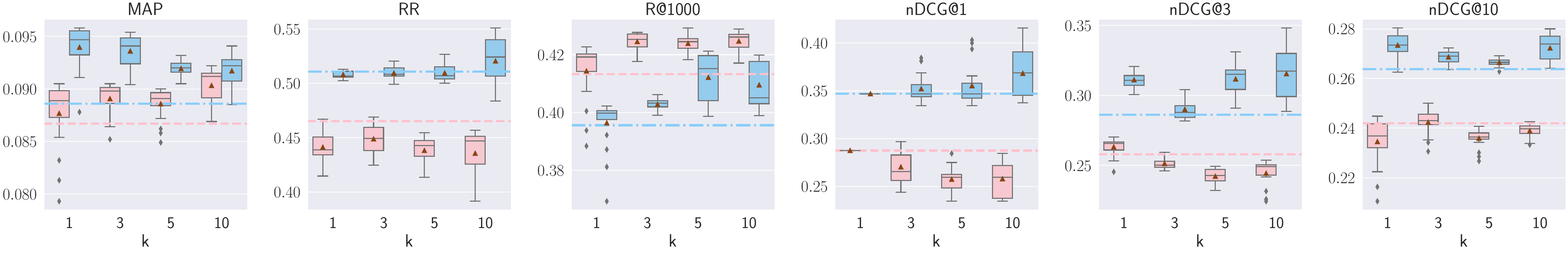}
	\caption{WebAP}
	\label{fig:repbert-prf-depth-webap}
	\end{subfigure}
	\begin{subfigure}{\columnwidth}
	\centering
	\includegraphics[width=\linewidth]{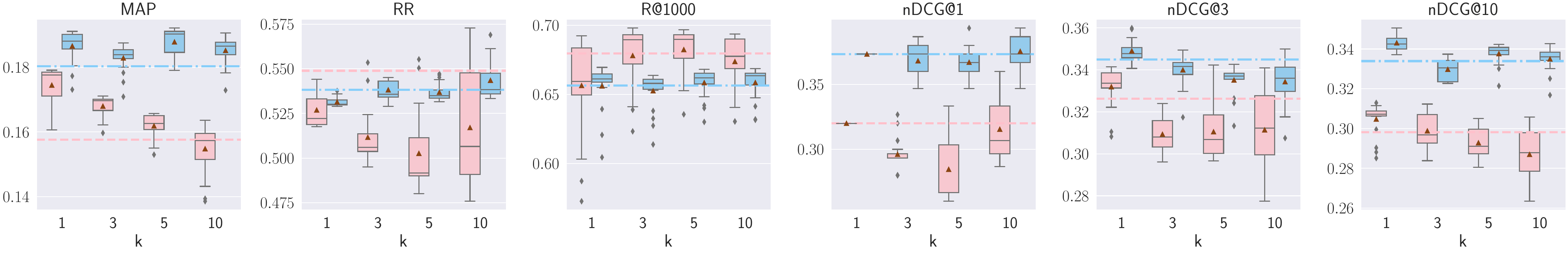}
	\caption{DL HARD}
	\label{fig:repbert-prf-depth-dl-hard}
	\end{subfigure}
	\caption{Impact of PRF depth on the effectiveness (y-axis) of RepBERT+PRF-RepBERT(R+PRF-R) and ANCE+PRF-ANCE(A+PRF-A) for the task of retrieval, $k$ represents different PRF depths. Baseline RepBERT(R) is marked with a dashed red line, ANCE(A) is marked with a dash-dot blue line. Increasing PRF depth tends to enhance the effectiveness over deep metrics (R@1000, nDCG@10 and MAP) for retrieval.}
	\label{fig:vector-base-prf-depth}
\end{figure}

\subsubsection{Retrieval with Different PRF Depths}

Results of \emph{vector-based} PRF (R+PRF-R and A+PRF-A) for retrieval are shown in Figure~\ref{fig:vector-base-prf-depth}. For deep evaluation metrics (MAP, nDCG@10 and R@1000), increased PRF depth is associated with significant improvements in effectiveness over the baseline dense retrievers across all datasets, with few exceptions for DL HARD. Increased PRF depth is associated with decreased RR values across all datasets, with few exceptions for A+PRF-A where PRF at depth of 10 is on par or marginally better. For shallow metrics such as nDCG@\{1, 3\}, mixed impact across datasets is witnessed with respect to changing PRF depths. For TREC DL 2019 and 2020, PRF of depth 1 is on par with the baselines. For TREC CAsT and WebAP, increased PRF depth is associated with significant increases  of effectiveness of A+PRF-A, while PRF of depth 1 enhances the effectiveness of R+PRF-R. For DL Hard, all PRF depths perform on par with the ANCE(A) baseline, while PRF of depth 1 performs on par with the RepBERT(R) baseline.

\subsubsection{Summary}

To summarize, increasing PRF depth tends to enhance the effectiveness of hybrid models over shallow metrics (RR, nDCG@\{1,3\}) for reranking, and deep metrics (R@1000, nDCG@10 and MAP) for retrieval. On the other hand, PRF depth negatively impacts the effectiveness of text-based reranking models. Vector-based PRF with BERT reranker does not seem to improve the metrics significantly except MAP, across datasets and PRF depths.

\subsection{Text Handling}

\textbf{RQ2: What is the impact of text handling techniques on the effectiveness of reranking and retrieval?} To answer this question, we vary the text handling techniques while displaying the distribution of results over other parameters (PRF depth and score estimation). We analyze the effectiveness under three text handling techniques: Concatenate and Truncate (CT), Concatenate and Aggregation (CA), and Sliding Window (SW); and two dense representations for text: RepBERT(R+PRF-R) and ANCE(A+PRF-A).

\begin{figure}
	\begin{subfigure}{0.87\columnwidth}
	\centering
	\includegraphics[width=\linewidth]{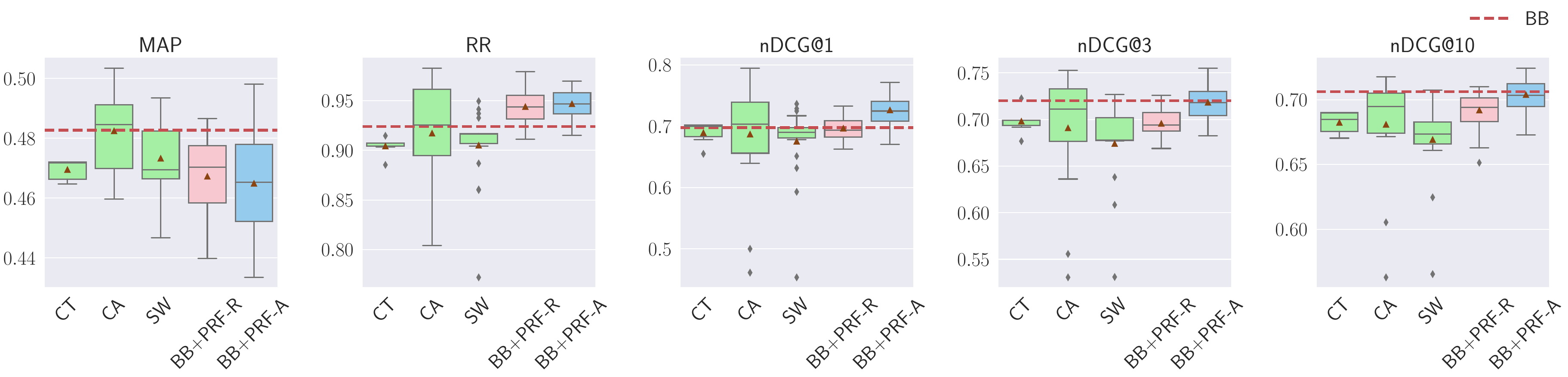}
	\caption{TREC DL 2019}
	\label{fig:diff-rep-trec-2019}
	\end{subfigure}
	\begin{subfigure}{0.87\columnwidth}
	\centering
	\includegraphics[width=\linewidth]{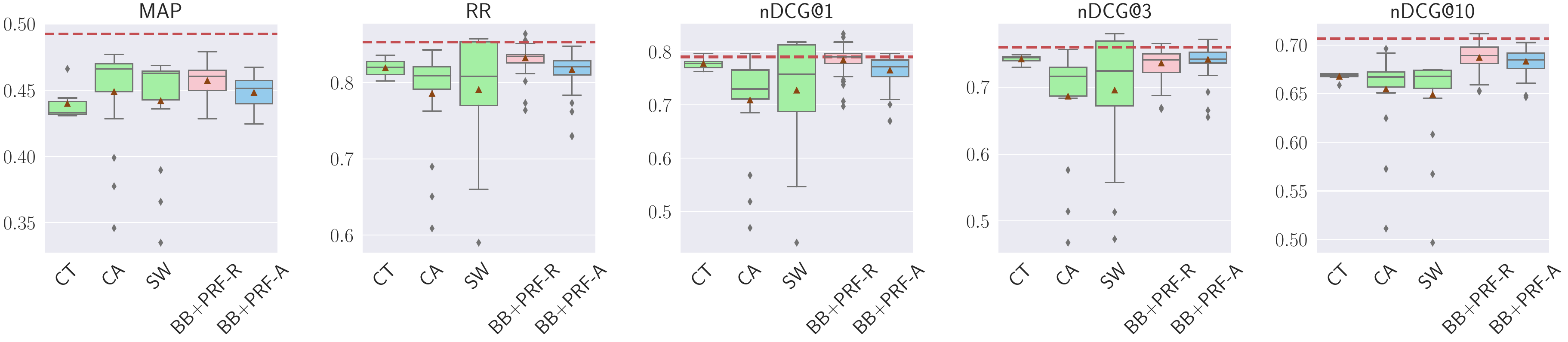}
	\caption{TREC DL 2020}
	\label{fig:diff-rep-trec-2020}
	\end{subfigure}
	\begin{subfigure}{0.87\columnwidth}
	\centering
	\includegraphics[width=\linewidth]{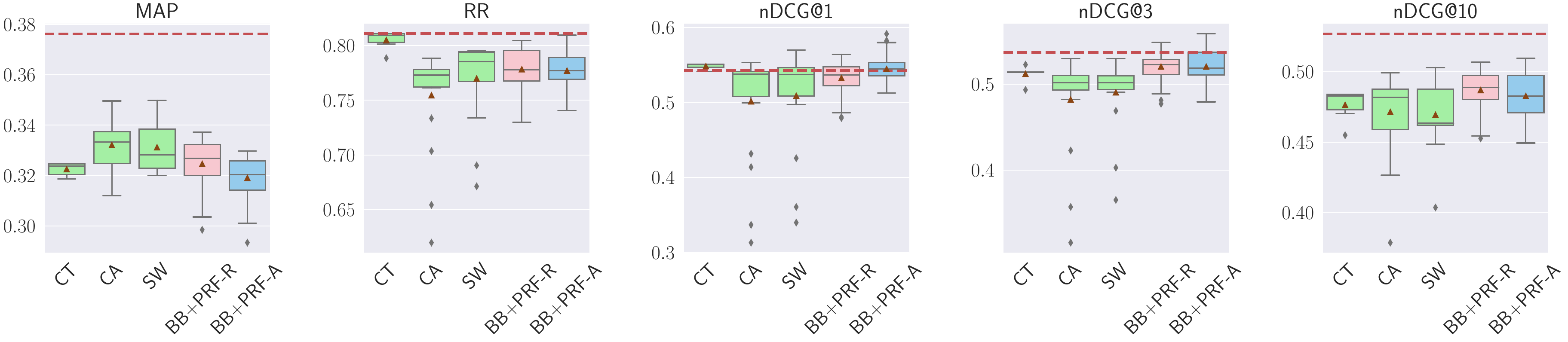}
	\caption{TREC CAsT}
	\label{fig:diff-rep-trec-cast}
	\end{subfigure}
	\begin{subfigure}{0.87\columnwidth}
	\centering
	\includegraphics[width=\linewidth]{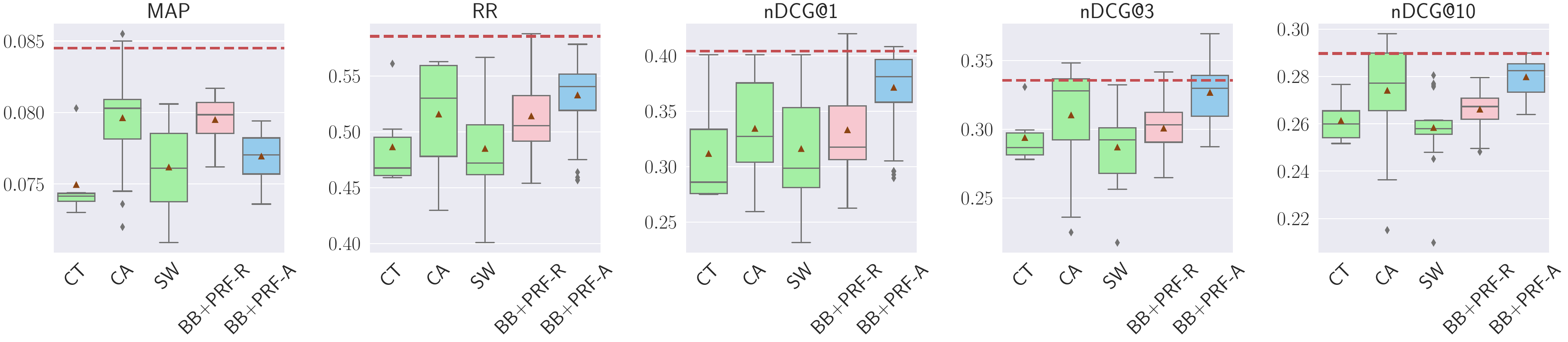}
	\caption{WebAP}
	\label{fig:diff-rep-webap}
	\end{subfigure}
	\begin{subfigure}{0.87\columnwidth}
	\centering
	\includegraphics[width=\linewidth]{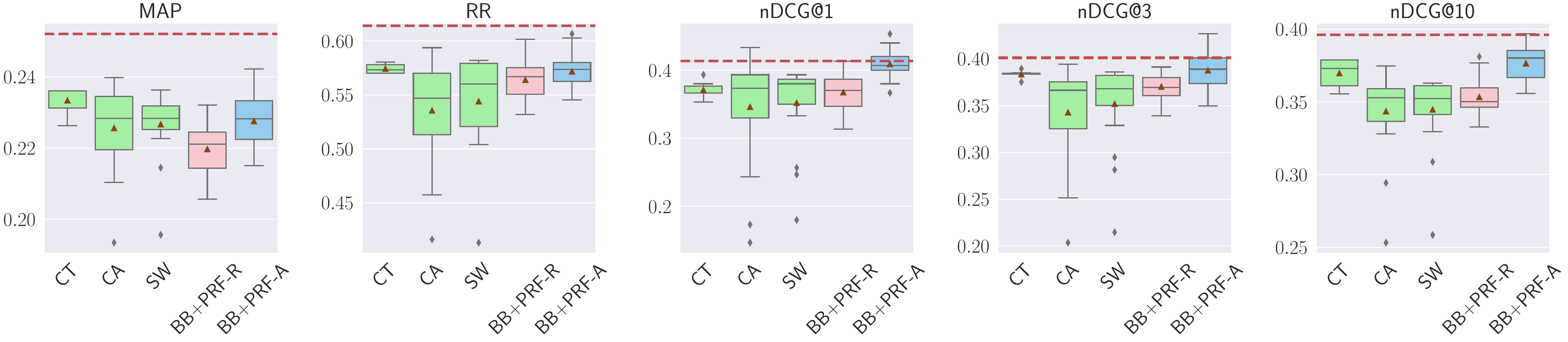}
	\caption{DL HARD}
	\label{fig:diff-rep-dl-hard}
	\end{subfigure}
	\caption{Impact of text handling on the effectiveness (y-axis) of PRF approaches for the task of reranking, where CT, CA and SW represent the text handling methods Concatenate and Truncate, Concatenate and Aggregate and Sliding Window, respectively, while BM25+BERT+PRF-RepBERT(BB+PRF-R) and BM25+BERT+PRF-ANCE(BB+PRF-A) are the dense representations for text. Baseline BM25+BERT(BB) is marked with a dashed red line. CA tends to improve more on MAP, but all other improvements are marginal and some significant losses can be observed.}
	\label{fig:rerank-diff-rep}
\end{figure}

\begin{figure}
	\begin{subfigure}{0.9\columnwidth}
	\centering
	\includegraphics[width=\linewidth]{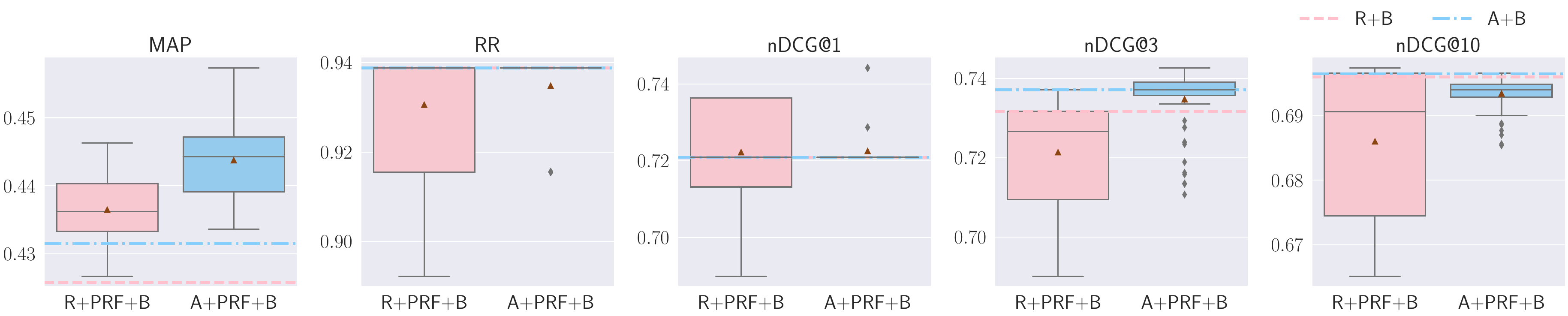}
	\caption{TREC DL 2019}
	\label{fig:vprf_bb-rep-trec-2019}
	\end{subfigure}
	\begin{subfigure}{0.9\columnwidth}
	\centering
	\includegraphics[width=\linewidth]{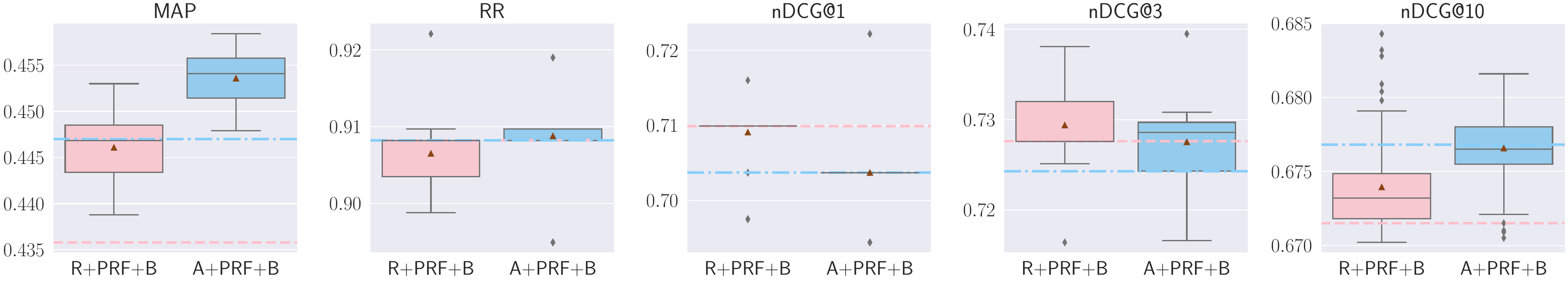}
	\caption{TREC DL 2020}
	\label{fig:vprf_bb-rep-trec-2020}
	\end{subfigure}
	\begin{subfigure}{0.9\columnwidth}
	\centering
	\includegraphics[width=\linewidth]{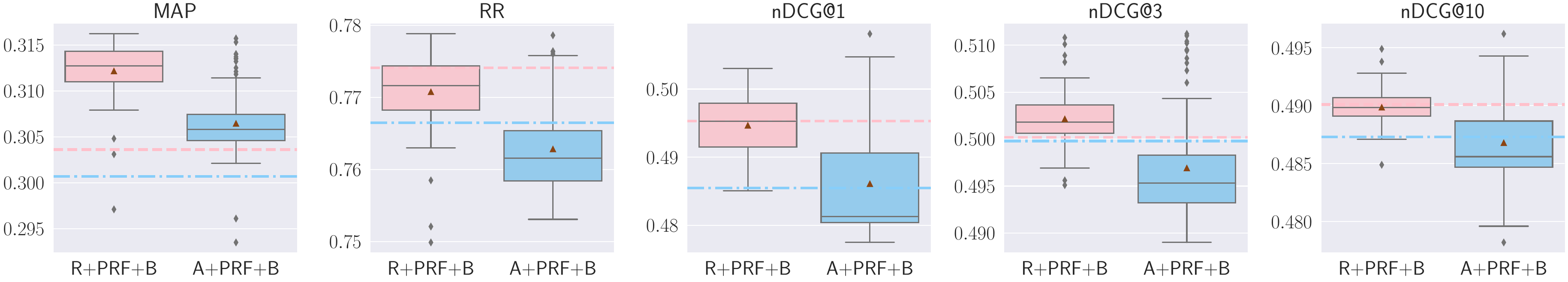}
	\caption{TREC CAsT}
	\label{fig:vprf_bb-rep-trec-cast}
	\end{subfigure}
	\begin{subfigure}{0.9\columnwidth}
	\centering
	\includegraphics[width=\linewidth]{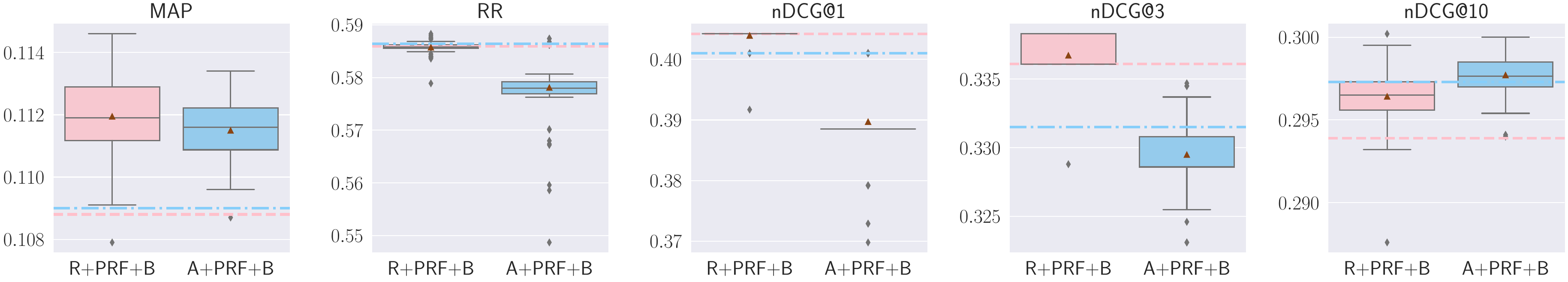}
	\caption{WebAP}
	\label{fig:vprf_bb-rep-webap}
	\end{subfigure}
	\begin{subfigure}{0.9\columnwidth}
	\centering
	\includegraphics[width=\linewidth]{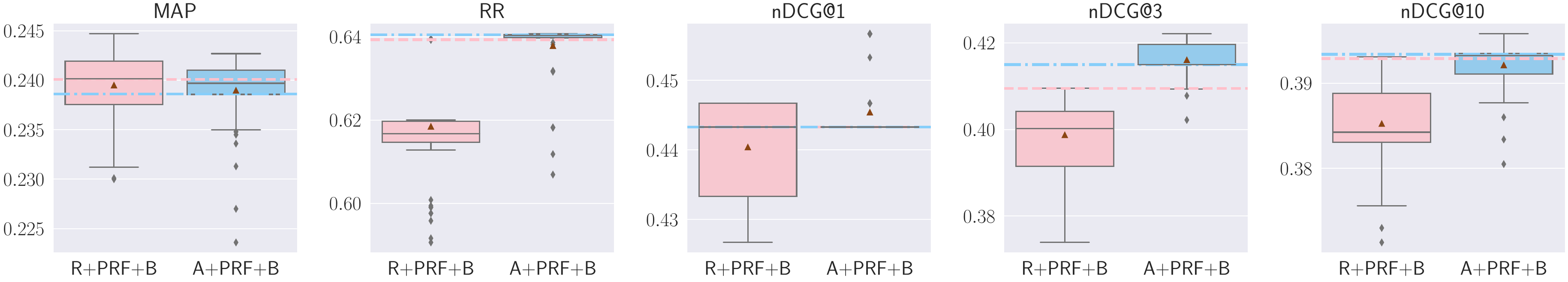}
	\caption{DL HARD}
	\label{fig:vprf_bb-rep-dl-hard}
	\end{subfigure}
	\caption{Impact of dense representations on the effectiveness (y-axis) of PRF approaches for the task of reranking, where R+PRF+B and A+PRF+B represents RepBERT+PRF+BERT and ANCE+PRF+BERT, respectively. Baseline ANCE+BERT(A+B) and RepBERT+BERT(R+B) are marked with a dash-dot blue line and a dashed red line respectively. When applying the BERT reranker after vector-based PRF, both ANCE-based and RepBERT-based improve MAP on all datasets, but there are no improvements nor losses on the remaining metrics across all datasets.}
	\label{fig:vprf_bb-diff-rep}
\end{figure}

\subsubsection{Reranking with Different Text Handling}

Results are shown in Figure~\ref{fig:rerank-diff-rep}. For TREC DL 2019, CA substantially improves MAP, RR, and nDCG@1, and marginally improves nDCG@3. BB+PRF-A and BB+PRF-R substantially improve RR, while BB+PRF-A also substantially improves nDCG@1. On the other hand, BB+PRF-R is on par with the baseline over $nDCG@1$, and BB+PRF-A is on par with nDCG@\{3, 10\}. All other methods do not improve effectiveness.
For TREC DL 2020, SW marginally improves nDCG@\{1, 3\}. BB+PRF-R is on par with the baseline for nDCG@1. All other methods do not improve over the baseline, and all methods, including SW and BB+PRF-R, hurt MAP.

For TREC CAsT 2019, unlike the previous datasets, no improvements can be observed for MAP and nDCG@10 across all  methods. CT is on par with the baseline in terms of RR, and marginal improvements are present for nDCG@1. BB+PRF-A is on par with the baseline for nDCG@1. All other metrics are not improved when employing different text handling methods.

For WebAP, no substantial improvements are found, regardless of the metric, with the exception of nDCG@3, for which BB+PRF-A is on par with the baseline.

For DL HARD, all methods hurt MAP and RR. BB+PRF-A is on par with the baseline for nDCG@1. No substantial improvements on other metrics can be observed for the remaining methods.

The results for \emph{vector-based} PRF models with BERT reranker are shown in Figure~\ref{fig:vprf_bb-diff-rep}. For TREC DL 2019, TREC DL 2020, ANCE-based is better than RepBERT-based PRF models over MAP, RR, nDCG@10. For other datasets except DL HARD, RepBERT-based is better than ANCE-based w.r.t all metrics except nDCG@10. However, the improvements only occur with MAP on all datasets, although marginal on DL HARD. Both ANCE-based and RepBERT-based either hurts or on par with baseline on all other metrics across all datasets.

\begin{figure}
	\begin{subfigure}{\columnwidth}
	\centering
	\includegraphics[width=\linewidth]{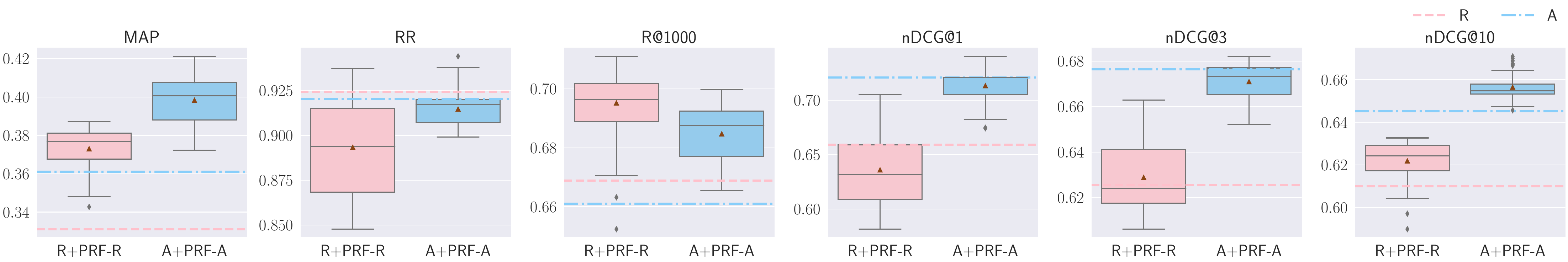}
	\caption{TREC DL 2019}
	\label{fig:vector-diff-rep-trec-2019}
	\end{subfigure}
	\begin{subfigure}{\columnwidth}
	\centering
	\includegraphics[width=\linewidth]{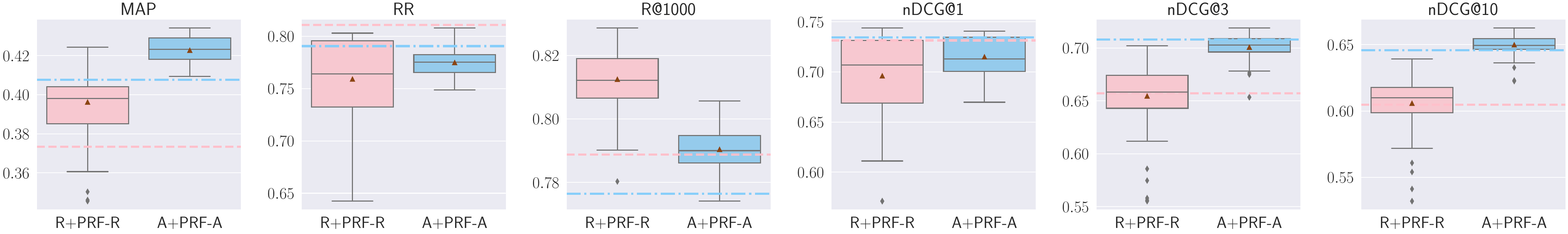}
	\caption{TREC DL 2020}
	\label{fig:vector-diff-rep-trec-2020}
	\end{subfigure}
	\begin{subfigure}{\columnwidth}
	\centering
	\includegraphics[width=\linewidth]{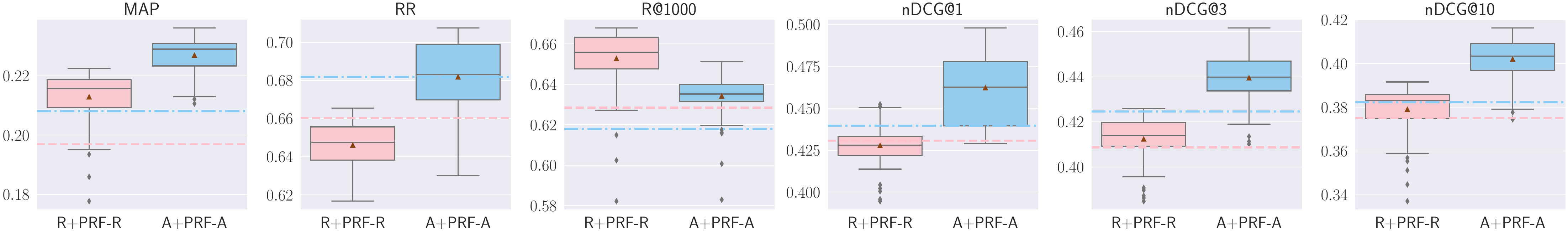}
	\caption{TREC CAsT}
	\label{fig:vector-diff-rep-trec-cast}
	\end{subfigure}
	\begin{subfigure}{\columnwidth}
	\centering
	\includegraphics[width=\linewidth]{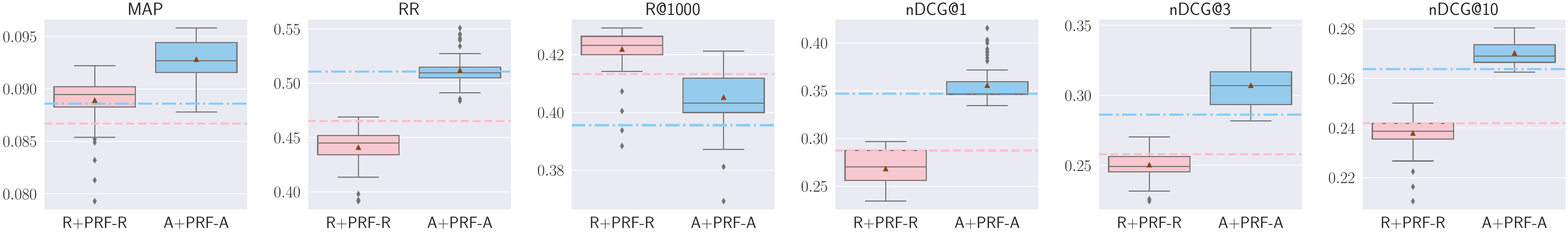}
	\caption{WebAP}
	\label{fig:vector-diff-rep-webap}
	\end{subfigure}
	\begin{subfigure}{\columnwidth}
	\centering
	\includegraphics[width=\linewidth]{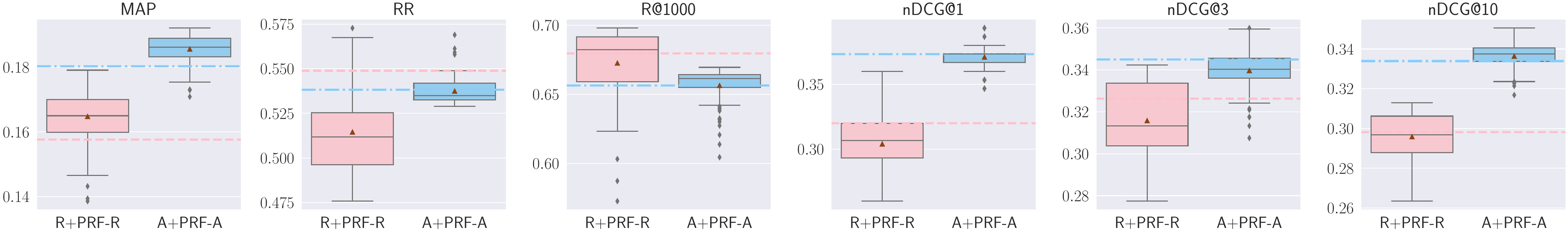}
	\caption{DL HARD}
	\label{fig:vector-diff-rep-dl-hard}
	\end{subfigure}
	\caption{Vector-based PRF retrieval effectiveness (y-axis) by using different dense retrieval models RepBERT(R+PRF-R) and ANCE(A+PRF-A). Baseline RepBERT(R) is marked with dashed red line, ANCE(A) is marked with dash-dot blue line. A+PRF-A is, overall, a better representation, as it improves all metrics and outperforms all baselines. R+PRF-R performs worse than A+PRF-A. This is because RepBERT(R) baseline is worse than ANCE(A) baseline across most metrics, causing the top ranked results to contain less relevant passages compared to A: hence, the PRF mechanism receives a noisier relevance signal from the feedback passages.}
	\label{fig:vector-based-prf-dense-model}
\end{figure}

\subsubsection{Retrieval with Different Text Handling}

Results are shown in Figure~\ref{fig:vector-based-prf-dense-model}. For TREC DL 2019, both methods substantially outperform their respective baselines in terms of MAP, R@1000, and nDCG@10. No improvement can be observed for RR and nDCG@1. A+PRF-A does not outperform the baseline in terms of nDCG@3, but R+PRF-R does.
For TREC DL 2020, both methods A+PRF-A and R+PRF-R substantially improve MAP and R@1000. On the other hand, they do not improve RR and nDCG@1. R+PRF-R is on par with the baseline for nDCG@\{3, 10\}. Marginal improvements can be observed for A+PRF-A in terms of nDCG@10.

For TREC CAsT 2019, both methods substantially improve the baseline in terms of MAP,  R@1000, and nDCG@\{3, 10\}.
Both A+PRF-A and R+PRF-R improve over the baseline in terms of nDCG@1, but A+PRF-A does so substantially; in addition A+PRF-A is on par with the baseline for RR. Both methods do not improve the baselines for other metrics.

For WebAP, similar trends can be observed for MAP, RR, and R@1000. R+PRF-R hurts the effectiveness over nDCG@\{1, 3\}, while A+PRF-A marginally improves nDCG@1 and substantially improves nDCG@3 and nDCG@10.

For DL HARD, both methods substantially improve MAP; R+PRF-R also substantially improves R@1000. A+PRF-A is on par with the baseline for RR, R@1000, nDCG@1, and marginally improves nDCG@10. No improvements are observed for the remaining metrics for either method.

\subsubsection{Summary}

When used for reranking, CA tends to improve more on MAP, BB+PRF-R tends to have more improvements for RR, and BB+PRF-A tends to improve more on nDCG@\{1, 3, 10\}. In general, all methods  tend to improve more nDCG than RR or MAP. When applying the BERT reranker after vector-based PRF, both ANCE-based and RepBERT-based improve MAP on all datasets, but there are no improvements nor losses on the remaining metrics across all datasets.

When used for retrieval, A+PRF-A is, overall, a better representation, as it improves all metrics and outperforms all baselines. R+PRF-R performs worse than A+PRF-A. This is because RepBERT(R) baseline is worse than ANCE(A) baseline across most metrics, causing the top ranked results to contain less relevant passages compared to A: hence, the PRF mechanism receives a noisier relevance signal from the feedback passages.

\subsection{Score Estimation}

\begin{figure}
	\begin{subfigure}{0.99\columnwidth}
	\centering
	\includegraphics[width=\linewidth]{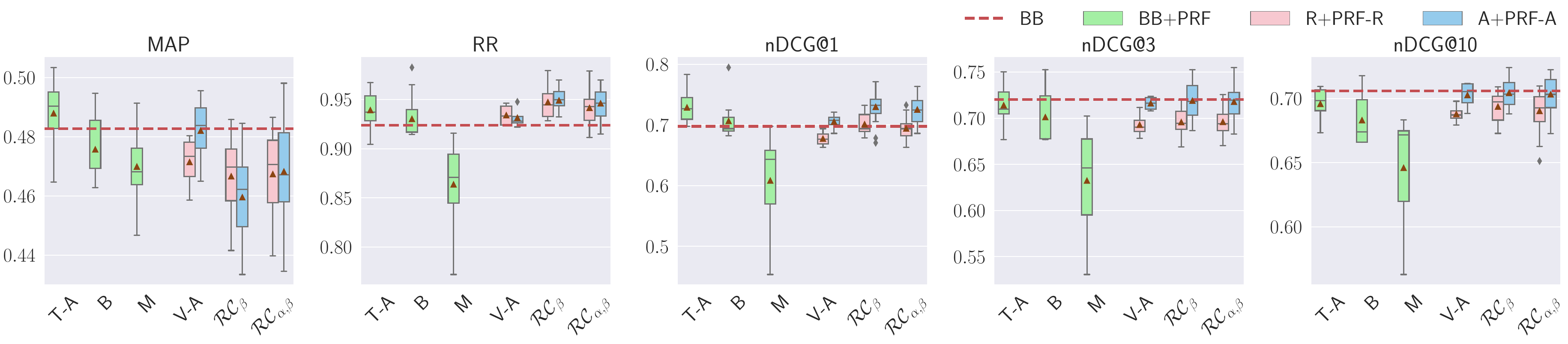}
	\caption{TREC DL 2019}
	\label{fig:diff-fusion-trec-2019}
	\end{subfigure}
	\begin{subfigure}{0.99\columnwidth}
	\centering
	\includegraphics[width=\linewidth]{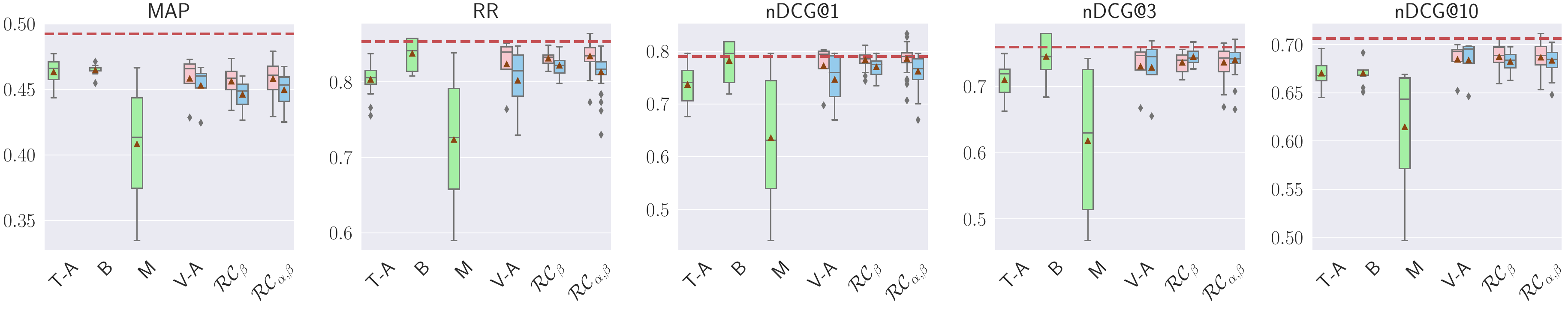}
	\caption{TREC DL 2020}
	\label{fig:diff-fusion-trec-2020}
	\end{subfigure}
	\begin{subfigure}{0.99\columnwidth}
	\centering
	\includegraphics[width=\linewidth]{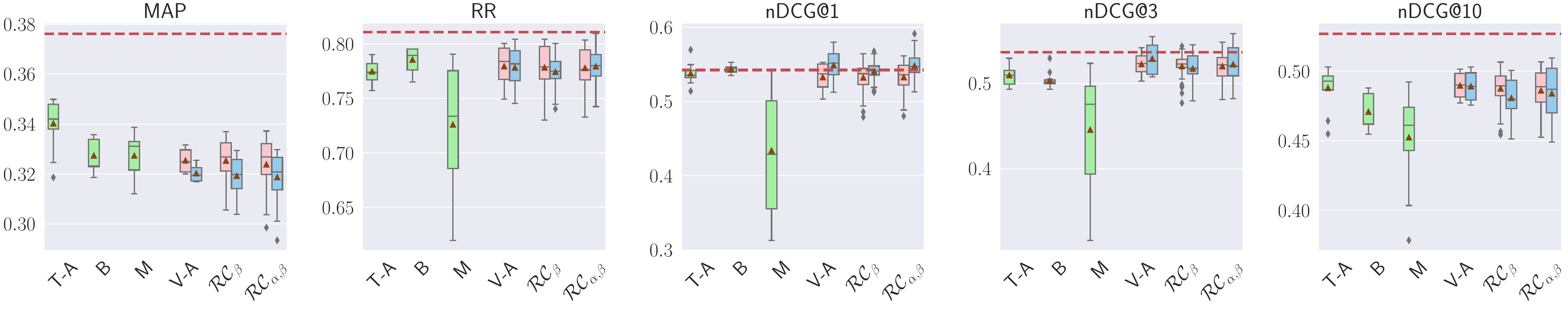}
	\caption{TREC CAsT}
	\label{fig:diff-fusion-trec-cast}
	\end{subfigure}
	\begin{subfigure}{0.99\columnwidth}
	\centering
	\includegraphics[width=\linewidth]{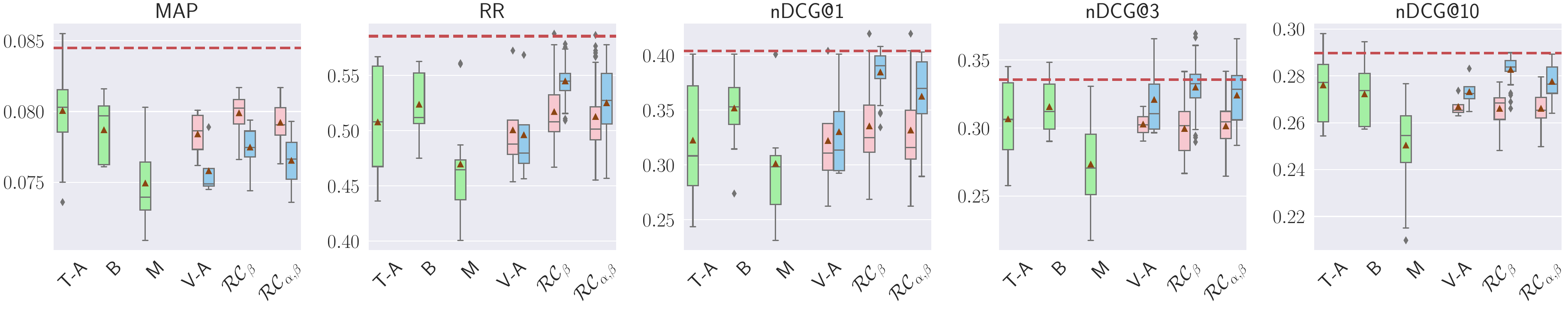}
	\caption{WebAP}
	\label{fig:diff-fusion-webap}
	\end{subfigure}
	\begin{subfigure}{0.99\columnwidth}
	\centering
	\includegraphics[width=\linewidth]{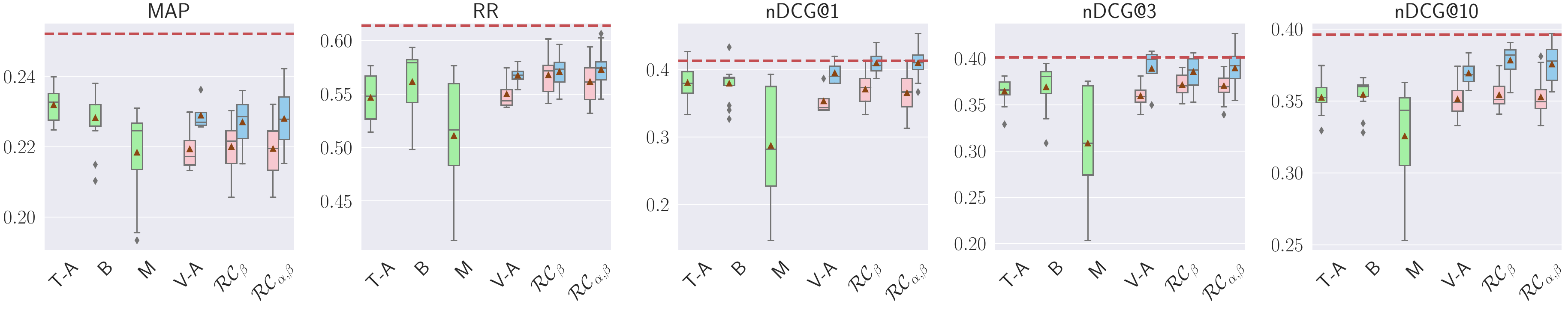}
	\caption{DL HARD}
	\label{fig:diff-fusion-dl-hard}
	\end{subfigure}
	\caption{Reranking effectiveness (y-axis) by using different score estimation methods. Where T-A is Text Average, B is Borda, M is Max, V-A is Vector Average, $\mathcal{RC}_\beta$ is Rocchio with fixed $\alpha$ value, and $\mathcal{RC}_{\alpha,\beta}$ is Rocchio with $\alpha$ and $\beta$. Baseline BM25+BERT(BB) is marked with dashed red line. $\mathcal{RC}_{\alpha,\beta}$ is found to perform considerably well across all the metrics and datasets. B, T-A, and $\mathcal{RC}_\beta$ also perform well across several metrics and all datasets. M performs poorly across all metrics and datasets.}
	\label{fig:rerank-agg-fusion}
\end{figure}

\begin{figure}
	\begin{subfigure}{\columnwidth}
	\centering
	\includegraphics[width=\linewidth]{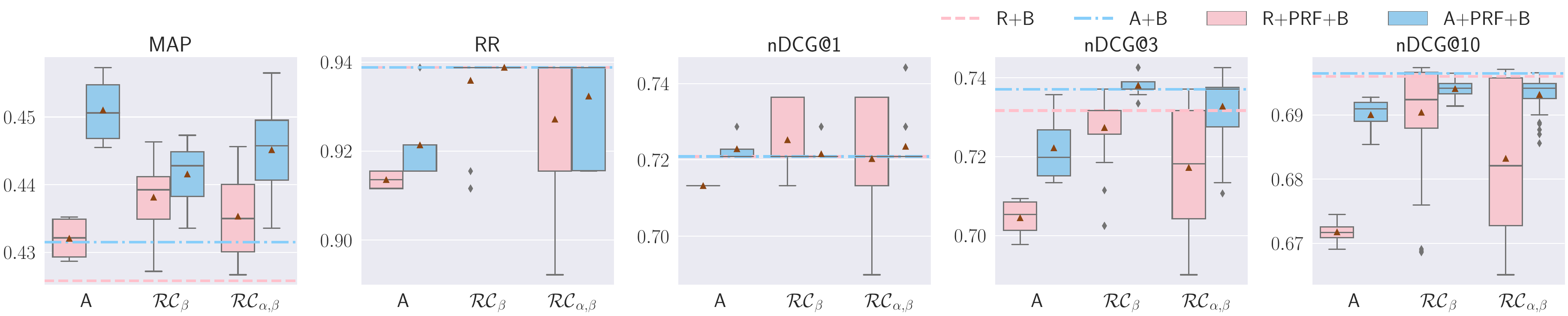}
	\caption{TREC DL 2019}
	\label{fig:vprf_bb-fusion-trec-2019}
	\end{subfigure}
	\begin{subfigure}{\columnwidth}
	\centering
	\includegraphics[width=\linewidth]{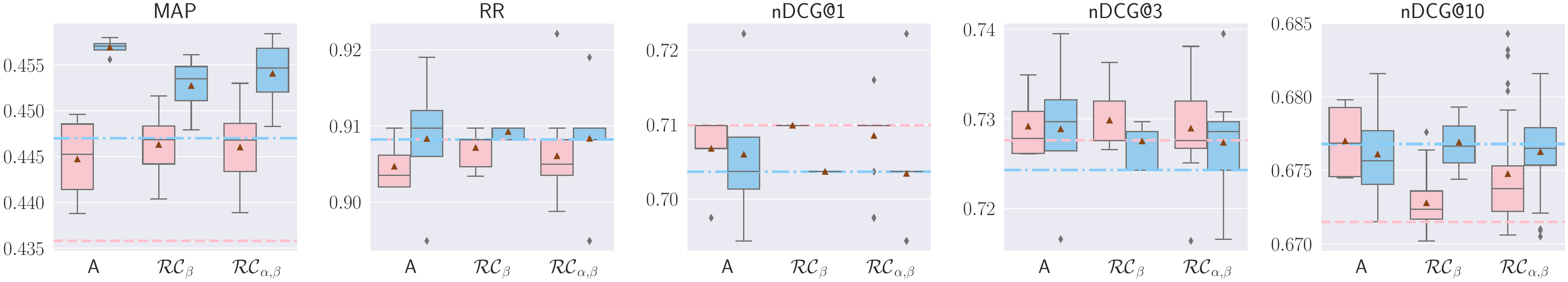}
	\caption{TREC DL 2020}
	\label{fig:vprf_bb-fusion-trec-2020}
	\end{subfigure}
	\begin{subfigure}{\columnwidth}
	\centering
	\includegraphics[width=\linewidth]{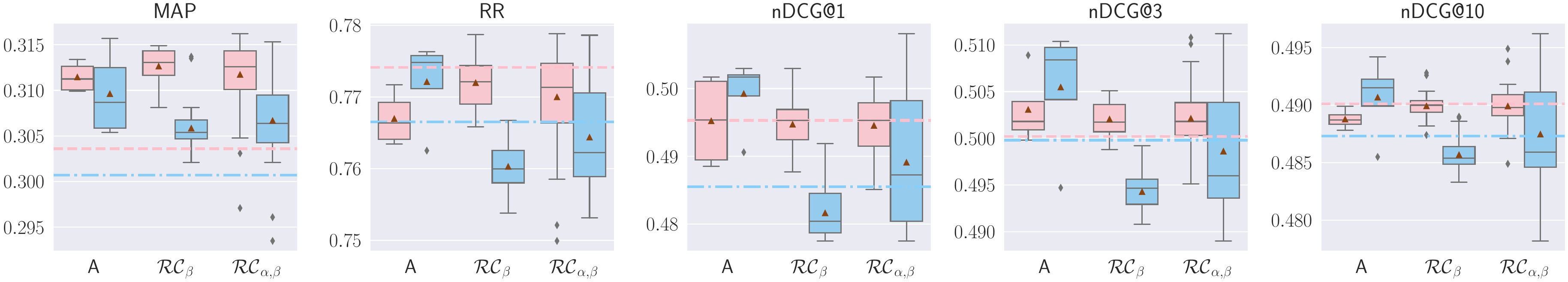}
	\caption{TREC CAsT}
	\label{fig:vprf_bb-fusion-trec-cast}
	\end{subfigure}
	\begin{subfigure}{\columnwidth}
	\centering
	\includegraphics[width=\linewidth]{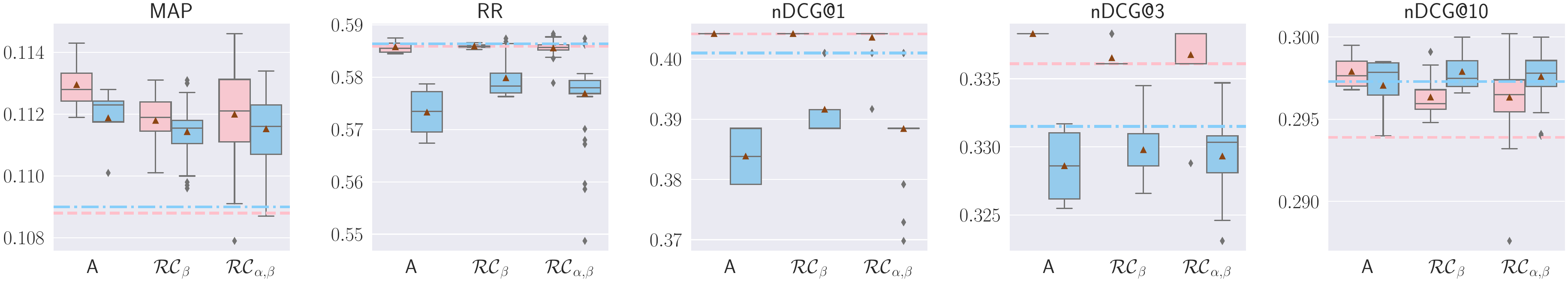}
	\caption{WebAP}
	\label{fig:vprf_bb-fusion-webap}
	\end{subfigure}
	\begin{subfigure}{\columnwidth}
	\centering
	\includegraphics[width=\linewidth]{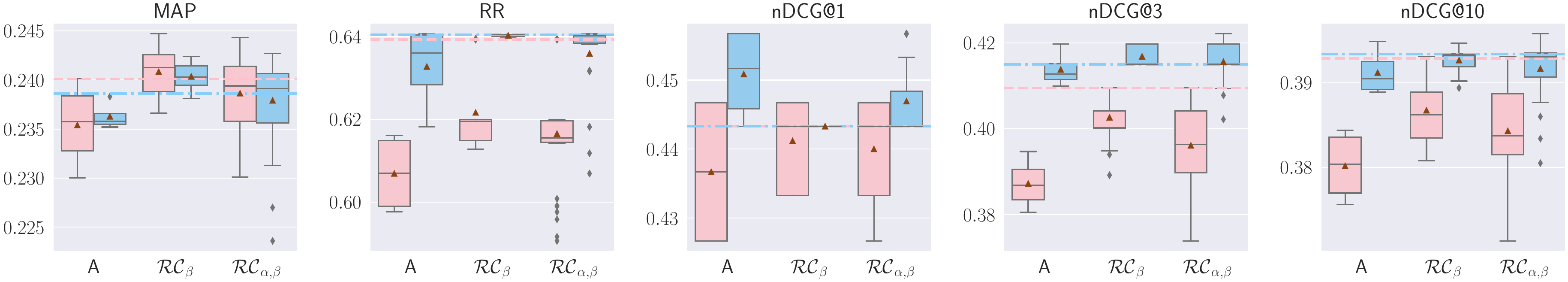}
	\caption{DL HARD}
	\label{fig:vprf_bb-fusion-dl-hard}
	\end{subfigure}
	\caption{Reranking effectiveness (y-axis) by using different score estimation methods. Where A is Vector Average, $\mathcal{RC}_\beta$ is Rocchio with fixed $\alpha$ value, and $\mathcal{RC}_{\alpha,\beta}$ is Rocchio with $\alpha$ and $\beta$. Baseline ANCE+BERT(A+B) and RepBERT+BERT(R+B) are marked with a dash-dot blue line and a dashed red line respectively. Significant improvements can be observed for MAP, with Average performing slightly better than the other two methods. Average also performs the best across the majority of the datasets and metrics, although the improvements where present are marginal compared to the baselines.}
	\label{fig:vprf_bb-agg-fusion}
\end{figure}

\textbf{RQ3: What is the impact of score estimation methods on the effectiveness of reranking and retrieval?} To answer this question, we vary the score estimation methods while displaying the distribution of results over other parameters (PRF depth and text handling). We analyze the effectiveness under three text-based score aggregation methods: Average (T-A), Borda (B) and Max (M); and three vector-based score fusion methods: Average (V-A), Rocchio with fixed $\alpha$ and varying $\beta$ ($\mathcal{RC}_\beta$), and Rocchio with varying $\alpha$ and $\beta$ ($\mathcal{RC}_{\alpha,\beta}$).

\subsubsection{Reranking with Text Score Estimation and Vector Fusion}

Results are shown in Figure~\ref{fig:rerank-agg-fusion}. For TREC DL 2019, T-A outperforms the baseline in terms of MAP, RR, and nDCG@1, while B is only on par with the baseline for RR, and $M$ hurts effectiveness across all metrics. BB+PRF-A with V-A is on par with the baseline across all metrics, except for marginal improvements found for RR. BB+PRF-R with V-A only improves RR marginally. Both BB+PRF-A and BB+PRF-R with $\mathcal{RC}_\beta$ and $\mathcal{RC}_{\alpha,\beta}$ substantially improve RR, while only BB+PRF-A with $\mathcal{RC}_{\alpha,\beta}$ substantially improves RR and nDCG@1, and is on par with the baseline A for nDCG@\{3, 10\}.

For TREC DL 2020, no score estimation method can outperform the baseline in terms of MAP. B, BB+PRF-R with $\mathcal{RC}_\beta$ and $\mathcal{RC}_{\alpha,\beta}$  are on par with the baseline for nDCG@\{1, 3\}. M is the worst estimation method for this dataset, as it only outperforms the baseline for nDCG@1, and all remaining methods decrease effectiveness.

For TREC CAsT 2019, M does not perform well across any metric, while T-A and B are on par with the baseline for nDCG@1. On the other hand, BB+PRF-A with V-A, $\mathcal{RC}_\beta$, and $\mathcal{RC}_{\alpha,\beta}$  is on par with the baseline for nDCG@1.

For WebAP, all methods are worse than the baseline in terms of MAP. For RR, only BB+PRF-A with $\mathcal{RC}_\beta$ is on par with the baseline for nDCG@3.

For DL HARD, all methods are worse than the baseline for MAP, RR, and nDCG@\{3, 10\}, except BB+PRF-A with $\mathcal{RC}_\beta$ and $\mathcal{RC}_{\alpha,\beta}$ , which is on par with the baseline for nDCG@1.

The results detailed where applying BERT reranker after the \emph{vector-based} PRF approaches are shown in Figure~\ref{fig:vprf_bb-agg-fusion}. Significant improvements for MAP over TREC DL 2019, TREC DL 2020, TREC CAsT, and WebAP can be observed with $\mathcal{RC}_\beta$, and $\mathcal{RC}_{\alpha,\beta}$. Average approach performs exceptionally well on TREC DL 2019 and TREC DL 2020 datasets. For all other metrics on all datasets, the majority of the improvements are marginal, and some statistically significant losses are observed. On the other hand, ANCE-based PRF with Average approach improves nDCG@1 significantly on TREC CAsT and DL HARD. 

\begin{figure}
	\begin{subfigure}{\columnwidth}
	\centering
	\includegraphics[width=\linewidth]{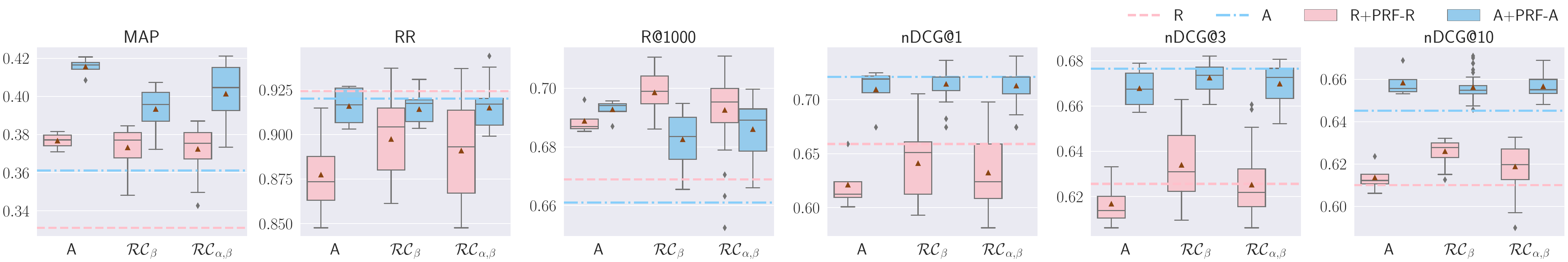}
	\caption{TREC DL 2019}
	\label{fig:vector-fusion-trec-2019}
	\end{subfigure}
	\begin{subfigure}{\columnwidth}
	\centering
	\includegraphics[width=\linewidth]{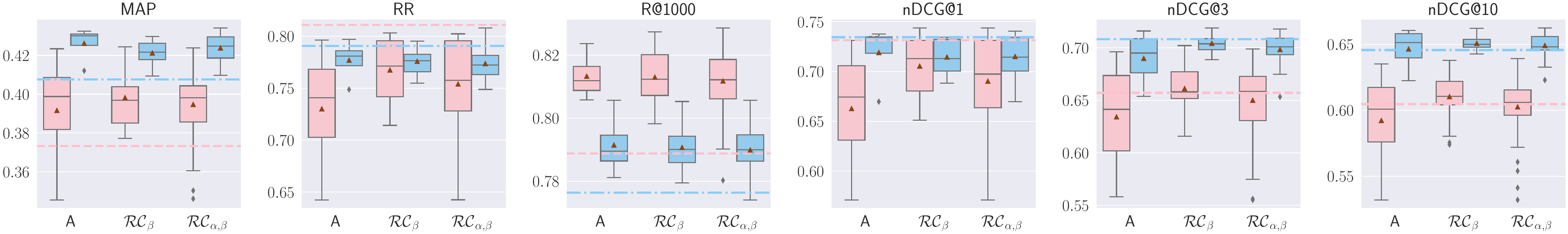}
	\caption{TREC DL 2020}
	\label{fig:vector-fusion-trec-2020}
	\end{subfigure}
	\begin{subfigure}{\columnwidth}
	\centering
	\includegraphics[width=\linewidth]{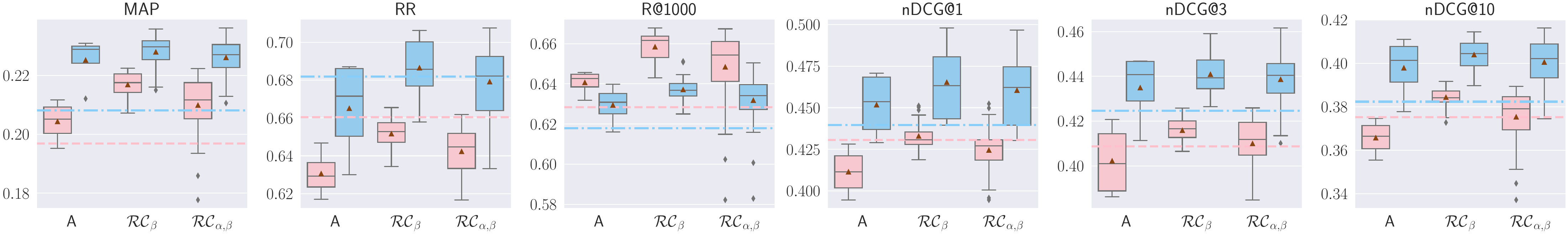}
	\caption{TREC CAsT}
	\label{fig:vector-fusion-trec-cast}
	\end{subfigure}
	\begin{subfigure}{\columnwidth}
	\centering
	\includegraphics[width=\linewidth]{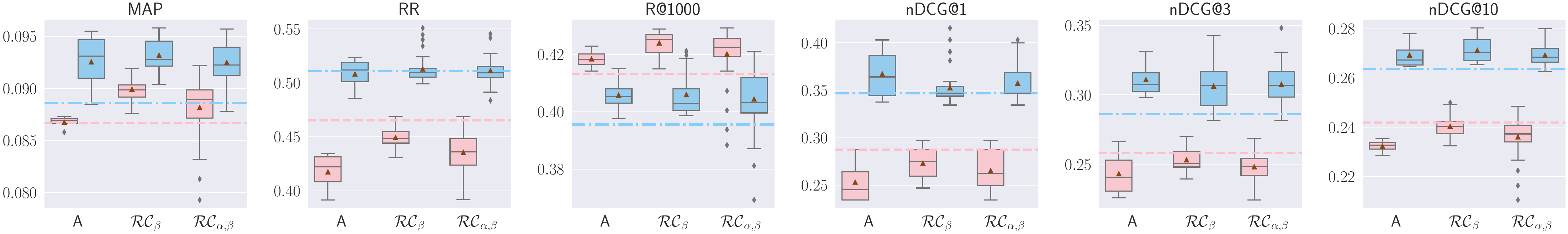}
	\caption{WebAP}
	\label{fig:vector-fusion-webap}
	\end{subfigure}
	\begin{subfigure}{\columnwidth}
	\centering
	\includegraphics[width=\linewidth]{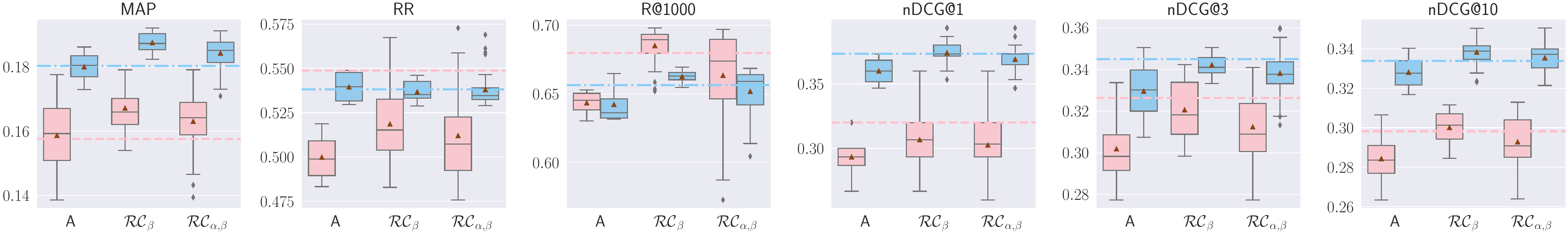}
	\caption{DL HARD}
	\label{fig:vector-fusion-dl-hard}
	\end{subfigure}
	\caption{Vector-based PRF retrieval effectiveness (y-axis) by using different vector fusion methods. Where A is Vector Average, $\mathcal{RC}_\beta$ is Rocchio with fixed $\alpha$ value, and $\mathcal{RC}_{\alpha,\beta}$ is Rocchio with $\alpha$ and $\beta$. Baseline RepBERT(R) is marked with dashed red line, ANCE(A) is marked with dash-dot blue line. $\mathcal{RC}_{\alpha,\beta}$ and $\mathcal{RC}_\beta$ perform the best in most circumstances. A+PRF-A with these methods is more likely to improve nDCG at early cut-offs, while R+PRF-R with these methods is more likely to improve deep recall.}
	\label{fig:vector-based-prf-retrieval-fusion}
\end{figure}

\subsubsection{Retrieval with Vector Fusion}

Results are shown in Figure~\ref{fig:vector-based-prf-retrieval-fusion}. For TREC DL 2019, overall, A+PRF-A outperforms the baseline across MAP, R@1000, and nDCG@10, while it is worse than the baseline for RR, and nDCG@\{1, 3\}. R+PRF-R also substantially improves MAP, R@1000, and nDCG@10, and it is on par with the baseline for nDCG@3 when $\mathcal{RC}_{\alpha,\beta}$ is used.

For TREC DL 2020, R+PRF-R performs exceptionally well in terms of R@1000, but both A+PRF-A and R+PRF-R do not outperform the respective baselines in terms of RR. On the other hand, A+PRF-A also outperforms the A+PRF-A baseline for R@1000, although the improvement is smaller than it was for R+PRF-R. All methods with A+PRF-A are on par with the baseline in terms of nDCG@10; a similar result is obtained  for R+PRF-R, except that marginal improvements can be observed with $\mathcal{RC}_\beta$.

For TREC CAsT 2019, both A+PRF-A and R+PRF-R substantially outperform the respective baselines in terms of MAP. A+PRF-A achieves substantial improvements in terms of nDCG@\{1, 3, 10\}. Both A+PRF-A and R+PRF-R, combined with any of V-A, $\mathcal{RC}_\alpha$ or $\mathcal{RC}_{\alpha,\beta}$, are either on par or worse than other metrics of the baseline and the dense retrievers without PRF (R and A).


For WebAP, both base models substantially improve MAP and R@1000, except R+PRF-R with A, which is on par with the baseline in terms of MAP. Overall, A+PRF-A is either on par with or improves the baseline across all metrics. On the other hand, R+PRF-R instead exhibits losses in terms of nDCG@\{1, 3, 10\}.

For DL HARD, A+PRF-A substantially improves MAP and nDCG@10 with $\mathcal{RC}_\beta$. R+PRF-R substantially improves MAP and R@1000 with $\mathcal{RC}_\beta$. All other metrics are either on par or worse than the baselines.

\subsubsection{Summary}

When the reranking task is considered, $\mathcal{RC}_{\alpha,\beta}$ is found to perform considerably well across all the metrics and datasets. B, T-A, and $\mathcal{RC}_\beta$ also perform well across several metrics and all datasets. M performs poorly across all metrics and datasets.

When adding the BERT reranker after the Vector-based PRF, significant improvements can be observed for MAP, with average performing slightly better than the other two methods. Average also performs the best across the majority of the datasets and metrics, although the improvements where present are marginal compared to the baselines.

When the retrieval task is considered, $\mathcal{RC}_{\alpha,\beta}$ and $\mathcal{RC}_\beta$ perform the best in most circumstances. A+PRF-A with these methods is more likely to improve nDCG at early cut-offs, while R+PRF-R with these methods is more likely to improve deep recall.

\subsection{Effectiveness of PRF}

\begin{table*}
	\centering
	\footnotesize
	\caption{Results of PRF approaches for the tasks of reranking and retrieval across different datasets. For each parametric method, the settings that achieve optimal effectiveness over all metrics are reported. Statistical significance (paired t-test) with $p<0.05$ between PRF models and BM25 is marked with $^a$, between PRF models and BM25+RM3 is marked with $^b$, between PRF and the corresponding baseline is marked with $^c$, between Vector-Based PRF+BERT and Dense Retriever+BERT is marked with $^d$ (For these two we do not compare with other baselines). Best results with respect to each dataset and each metric are highlighted in \textbf{bold}.}
	\resizebox{!}{.316\paperheight}{%
	\begin{tabular}{lp{5cm}p{1.2cm}p{1.2cm}p{1.2cm}p{1.2cm}p{1.2cm}p{1.2cm}}
		\toprule
		                                                                  & \textbf{Model}                     & \textbf{MAP}       & \textbf{RR}        & \textbf{nDCG@1}    & \textbf{nDCG@3}    & \textbf{nDCG@10}  & \textbf{R@1000}    \\ \midrule
\multirow{14}{*}{\rotatebox[origin=c]{90}{\textsc{Trec DL 2019}}} & BM25                               & .3773             & .8245             & .5426             & .5230             & .5058            & .7389             \\
		                                                                  & BM25+RM3                           & .4270             & .8167             & .5465             & .5195             & .5180            & \bf .7882             \\
		                                                                  & BM25+BERT (BB)                     & .4827             & .9240             & .6977             & .7203             & .7061            & .7389             \\
		                                                                  & RepBERT+BERT (R+B)                 & .4258             & .9388             & .7209             & .7317             & .6960            & .6689             \\
		                                                                  & ANCE+BERT (A+B)                    & .4315             & .9388             & .7209             & .7371             & .6965            & .6610             \\
		                                                                  & RepBERT (R)                       & .3311             & .9243             & .6589             & .6256             & .6100            & .6689             \\
		                                                                  & ANCE (A)                          & .3611             & .9201             & .7209             & .6765             & .6452            & .6610             \\
		\cmidrule{2-8}                                                    & BB+PRF($k=10$,CA,BORDA)               & .4947$^{ab}$      & \bf .9826$^{abc}$ & \bf .7946$^{abc}$ & \bf .7528$^{abc}$ & .7178$^{ab}$ & --                 \\
		                                                                  & BB+PRF-R($k=10,\beta=.3$)            & .4705$^{a}$       & .9793$^{ab}$      & .7326$^{ab}$      & .6963$^{ab}$      & .6993$^{ab}$     & --                 \\
		                                                                  & BB+PRF-A($k=10,\alpha=.3,\beta=.7$) & \bf .4955$^{ab}$  & .9690$^{ab}$      & .7519$^{ab}$      & .7385$^{ab}$  & \bf .7210$^{ab}$ & --                 \\
		\cmidrule{2-8}                                                    & R+PRF+B($k=5,\beta=.5$)              & .4463$^{d}$      & .9388             & .7209             & .7371         & .6968            & .7097$^{d}$                 \\
		                                                                  & A+PRF+B($k=5,\alpha=.4,\beta=.6$)    & .4514$^{d}$      & .9388             & .7209             & .7390         & .6953            & .6997$^{d}$                 \\
		\cmidrule{2-8}                                                    & R+PRF-R($k=10,\beta=.3$)              & .3669$^{c}$       & .9368             & .7054$^{c}$       & .6559$^{abc}$ & .6252$^{ab}$ & .7012$^{bc}$  \\
		                                                                  & A+PRF-A($k=10,\alpha=.4,\beta=.6$)   & .4151$^{c}$       & .9440$^{a}$       & .7403$^{ab}$      & .6807$^{ab}$  & .6629$^{ab}$ & .6962$^{bc}$  \\ 
		                                                                  \midrule \midrule
		\multirow{14}{*}{\rotatebox[origin=c]{90}{\textsc{Trec DL 2020}}} & BM25                               & .2856             & .6585             & .5772             & .5021             & .4796            & .7863             \\
		                                                                  & BM25+RM3                           & .3019             & .6360             & .5648             & .4740             & .4821            & \bf .8217             \\
		                                                                  & BM25+BERT (BB)                     & \bf .4926             & .8531             & .7901             & .7598             & .7064            & .7863             \\
		                                                                  & RepBERT+BERT (R+B)                 & .4358                 & .9082         & .7099             & .7276             & .6715            & .6593             \\
		                                                                  & ANCE+BERT (A+B)                    & .4470                 & .9082         & .7037             & .7243             & .6768            & .6819        \\
		                                                                  & RepBERT (R)                       & .3733             & .8109             & .7315             & .6572             & .6047            & .7888             \\
		                                                                  & ANCE (A)                          & .4076             & .7907             & .7346             & .7082             & .6458            & .7764             \\
		\cmidrule{2-8}                                                    & BB+PRF($k=3$,SW,BORDA)                & .4644$^{ab}$      & .8575$^{ab}$  & .8179$^{ab}$  & \bf .7798$^{ab}$  & .6739$^{ab}$     & --                 \\
		                                                                  & BB+PRF-R($k=5,\alpha=.4,\beta=.6$)  & .4778$^{ab}$      & .8638$^{ab}$  & \bf .8333$^{ab}$  & .7544$^{ab}$      & \bf .7111$^{ab}$ & --                 \\
		                                                                  & BB+PRF-A($k=1,\alpha=.5,\beta=.5$)  & .4606$^{abc}$     & .8476$^{ab}$      & .7963$^{ab}$  & .7691$^{ab}$  & .6984$^{ab}$ & --                 \\
		\cmidrule{2-8}                                                    & R+PRF+B($k=3,\alpha=.4,\beta=.6$)   & .4530$^{d}$       & .9050             & .7099         & .7320         & .6750        & .7022$^{d}$        \\
 																		  & A+PRF+B($k=3,\alpha=.4,\beta=.6$)   & .4584$^{d}$       & \bf .9097             & .7037     & .7297         & .6791        & .7019$^{d}$        \\
 		\cmidrule{2-8}                                                    & R+PRF-R($k=1,\alpha=.6,\beta=.4$)    & .4239$^{abc}$ & .7951$^{ab}$      & .7315$^{ab}$  & .6991$^{ab}$  & .6393$^{ab}$ & .8159$^{c}$   \\
		                                                                  & A+PRF-A($k=3,\alpha=.4,\beta=.6$)    & .4341$^{abc}$ & .8079$^{ab}$  & .7407$^{ab}$  & .7117$^{ab}$  & .6598$^{ab}$ & .7948         \\ \midrule \midrule
		\multirow{14}{*}{\rotatebox[origin=c]{90}{\textsc{Trec CAsT}}}    & BM25                               & .2936             & .6502             & .3631             & .3542             & .3526            & \bf .8326             \\
		                                                                  & BM25+RM3                           & .3132             & .6556             & .3971             & .3829             & .3817            & .8246             \\
		                                                                  & BM25+BERT (BB)                     & \bf .3762             & \bf .8108             & .5425             & .5366             & \bf .5269            & \bf .8326  \\
		                                                                  & RepBERT+BERT (R+B)                 & .3036                 & .7741         & .4953             & .5002             & .4901            & .6284              \\
		                                                                  & ANCE+BERT (A+B)                    & .3007                 & .7665         & .4855             & .4998             & .4890            & .6179        \\
		                                                                  & RepBERT (R)                       & .1969             & .6604             & .4307             & .4087             & .3752            & .6284             \\
		                                                                  & ANCE (A)                          & .2081             & .6819             & .4396             & .4246             & .3823            & .6179             \\
		\cmidrule{2-8}                                                    & BB+PRF($k=10$,CC)                     & .3247$^{ac}$      & .8106$^{ab}$      & .5510$^{abc}$ & .5140$^{ab}$      & .4838$^{abc}$    & --                 \\
		                                                                  & BB+PRF-R($k=3,\alpha=.5,\beta=.5$)  & .3372$^{ac}$      & .7985$^{ab}$      & .5480$^{ab}$  & .5468$^{ab}$  & .5067$^{abc}$    & --                 \\
		                                                                  & BB+PRF-A($k=3,\alpha=.3,\beta=.7$)  & .3274$^{ac}$      & .8093$^{ab}$      & \bf .5914$^{ab}$  & \bf .5583$^{ab}$  & .5055$^{abc}$    & --                 \\
		\cmidrule{2-8}                                                    & R+PRF+B($k=5,\alpha=.4,\beta=.6$)   & .3162$^{d}$       & .7722             & .4915             & .5023             & .4909            & .6635$^{d}$     \\
																		  & A+PRF+B($k=3,\alpha=.3,\beta=.7$)   & .3153$^{d}$       & .7725             & .4991             & .5043             & .4939            & .6432$^{d}$     \\
		\cmidrule{2-8}                                                    & R+PRF-R($k=10,\alpha=.8,\beta=.2$)   & .2150$^{abc}$ & .6618         & .4498$^{a}$   & .4146$^{a}$   & .3844$^{c}$  & .6566$^{abc}$ \\
		                                                                  & A+PRF-A($k=3,\beta=.9$)               & .2347$^{abc}$ & .6826         & .4626$^{a}$   & .4434$^{abc}$ & .4138$^{ac}$ & .6508$^{abc}$ \\ \midrule \midrule
		\multirow{14}{*}{\rotatebox[origin=c]{90}{\textsc{WebAP}}}        & BM25                               & .0436             & .3099             & .1667             & .1604             & .1404            & .2944             \\
		                                                                  & BM25+RM3                           & .0536             & .2767             & .1344             & .1316             & .1376            & .3472             \\
		                                                                  & BM25+BERT (BB)                     & .0845             & .5856             & .4042             & .3356             & .2897            & .2944             \\
		                                                                  & RepBERT+B (R+B)                    & .1088			   & .5859				& .4042				&.3361				& .2939			& .4133      \\
		                                                                  & ANCE+BERT (A+B)                    & .1090				& .5846				& .4010				& .3315				& .2973			& .3956   \\
		                                                                  & RepBERT (R)                       & .0867             & .4653             & .2875             & .2580             & .2419            & .4133             \\
		                                                                  & ANCE (A)                          & .0886             & .5107             & .3469             & .2863             & .2638            & .3956             \\
		\cmidrule{2-8}                                                    & BB+PRF($k=15$,CA,AVG)                 & .0855$^{ab}$  & .5459$^{ab}$      & .3271$^{ab}$      & \bf .3444$^{ab}$  & .2980$^{ab}$ & --                 \\
		                                                                  & BB+PRF-R($k=1,\alpha=.7,\beta=.3$)  & .0809$^{ab}$      & .5866$^{ab}$  & \bf .4198$^{ab}$  & .3418$^{ab}$  & .2708$^{ab}$     & --                 \\
		                                                                  & BB+PRF-A($k=3,\beta=.8$)             & .0790$^{abc}$     & .5502$^{ab}$      & .4083$^{ab}$  & .3394$^{ab}$  & .2842$^{ab}$     & --                 \\
		\cmidrule{2-8}                 									  & R+PRF+B($k=3,\alpha=.3,\beta=.7$)    & \bf .1146$^{d}$ 		 & \bf .5880 			& .4042 	& .3383			& .2995$^{d}$		& \bf .4253 \\
																	      & A+PRF+B($k=5,\alpha=.4,\beta=.6$)    & .1134$^{d}$		& .5790					& .3885		& .3308		& \bf .2996					& .4202 \\
		\cmidrule{2-8}                                                    & R+PRF-R($k=3,\beta=.1$)               & .0887$^{abc}$ & .4690$^{ab}$  & .2969$^{ab}$  & .2594$^{ab}$  & .2433$^{ab}$ & .4206$^{abc}$ \\
		                                                                  & A+PRF-A($k=3,\beta=.9$)               & .0953$^{abc}$ & .5134$^{ab}$  & .3563$^{ab}$  & .2928$^{ab}$  & .2710$^{ab}$ & .4027$^{ab}$  \\ \midrule \midrule
		\multirow{14}{*}{\rotatebox[origin=c]{90}{\textsc{DL-Hard}}}      & BM25                               & .1845             & .5422             & .3533             & .3137             & .2850            & .6288             \\
		                                                                  & BM25+RM3                           & .1925             & .4381             & .2467             & .2508             & .2555            & .6522             \\
		                                                                  & BM25+BERT (BB)                     & \bf .2521             & .6139             & .4133             & .4012             & .3962            & .6288             \\
		                                                                  & RepBERT+B (R+B)                    & .2401				& .6393					& .4433 		& .4095				& .3929				& .6797 \\
		                                                                  & ANCE+BERT (A+B)                    & .2386				& \bf .6405				& .4433				& .4150				& .3934				& .6564  \\
		                                                                  & RepBERT (R)                       & .1576             & .5489             & .3200             & .3263             & .2982            & .6797             \\
		                                                                  & ANCE (A)                          & .1803             & .5382             & .3733             & .3450             & .3339            & .6564             \\
		\cmidrule{2-8}                                                    & BB+PRF($k=3$,CA,BORDA)                & .2380$^{a}$       & .5937$^{b}$       & .4333$^{b}$   & .3944$^{b}$       & .3550$^{abc}$    & --                 \\
		                                                                  & BB+PRF-R($k=5,\alpha=.8,\beta=.2$)  & .2255$^{c}$       & .5843$^{ab}$      & .3867$^{b}$       & .3861$^{b}$       & .3646$^{abc}$    & --                 \\
		                                                                  & BB+PRF-A($k=5,\alpha=.4,\beta=.6$)  & .2422$^{a}$       & .5904$^{ab}$      & .4333$^{b}$   & \bf .4267$^{ab}$  & \bf .3968$^{ab}$ & --                 \\
		\cmidrule{2-8}                 									  & R+PRF+B($k=5,\beta=.2$) 				& .2439$^{d}$		& .6394 		& \bf .4433				& .4095			& .3926 		& \bf .6968$^{d}$		\\
																	      & A+PRF+B($k=5,\alpha=.8,\beta=.2$)   & .2419$^{d}$		& \bf .6405		& \bf .4433				&  .4150			& .3941  	& .6683$^{d}$  \\
		\cmidrule{2-8}                                                    & R+PRF-R($k=5,\alpha=.9,\beta=.1$)    & .1654         & .5504$^{b}$   & .3333         & .3368         & .3030        & .6929$^{c}$   \\
		                                                                  & A+PRF-A($k=10,\beta=.4$)              & .1865         & .5426         & .3933$^{b}$   & .3453         & .3380$^{b}$  & .6681$^{c}$   \\ \bottomrule
	\end{tabular}}
	\label{tab:effectiveness}
\end{table*}

\textbf{RQ4: What is the impact of PRF models on the effectiveness of reranking and retrieval?} To answer this question, we consider only our best performing PRF models with the optimal values for all the parameters combined. Results are presented in Table~\ref{tab:effectiveness}. For each dataset, the middle three rows represent PRF rerankers(BB+PRF, BB+PRF-R and BB+PRF-A), and the last two rows represent PRF retrievers (R+PRF-R and A+PRF-A). BB, R, and A are abbreviations for BM25+BERT, RepBERT, and ANCE, respectively. R@1000 is considered only for the evaluation of retrieval, mainly where it is infeasible to employ BERT for retrieval.

\subsubsection{Reranking with Text-Based PRF (BB+PRF)}

For TREC DL 2019, our model improves effectiveness over all metrics, with statistical significance mainly for shallow metrics (RR, nDCG@\{1,3\}). For TREC DL 2020, we observe improvements over shallow metrics only, although statistically not significant.

For TREC CAsT 2019, BB+PRF does not improve the effectiveness of BM25+BERT, except for nDCG@1. We speculate this is because BM25+BERT is trained with short passages, so it performs the best on CAsT (which consists of short passages).

For WebAP, improvements are observed over MAP and nDCG@\{3,10\}. Again, we believe this to be associated with the length of the passages in the dataset: here passages are longer and thus BM25+BERT (trained/fine-tuned on short passages) does not perform well.

For DL HARD, the improvement is only on nDCG@1, yet not significant, while nDCG@10 is significantly worse than the baseline. We speculate this is due to the poor relevance signals received by the PRF mechanism. Note that shallow metrics values on DL HARD are far below those in TREC DL 2019 and 2020 (which share the same passages): this means that the passages used for PRF are likely not relevant, thus possibly causing query drift.

To summarize, our proposed BB+PRF approach achieves substantially better results than BM25 and BM25+RM3. However, the improvements over BM25+BERT are more patchy, and are mostly achieved for shallow metrics. We put this down to the length of the text passages formed by the PRF methods: these are substantially longer than the passages used to train/fine-tune the BERT reranker.

\subsubsection{Reranking with Vector-Based PRF (BB+PRF-R, BB+PRF-A)}

When RepBERT is used as the base model (BB+PRF-R), for TREC DL 2019, improvements are obtained for RR and nDCG@1, while no improvements are obtained on the remaining metrics. For TREC DL 2020, we observe improvements over shallow metrics only (RR, nDCG@\{1, 10\}).
For TREC CAsT 2019, the improvements are observed at nDCG@\{1, 3\}, but nDCG@10 is significantly worse than the baseline, and so is MAP.
For WebAP, we observe improvements in shallow metrics as well (RR, nDCG@\{1, 3\}).
For DL HARD, there are no improvements over all reported metrics; on the contrary, it performs significantly worse than the baseline on MAP and nDCG@10.

With ANCE as the base model (BB+PRF-A), for TREC DL 2019, all shallow metrics (RR, nDCG@\{1, 3, 10\}) and MAP are improved.
For TREC DL 2020 and DL HARD, improvements are found at nDCG@\{1, 3, 10\}.
For TREC CAsT and WebAP, we observe improvements over nDCG@\{1, 3\} for both datasets.

To summarize, the proposed vector-based PRF as reranker (BB+PRF-R, BB+PRF-A): (1) it improves the effectiveness over BM25+BERT across several metrics and for all datasets,  (2) it achieves substantially better results than BM25, BM25+RM3, and RepBERT/ANCE, except on DL HARD, (3) it provides mixed results when compared with BM25+BERT, with no clear pattern of improvements (or deficiencies) across measures and datasets.

\subsubsection{Reranking with BERT on Top of Vector-Based PRF (R+PRF+B, A+PRF+B)}

R+PRF+B significantly improves MAP and R@1000 on all datasets except WebAP. All other results are either on par or slightly better (not significant) than the baseline. Moreover, R+PRF+B significantly improves nDCG@10 on WebAP compared to the baseline. However, the gain from R@1000 is mainly due to the gain obtained where moving from R to R+PRF: the reranking step does not contribute to this gain. 

The trend for A+PRF+B is similar to that of R+PRF+B: it significantly improves MAP and R@1000 across all datasets except WebAP. A+PRF+B achieves the best effectiveness for nDCG@10 on WebAP, although effectiveness are not significant. All other results are either on par or slightly better than the baseline.

To summarize, the use of the BERT reranker on top of \emph{vector-based} PRF significantly improves MAP, but for other metrics, improvements are not statistically significant. 

\subsubsection{Retrieval with Vector-Based PRF (R+PRF-R, A+PRF-A)}

For R+PRF-R, results show similar trends on all datasets: improvements can be observed on all reported metrics (MAP, RR, R@1000, and nDCG@\{1, 3, 10\}, except on TREC DL 2020, where PRF performs  worse than the RepBERT baseline for RR.

For A+PRF-A, PRF performs better than ANCE baseline on all evaluation metrics and across all datasets. The improvements in MAP are significant in TREC DL 2019, TREC DL 2020, TREC CAsT, and WebAP; the improvements for R@1000 are significant in TREC DL 2019, TREC CAsT, and DL HARD. Significant improvements in nDCG@\{3, 10\} are found only in TREC CAsT. Overall, A+PRF-A achieve higher effectiveness than R+PRF-R: ANCE per se is a stronger model than RepBERT, thus encoding more relevant information from the text. Hence, when PRF uses ANCE, it can better encode the additional relevance signals, leading to enhanced effectiveness.

To summarize, our proposed A+PRF-A and R+PRF-R models work well across all datasets and metrics. They also achieve substantial improvements over BM25 and BM25+RM3 baselines across almost all metrics, and they outperform the BM25+BERT baseline on several metrics.

\subsubsection{Generalizability to Other Dense Retrievers}


The results shown in Table~\ref{tab:other-dense} demonstrate that vector-based PRF consistently improves the effectiveness even where dense retrievers more effective than ANCE and RepBERT are used. Vector-based PRF tends to improve nDCG@3,10 (for the majority of the dense retrievers), MAP, nDCG@100, and R@1000 for all models in both TREC DL 2019 and TREC DL 2020. However, vector-based PRF does not improve RR and nDCG@1 in a consistent manner.


\begin{table*}
	\centering
	\footnotesize
	\caption{Results of vector-based PRF for the task of retrieval, using dense retrievers more effective than ANCE and RepBERT. We randomly choose and fix the parameters for Rocchio and Average in all the experiments in this table, where Average PRF depth is $3$, Rocchio PRF depth is $5$. We also fix $\alpha$ and $\beta$ for Rocchio to be 0.4 and 0.6 respectively. The best results for each model are marked in \textbf{Bold}.}
		\resizebox{!}{.20\paperheight}{%
\begin{tabular}{ll|llllllll}
	\toprule
	                        & Model                                & Method   & MAP             & RR              & nDCG@1          & nDCG@3          & nDCG@10         & nDCG@100        & R@1000          \\ \midrule
	\multirow{15}{*}{\rotatebox[origin=c]{90} {\textsc{TREC DL 2019}}} & \multirow{3}{*}{TCT-ColBERT V1}      & Original & 0.3864          & \textbf{0.9512} & \textbf{0.7326} & 0.6874          & 0.6700          & 0.5730          & 0.7207          \\
	                               &                                      & Average  & 0.4457          & 0.8999          & 0.6705          & 0.6779          & 0.6639          & 0.6119          & 0.7570          \\
	                               &                                      & Rocchio  & \textbf{0.4479} & 0.9368          & 0.7093          & \textbf{0.7083} & \textbf{0.6875} & \textbf{0.6143} & \textbf{0.7720} \\ \cmidrule{2-10}
	                               & \multirow{3}{*}{TCT-ColBERT V2 HN+}  & Original & 0.4626          & \textbf{0.9767} & \textbf{0.8023} & 0.7410          & 0.7204          & 0.6318          & 0.7603          \\
	                               &                                      & Average  & 0.5123          & \textbf{0.9767} & 0.7713          & \textbf{0.7454} & \textbf{0.7312} & \textbf{0.6719} & 0.8115          \\
	                               &                                      & Rocchio  & \textbf{0.5161} & 0.9244          & 0.7248          & 0.7129          & 0.7111          & 0.6684          & \textbf{0.8147} \\ \cmidrule{2-10}
	                               & \multirow{3}{*}{DistilBERT KD}       & Original & 0.3759          & 0.9306          & \textbf{0.7558} & \textbf{0.7370} & 0.6994          & 0.5765          & 0.6853          \\
	                               &                                      & Average  & 0.4362          & 0.9253          & 0.7481          & 0.7241          & \textbf{0.7096} & \textbf{0.6217} & 0.7180          \\
	                               &                                      & Rocchio  & \textbf{0.4378} & \textbf{0.9345} & 0.7442          & 0.7286          & 0.7052          & 0.6189          & \textbf{0.7291} \\ \cmidrule{2-10}
	                               & \multirow{3}{*}{DistilBERT Balanced} & Original & 0.4761          & \textbf{0.9510} & \textbf{0.7558} & \textbf{0.7494} & 0.7210          & 0.6360          & 0.7826          \\
	                               &                                      & Average  & 0.5057          & 0.9458          & 0.7364          & 0.7383          & 0.7190          & 0.6526          & 0.8054          \\
	                               &                                      & Rocchio  & \textbf{0.5249} & 0.9359          & 0.7364          & 0.7386          & \textbf{0.7231} & \textbf{0.6684} & \textbf{0.8352} \\ \cmidrule{2-10}
	                               & \multirow{3}{*}{SBERT}               & Original & 0.4097          & \textbf{0.9767} & \textbf{0.8372} & \textbf{0.7642} & 0.6930          & 0.5985          & 0.7201          \\
	                               &                                      & Average  & 0.4565          & 0.9413          & 0.7403          & 0.7326          & \textbf{0.7001} & \textbf{0.6149} & 0.7357          \\
	                               &                                      & Rocchio  & \textbf{0.4578} & 0.9355          & 0.7558          & 0.7448          & 0.6952          & \textbf{0.6149} & \textbf{0.7405} \\ \midrule \midrule
	\multirow{15}{*}{\rotatebox[origin=c]{90} {\textsc{TREC DL 2020}}} & \multirow{3}{*}{TCT-ColBERT V1}      & Original & 0.4290          & 0.8183          & 0.7500          & 0.7245          & 0.6678          & 0.5826          & 0.8181          \\
	                               &                                      & Average  & \textbf{0.4725} & 0.8220          & 0.7346          & 0.7253          & \textbf{0.6957} & \textbf{0.6101} & \textbf{0.8667} \\
	                               &                                      & Rocchio  & 0.4625          & \textbf{0.8392} & \textbf{0.7840} & \textbf{0.7410} & 0.6945          & 0.6056          & 0.8576          \\ \cmidrule{2-10}
	                               & \multirow{3}{*}{TCT-ColBERT V2 HN+}  & Original & 0.4754          & \textbf{0.8392} & \textbf{0.7932} & 0.7199          & \textbf{0.6882} & 0.6206          & 0.8429          \\
	                               &                                      & Average  & 0.4811          & 0.8212          & 0.7870          & \textbf{0.7386} & 0.6836          & 0.6228          & \textbf{0.8579} \\
	                               &                                      & Rocchio  & \textbf{0.4860} & 0.8154          & 0.7685          & 0.7273          & 0.6804          & \textbf{0.6254} & 0.8518          \\ \cmidrule{2-10}
	                               & \multirow{3}{*}{DistilBERT KD}       & Original & 0.4159          & \textbf{0.8215} & \textbf{0.7284} & \textbf{0.7113} & \textbf{0.6447} & 0.5728          & 0.7953          \\
	                               &                                      & Average  & \textbf{0.4214} & 0.7715          & 0.7130          & 0.6911          & 0.6316          & 0.5755          & 0.8403          \\
	                               &                                      & Rocchio  & 0.4145          & 0.7703          & 0.7037          & 0.6823          & 0.6289          & \textbf{0.5760} & \textbf{0.8433} \\ \cmidrule{2-10}
	                               & \multirow{3}{*}{DistilBERT Balanced} & Original & 0.4698          & 0.8350          & 0.7593          & 0.7426          & 0.6854          & 0.6346          & 0.8727          \\
	                               &                                      & Average  & \textbf{0.4887} & 0.8380          & 0.7809          & 0.7510          & \textbf{0.7086} & 0.6449          & \textbf{0.9030} \\
	                               &                                      & Rocchio  & 0.4879          & \textbf{0.8641} & \textbf{0.8056} & \textbf{0.7564} & 0.7083          & \textbf{0.6470} & 0.8926          \\ \cmidrule{2-10}
	                               & \multirow{3}{*}{SBERT}               & Original & 0.4124          & \textbf{0.7995} & \textbf{0.7346} & 0.6870          & 0.6344          & 0.5734          & 0.7937          \\
	                               &                                      & Average  & 0.4258          & 0.7619          & 0.6728          & 0.6723          & 0.6412          & 0.5781          & 0.8169          \\
	                               &                                      & Rocchio  & \textbf{0.4342} & 0.7941          & 0.7160          & \textbf{0.7032} & \textbf{0.6559} & \textbf{0.5851} & \textbf{0.8226} \\ \bottomrule
\end{tabular}}
\label{tab:other-dense}
\end{table*}

\subsubsection{Summary}

To answer RQ4, our results suggest that PRF used for either reranking or retrieval can improve effectiveness as measured across several metrics. More specifically, compared to the respective baselines, R+PRF-R and A+PRF-A tend to deliver improvements across all metrics and datasets, except for RR on TREC DL 2020 with RepBERT as base model. On the other hand, BB+PRF tends to only improve shallow metrics, especially when compared to BM25, BM25+RM3, and dense retriever baselines; BB+PRF-R and BB+PRF-A exhibit similar trends. Applying the BERT reranker on top of vector-based PRF significantly improves MAP, and outperforms the baseline on all metrics, but not significantly for the remaining metrics on all datasets with both R+PRF+B, A+PRF+B, and nDCG@10 on WebAP with R+PRF+B.

\subsection{Efficiency of PRF}
\label{sec:efficiency}

\textbf{RQ5: What is the impact of PRF models on the efficiency of reranking and retrieval?}

\begin{table}
	\centering
	\footnotesize
	\caption{Query latency of the investigated methods on TREC DL 2019: the lower latency, the better (faster).}
	\resizebox{0.6\columnwidth}{!}{
	\begin{tabular}{l|lr}
	    \toprule
		& \textbf{Models} & \textbf{Latency (ms/q)} \\ \midrule
		\multirow{6}{*}{\shortstack{Baselines}} & BM25 (Anserini) & 81 \\
		& BM25 + RM3 (Anserini) & 140 \\
		& RepBERT(R) & 93 \\
		& ANCE(A) & 94 \\ 
		& RepBERT+BERT(R+B) & 3,324 \\ 
		& ANCE+BERT(A+B) & 3,327 \\ \midrule 
		\multirow{4}{*}{\shortstack[l]{Vector-based PRF\\Retriever}} & R+PRF-R-Average & 163 \\
		& R+PRF-R-Rocchio & 163 \\
		& A+PRF-A-Average & 173 \\
		& A+PRF-A-Rocchio & 174 \\ \midrule
		\multirow{4}{*}{\shortstack[l]{Vector-based PRF\\Reranker}} & BB+PRF-R-Average & 3,411 \\
		& BB+PRF-R-Rocchio & 3,414 \\
		& BB+PRF-A-Average & 3,409 \\
		& BB+PRF-A-Rocchio & 3,414 \\ \midrule
		\multirow{3}{*}{\shortstack[l]{Text-based PRF\\Reranker}} & BB+PRF($k=5$)-CT & 6,889 \\
		& BB+PRF($k=5$)-CA & 17,266 \\
		& BB+PRF($k=5$)-SW & 22,314 \\ \midrule
		\multirow{4}{*}{\shortstack[l]{Vector-based PRF \\with BERT Reranker}} & R+PRF(Average)+B & 3,395 \\
		& R+PRF(Rocchio)+B & 3,397 \\
		& A+PRF(Average)+B & 3,419 \\ 
		& A+PRF(Rocchio)+B & 3,421 \\ \midrule
 		\multirow{2}{*}{\shortstack[l]{BERT Reranker}} & BM25 + BERT(BB) & 3,246 \\
		& BM25 + BERT Large & 9,209 \\ \bottomrule
	\end{tabular}
	}
	\label{table:efficiency}
\end{table}

\begin{figure}[t!]
	\centering
	\includegraphics[width=0.9\columnwidth]{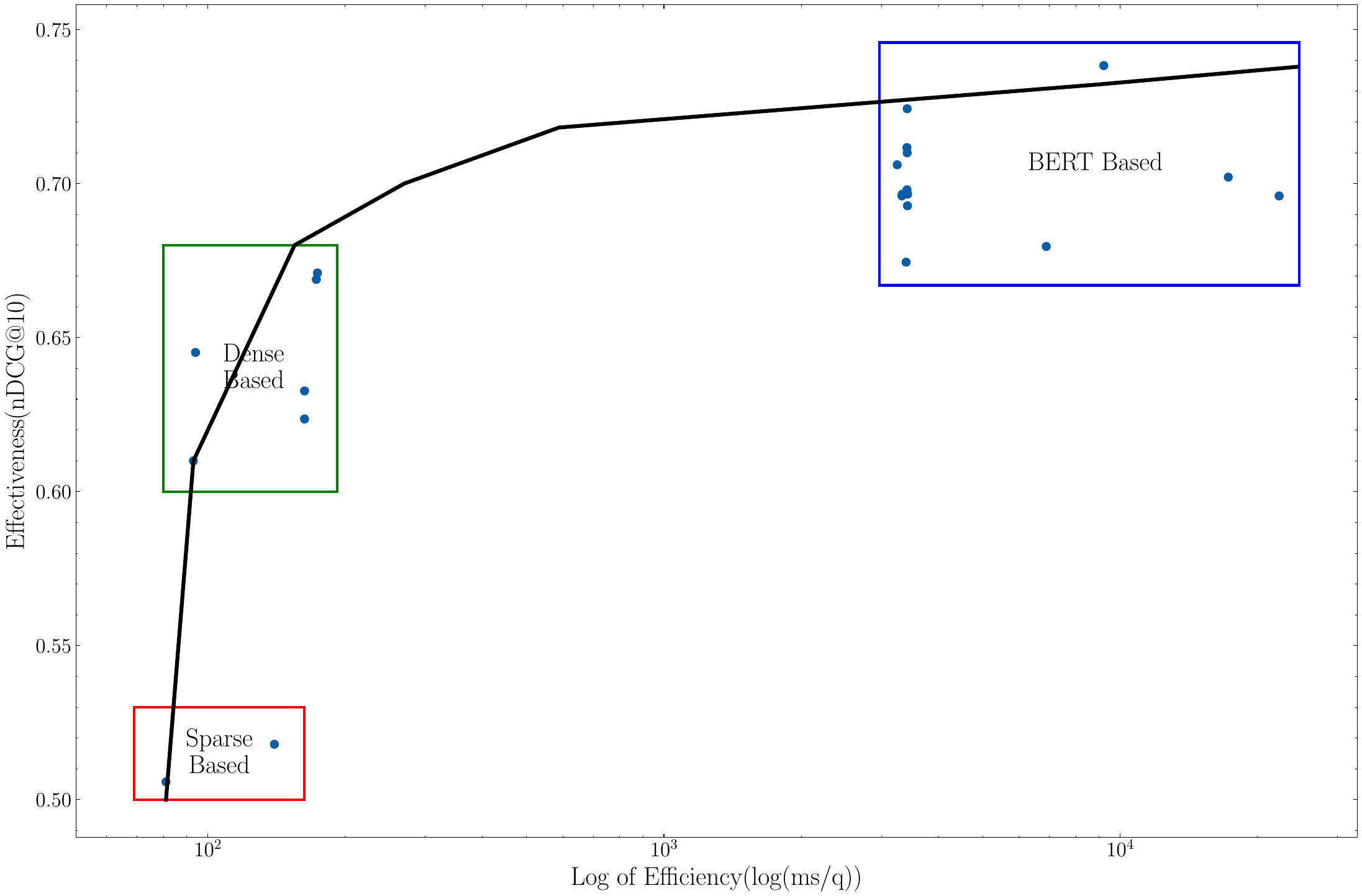}
	\caption{Trade-off between effectiveness and efficiency for all methods in our experiments. Effectiveness is measured using nDCG@10, and efficiency is measured using $\log(ms/q)$. The sparse baselines (BM25 and BM25+RM3) cluster on the left bottom corner (red box). Dense based approaches, including ANCE, RepBERT, and our Vector-based PRF approaches cluster on the center left (green box). All rerankers, i.e., BM25+BERT(base/large), Text-based PRF, BM25+BERT+Vector-based PRF, Vector-based PRF+BERT reranker, cluster on the top right side (blue box); these methods present the worst efficiency compared to others. The black line shows the trade-off trend between effectiveness and efficiency.}
	\label{fig:trade-off-figure}
\end{figure}

To answer this question, we study the efficiency of the PRF approaches and the baseline models. Low query latency -- the time required for a ranker to produce a ranking in answer to a query -- is an essential feature for the deployment of retrieval methods into real-time search engines.
The query latency of the investigated methods is summarised in Table~\ref{table:efficiency} and Figure~\ref{fig:trade-off-figure}\footnote{We have produced a fully annotated version of this image that includes each model name and made it available for online consultation at: \url{https://github.com/ielab/Neural-Relevance-Feedback-Public/blob/master/figures/trade-off-with-label.pdf}}.

The dense retrievers studied in this work (R and A) have a comparable query latency to BM25, with the latter being 10ms faster. Applying vector-based PRF to dense retrievers (R+PRF-R and A+PRF-A) has a comparable impact on query latency to BM25+RM3, with the latter being 23--34ms faster. The latency values measured in our experiments are compatible with the requirements of real-time search engines.
On the other hand, applying vector-based PRF to BM25+BERT, as a reranking stage(BB+PRF-R and BB+PRF-A), has a high query latency, similar to that of BM25+BERT. Lastly, we found the two-stage BM25+BERT-Large to be the least efficient (up to 2 orders of magnitude slower than other methods) except the text-based PRF approaches(BB+PRF). While BERT and BERT-Large reranking models consider only the top 1,000 passages from BM25, their query latency remains impractical for real-time search engines, the BB+PRF approaches actually creates more queries from one original query, hence leads to worse query latency overall.

In terms of applying BERT on top of Vector-based PRF approaches, the efficiency is similar to the Vector-based PRF reranker and the BERT reranker, which is much lower than the Vector-based PRF approaches because of the additional BERT inference time. As for BERT, this approach is also impractical for real-time search engines.

We also analysed the relationship between query length (either original query, or query plus PRF signal) and latency. For BM25 and BM25+RM3, query latency increases with the increase of query length: the longer the query (including the PRF component), in fact, the more posting lists need to be traversed.
 On the other hand, query length does not affect the query latency of ANCE or RepBERT\footnote{If not just noticeably because more tokens need to be passed through the tokenizer.}, because the query is converted to fixed length vectors: no matter how many words in the query, the generated query vector is always of the same length. This same reasoning applied for the vector-based PRF approaches: even when increasing the number of PRF passages considered ($k$), the query latency remains unchanged. As for the text-based PRF, we cut-off the query (including the revised query after PRF) to the length of 256 tokens, and before the query/passage pair is passed to BERT, the pair is padded to be of a total length of 512 tokens. Thus, no matter how long the query is (including possibly PRF), the sequence passed to BERT is always 512 tokens long. The reason for padding the input for BERT is that for each batch of query-passage pairs passed to BERT, the pairs in a batch need to be of the same length. The batching mechanism is useful for efficiently exploiting the GPU processing. In such cases, the query latency of the text-based PRF does not change with the increase of query length. A factor instead that does greatly affect the efficiency of text-based PRF with CA or SW is the depth of PRF. The more PRF passages, in fact, the more BERT inferences are required at run time, and thus the higher the query latency. For example, one original query with PRF depth at 5, the CA will generate 5 new queries, each for one feedback passage. So for each original query, it only needs one inference for BM25+BERT, but with the new queries from CA, it needs 5 more inferences, combined with BM25+BERT, it requires 6 inferences in total, similar for SW. (Table~\ref{table:efficiency})

\section{Use-case Analysis}
\label{sec:use}

To complement our analysis of the results and separate from our core research questions, we further analysed the results for a subset of the queries to gain a qualitative understanding of when, and possibly why, the PRF methods work -- or don't. For this, we limit our analysis to the retrieval task and the ANCE and Vector-based PRF with Rocchio with $k=5$. 

\begin{figure}[t!]
	\centering
	\includegraphics[width=1\columnwidth]{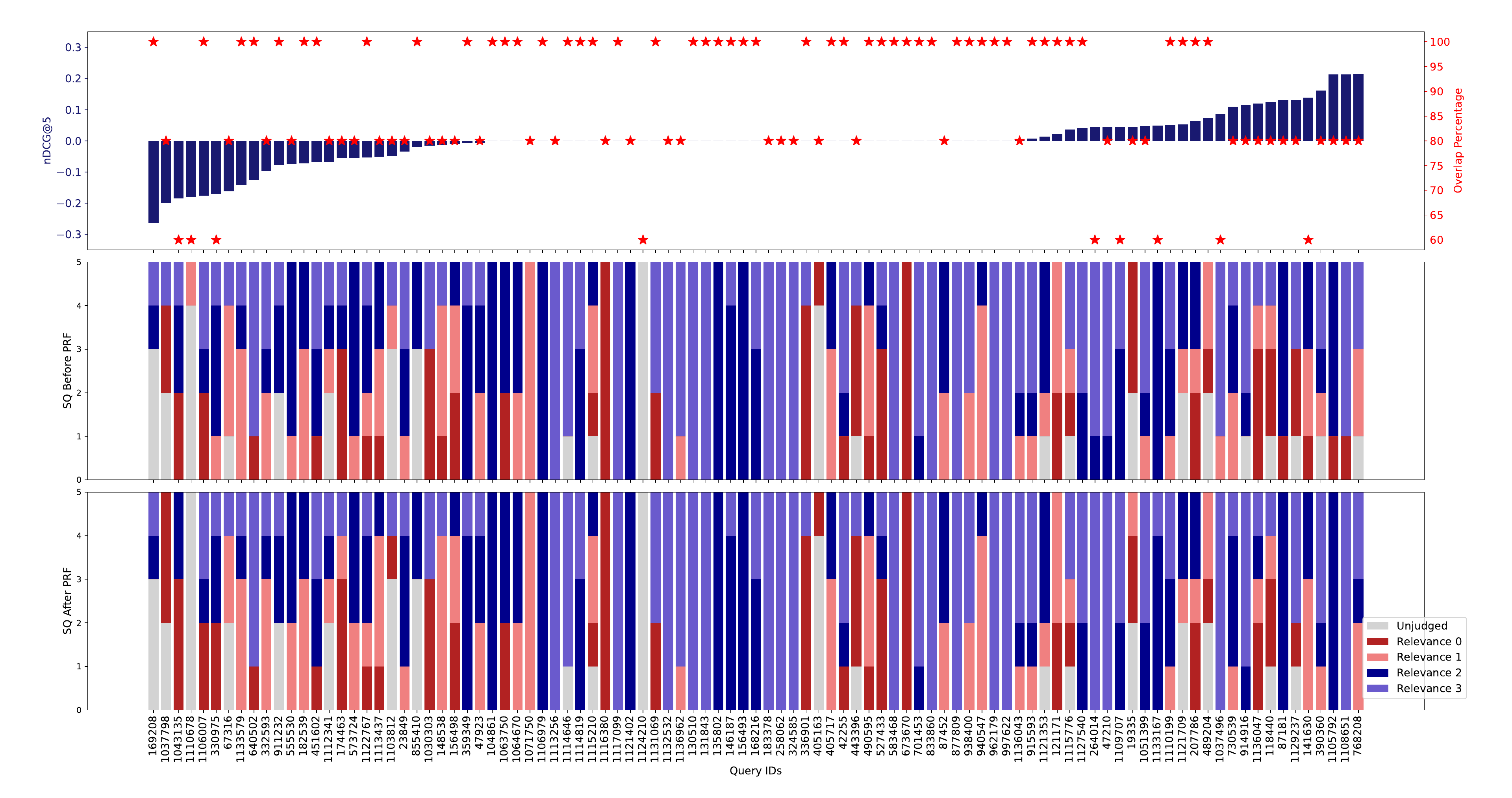}
	\caption{The query-by-query analysis on the combined set of TREC DL 2019 and 2020 queries of (1) the gain/loss obtained by  Vector-based PRF(Rocchio) with respect to ANCE, as represented by the barplot at the top, (2) the retrieved passages overlap, as represented by the red dots in the top plot, (3) and the analysis of the passage relevance for the feedback signal provided as input to the PRF (top $k=5$ passages from ANCE -- mid plot) and the top 5 passages retrieved by the PRF method (bottom plot).}
	\label{fig:gain-loss-figure}
\end{figure}

Figure~\ref{fig:gain-loss-figure}(top) presents a query-by-query analysis of the gains and losses of Vector-based PRF (Rocchio) with respect to ANCE on the combined set of TREC DL 2019 and 2020 queries. 
The effectiveness measure used is nDCG@5 and gains and losses are represented by bars in the figure. We used nDCG@5 for this analysis and not the rank cut-offs we have reported for other analyses to align the evaluation cut-off with the cut-off used for the feedback signal, i.e. the PRF depth $k$, which was also 5. The observations that follow, based on nDCG@5, are also found when other rank cut-offs are considered.

As one would expect, there are a number of queries for which the PRF method experiences losses, and other queries for which gains are observed; interestingly there is a considerable amount of queries for which neither gains nor losses are found. We further verified whether the encoding of the query obtained via PRF is different from that of the original query. This is done especially for the cases in which neither gains nor losses are observed: does this happen because there is no difference between the representation of the original query and of the query after PRF?
We performed this analysis by comparing the two dense representations using the inner product between the encoding vectors. We found that the similarity between the initial query encoding and the one obtained with PRF varies across queries and that queries for which neither gains nor losses are obtained, are no different in terms of difference between query representations compared to queries that exhibit gains or losses (figure not shown here, but available in the online appendix\footnote{ \url{https://github.com/ielab/Neural-Relevance-Feedback-Public}}). In other words, the amount by which PRF changes the initial query representation is not directly associated with a gain or a loss, and queries for which no difference in effectiveness is found are often characterised by differences between the original and PRF representations of the query.

For each query, Figure~\ref{fig:gain-loss-figure} (top) also reports the amount of \textit{overlap} between the top $k$ passages retrieved by the first stage of retrieval (ANCE) -- this is effectively the signal that is used as input to the PRF method -- and the top $k$ passages retrieved by the PRF method. The overlap is reported in percentage of the top $k$ passages that are in common between the two sets; $k=5$ and thus an overlap of $80\%$ means that 4 out of 5 passages are in common.
 This statistic helps us to analyse whether the PRF method ended up just re-ranking the top $k$ passages from the first stage, or it did push into the top $k$ ranks passages that were not there before. Losses in effectiveness for the PRF method tend to be associated with re-ranking of the top $k$ passages from the first stage of retrieval: the average overlap  for these queries is 85.2\%. And thus, these losses are not caused by retrieving less of the relevant passages, or passages that are less relevant (e.g. marginally relevant instead of highly relevant); they are instead caused by re-ordering the passages from the first stage in a less effective manner.
 On the other hand, gains instead have a higher tendency to be associated with the ability of the PRF method to bring into the top $k$ additional, new relevant passages compared to what the first stage of retrieval could do: the average overlap  for these queries is 82.8\%. Differences, however, are not statistically significant (Pearson's Chi-squared test).


Figure~\ref{fig:gain-loss-figure} (mid) presents an analysis of the relevancy of the top $k$ passages in the first stage of retrieval, while Figure~\ref{fig:gain-loss-figure} (bottom) presents an analysis of the relevancy of the top $k$ passages after PRF. Remember that the relevancy of the top $k$ passages in the first stage of retrieval in all effects represents the quality of the feedback signal that is passed as input to the PRF method. Each stacked column refers to a specific query, and the column is aligned to the gain-loss plot (Figure~\ref{fig:gain-loss-figure} (top)) so that it is possible to directly compare the relevancy of the PRF signal and the gain or loss produced by the PRF method. No obvious pattern appears at first sight. PRF signals that contain non-relevant documents produce results that can go either way: at times losses are obtained, but it is also common to obtain gains (or even have no effect at all). However, a more scrupulous examination of this signal leads to an interesting observation. We do this by examining the five queries with largest losses, and then the five queries with the largest gains for which there is at least one non-relevant passage. We observe the following. Non-relevant passages for queries that display losses are largely off topic: they have a very weak relationship to the query, if at all. For example, for query \texttt{1110678: what is the un fao}, the PRF signal includes 4 non-relevant passages, and these passages are related to the card game Uno; they have nothing to do with the United Nations FAO program. A subset of these passages (along with other examples) are presented in Table~\ref{tab:irrelevant-examples}.
Another similar example is query \texttt{169208:  does mississippi have an income tax}, for which the PRF signal includes 3 non-relevant passages and, while they are related to tax and the state of Mississipi, they are about unemployment or start-up tax and thus not relevant to the need of information about the more general income tax. When this type of feedback is used in the PRF mechanism, the new query representation displays \textit{query drift}, and it does score higher passages that are non-relevant, but are related to those in the feedback signal (e.g., for query \textit{1110678} many top retrieved passages by PRF are related to Uno, even beyond the top $k$). This situation instead does not occur when we examine the PRF signal of queries for which PRF provides a gain. In particular, while this signal does still contain non-relevant passages, the non-relevancy nature of these passages is different from that of passages for which losses are found. Let us unwrap this with actual examples. The PRF method exhibits a large gain for query \texttt{1136047: difference between a company’s strategy and business model is}, despite the PRF signal for this query displays several non-relevant passages. These non-relevant passages, shown in Table~\ref{tab:irrelevant-examples}, however, do display a strong aboutness relation~\cite{maron1977indexing} with the query, for example by mentioning the key terms of the query -- and when this is used in the PRF mechanism it would help reinforce the query representation. We also note that the non-relevant label assigned to some of these passages by the original TREC assessors may be somewhat questioned: for example passage \texttt{8724036} for query \texttt{1136047} appears to us as at least marginally relevant.
Other example queries that display a pattern similar to that of query \texttt{1136047} are included in Table~\ref{tab:irrelevant-examples}.

\begin{table}[t]
\centering
\footnotesize
\caption{Example queries for which the Vector-based PRF (Rocchio) method produces losses/gains, along with example non-relevant passages ranked in the top $k$ from the first stage ranker (and thus used as input to PRF).}
\begin{tabular}{p{0.03\linewidth}|p{0.08\linewidth}|p{0.25\linewidth}|p{0.55\linewidth}}
 &
  \textbf{Query ID} &
  \textbf{Query} &
  \textbf{Passage} \\ \toprule
\multirow{5}{*}[-5em]{\rotatebox[origin=c]{90}{\textsc{Queries for which PRF produces a loss}}} &
  \multirow{1}{*}[-2em]{\textbf{169208}} &
  \multirow{1}{*}[-2em]{does mississippi have an income tax\footnote{Two other, similar non-relevant passages appearing in the top $k=5$ for this topic are omitted from the table for space reasons.}} &
  \textbf{2888361}: Unemployment Tax Rates. Reporting and Filing. In Mississippi, the tax rate for a start-up business is 1.00\% the first year of liability, 1.10\% the second year of liability and 1.20\% the third and subsequent years of liability until the employer is eligible for a modified rate \\ \cmidrule{2-4}
 &
  \multirow{2}{*}[-5em]{\textbf{1037798}} &
  \multirow{2}{*}[-5em]{who is robert gray} &
  \textbf{2868740}: (Redirected from Gary Leroi Gray) Gary LeRoi Gray (born February 12, 1987) is an actor and voice actor involved with movies, television, and animation. He is most recognized for his childhood role as Nelson Tibideaux, the son of Sondra Huxtable Tibideaux and Elvin Tibideaux on the NBC sitcom The Cosby Show. He appeared on the series during its eighth and final season (1991-1992). \\ \cmidrule{4-4}
 &
   &
   &
  \textbf{2866248}: Matthew Gray Gubler. Matthew Gray Gubler is an Emmy award winning actor, director, producer, painter, and voice over actor from Las Vegas Nevada. While studying film directing at NYU he interned for Wes Anderson who gave him his first feature film role as Bill Murray's loyal intern Nico in The Life Aquatic with Steve Zissou (2004). For the past eleven years Gubler ... \\ \cmidrule{2-4}
 &
  \multirow{2}{*}[-4em]{\textbf{1110678}} &
  \multirow{2}{*}[-4em]{what is the un fao$^7$
  } &
  \textbf{5253767}: Uno (/EuEnoE/; from Italian and Spanish for ‘one’) (stylized as UNO) is an American card game that is played with a specially printed deck (see Mau Mau for an almost identical game played with normal playing cards). The game was originally developed in 1971 by Merle Robbins in Reading, Ohio, a suburb of Cincinnati. \\ \cmidrule{4-4}
 &
   &
   &
  \textbf{3386130}: Uno (card game) For the video game adaptation, see Uno (video game). Uno (/EuEnoE/; from Italian and Spanish for ‘one’) (stylized as UNO) is an American card game that is played with a specially printed deck (see Mau Mau for an almost identical game played with normal playing cards). The game was originally developed in 1971 by Merle Robbins in Reading, Ohio, a suburb of Cincinnati. \\ \midrule \midrule
\multirow{6}{*}[-6em]{\rotatebox[origin=c]{90}{\textsc{Queries for which PRF produces a gain}}} &
  \multirow{2}{*}[-3em]{\textbf{1129237}} &
  \multirow{2}{\linewidth}[-3em]{hydrogen is a liquid below what temperature} &
  \textbf{8588226}: Hydrogen is a liquid below what temperature? was asked by Shelly Notetaker on May 31 2017. 426 students have viewed the answer on StudySoup. View the answer on StudySoup. Sign Up Login \\ \cmidrule{4-4}
 &
   &
   &
  \textbf{8588222}: Answer to: Hydrogen is a liquid below what temperature? By signing up, you'll get thousands of step-by-step solutions to your homework questions.... for Teachers for Schools for Companies \\ \cmidrule{2-4}
 &
  \multirow{1}{*}[-2em]{\textbf{87181}} &
  \multirow{1}{*}[-2em]{causes of left ventricular hypertrophy} &
  \textbf{47203}: Causes of Right Ventricular Hypertrophy. There are four usual causes of right ventricular hypertrophy. The first one is pulmonary hypertension. As stated earlier, pulmonary hypertension is a condition wherein the blood pressure increases in the pulmonary artery. And this can lead to shortness of breath, dizziness and fainting. \\ \cmidrule{2-4}
 &
  \multirow{3}{*}[-8em]{\textbf{1136047}} &
  \multirow{3}{\linewidth}[-8em]{difference between a company’s strategy and business model is} &
  \textbf{8724032}: 6. The difference between a company's business model and a company's strategy is that: a. a company's business model is - Answered by a verified Business Tutor \\ \cmidrule{4-4}
 &
   &
   &
  \textbf{8724038}: Now, we address our second question. What is the difference between a strategy and a business model? A strategy is about the external logic of a business. How are we going to compete? Check out my earlier post on the elements of a strategy for more about strategies. A business model is about the internal logicâ¦ \\ \cmidrule{4-4}
 &
   &
   &
  \textbf{8724036}: Now, we address our second question. What is the difference between a strategy and a business model? A strategy is about the external logic of a business. How are we going to compete? Check out my earlier post on the elements of a strategy for more about strategies. A business model is about the internal logic of the business. See my post on the elements of a business model. Does it make operational and economic sense? Do all of the pieces fit together? A business model is a tool that complements both a business strategy and a business plan. It is an important tool because 1) it ensures that you understand the logic of your business; and, 2) it helps you communicate the logic of your business. Of course, this begs the question, what is a business plan? \\ \bottomrule
\end{tabular}
\label{tab:irrelevant-examples}
\end{table}

\section{Discussion}
\label{sec:dis}

Integrating PRF with deep language models has two main challenges: computational cost and input size limit, which can be roughly mapped to efficiency and effectiveness, respectively. Previous research has focused mainly on effectiveness, and applying PRF as a second stage that considers a subset of the data collection, to mitigate the computational cost. For instance,~\citet{zheng-etal-2020-bert} proposed a PRF framework which is divided into three phases, each involving BERT, to rerank passages and chunks of text after the initial retrieval. Similarly,~\citet{wang2020pseudo} applied PRF to the second stage retrieval, utilising BERT for reranking sentences. Other work that applied PRF to the first stage retrieval has limited the second phase of retrieval to a subset of the collection, either to depth of 500~\cite{yu2021pgt} or 1000~\cite{li2018nprf}. We argue that the main limitation with these approaches is related to the first stage retrieval model they employed --- if the first stage is basic bag-of-words then the overall method inherits many of this models limitations.


To address the applicability, effectiveness, and efficiency of text-based, vector-based and hybrid PRF approaches, we compared them over two different tasks, retrieval for vector-based PRF and reranking for both text-based and hybrid PRF. While the text-based and hybrid PRF approaches handle the input size limit, we showed that they enhance the ranking effectiveness only marginally. This finding aligns with previous research, but we demonstrated it happens at the cost of efficiency. The BERT model is slow (ranking 1,000 passages for each query takes 9,209 ms), rendering text-based and hybrid PRF approaches infeasible and inapplicable to real-time search engines for this task,  as shown in Figure~\ref{fig:trade-off-figure}. On the other hand, our proposed vector-PRF approach is substantially more efficient: it only takes 163-174 ms to process one query. In addition, it does so while also improving the effectiveness of dense retrievers in terms of all metrics compared to ANCE and RepBERT (for retrieval), and in terms of shallow metrics compared to BERT (for reranking).

We also apply the BERT reranker on top of Vector-based PRF approaches. Although this indeed improves effectiveness across all evaluation metrics, only improvements on MAP are significant on all datasets, and R+PRF+B significantly improves nDCG@10 on WebAP only. On the other hand, R@1000 is significantly improved across several datasets, but this improvement comes from the gain produced by the Vector-based PRF compared to the respective dense retrievers.

The BERT reranker used as a strong baseline in our experiments is an off-the-shelf model, originally produced by~\citet{nogueira2019passage}. This model is fine-tuned on the training set of the MS MARCO Passage Retrieval Dataset. The query length in this training set is much shorter than the newly formed queries from the text-based PRF approaches. This causes a mismatch between training and testing data. Theoretically, if the testing data is significantly different from the trained data data used to fine-tune the model, the model itself will be less effective. Despite this, in our experiments we found that the formed long PRF queries still significantly improve results for some metrics across several datasets (although improvements are often only marginal) -- but in the majority of cases their effectiveness is worse than that of BM25+BERT. It will then be interesting to investigate whether the effectiveness of text-based PRF can be further improved by fine-tuning the BERT model with long PRF queries, hence eliminating the training/testing data mismatch. 
On the other hand, vector-based PRF is not affected by this issue, because the query vector is of fixed length: no matter how many PRF passages are added to the original query vector, the query vector formed via PRF maintains the same length, and thus no training/testing data mismatch occurs. 
Another advantage of vector-based PRF over the text-based is that, although its effectiveness might be improved significantly after fine-tuning on long PRF queries, the text-based PRF approach is still inefficient and it is infeasible to apply it to real-time settings.

As complementary experiments, we also executed the retrieval task and re-ranking task on the MS MARCO dev queries. We omit the detailed results for these experiments from the manuscript in the interest of space and focus; however we make them available in the online appendix\footnote{\url{https://github.com/ielab/Neural-Relevance-Feedback-Public/blob/master/Vector_Based_PRF/README.md}} and we provide a summary of the observations next. 

For both retrieval and re-ranking tasks on the MS MARCO dev queries, we found no improvement over the non-PRF baselines. This is unlike in the TREC DL datasets, where improvements are observed. We believe that the reason for this difference is to be found on the largely sparse annotated nature of MS MARCO vs. the more complete relevance annotations in TREC DL. One may argue that other neural PRF methods, such as ANCE-PRF, have shown improvements on MS MARCO too, and not just on TREC DL. However, we stress an important difference. Both the baseline dense retrievers and the \emph{learnt} PRF methods (i.e. ANCE-PRF and the likes) are trained so as to build representations of the single ``gold'' (i.e. relevant) MS MARCO passage that are close to the query (for ANCE-PRF: representations of the query after PRF that are close to the gold MS MARCO passage). Thus, in the context of trained PRF methods like ANCE-PRF, the query representation is changed to become closer to the single MS MARCO target gold passage. 
Our PRF method changes the query representation (and consequently perturbs the ranking) based on the representation of the top $k$ passages alone: there is no notion of a gold passage towards which such new representation should be moved. This means these \emph{non-learnt} vector-based PRF methods are not suitable to situations in which a single golden passage exists, for example for the task of known item retrieval -- and this is shown by the MS MARCO results too. However, this does not mean these methods are not suitable when there is an array of relevant passages to be retrieved; and this in fact is what the TREC DL results show.

\section{Conclusion}

This article investigated the integration of PRF within transformer-based deep language models and dense retrievers for retrieval and reranking. In this context, a text-based PRF approach, applicable to the reranking task only, and a vector-based PRF approach, applicable to both retrieval and reranking tasks, were proposed to leverage the relevance signals from the feedback passages. The proposed approaches are based on the BERT reranker and the dense retrievers ANCE and RepBERT. In this context, we studied the impact of vector representation (for vector-based PRF), PRF depth, text handling techniques (for text-based PRF), and different score estimation (vector fusion) methods (for vector-based PRF) . 

When analysing results for the vector-based PRF approaches, we found they differed depending on whether we considered the task of retrieval or reranking.

In terms of \textit{which vector-based representation is most effective} for retrieval, we empirically found that performing PRF using ANCE as a representation (A+PRF-A) is better than when using RepBERT (R+PRF-R) across metrics and datasets, with the exception of R@1000. Indeed, our results show that R+PRF-R tends to improve deep metrics, while A+PRF-A tends to improve shallow metrics.
When considering the reranking task, substantial improvements only occurred on a few datasets and for a limited amount of metrics, and overall ANCE-based PRF performs better than RepBERT-based.

In terms of the \textit{impact of depth of PRF signal} for vector-based PRF, we found that when retrieval is considered, substantial improvements are recorded when 3 to 5 passages are given as feedback, while deeper feedback (10 passages) hurts effectiveness most of the times. When reranking is considered, PRF depth is not a factor substantially influencing effectiveness in a consistent manner across datasets and metrics.

In terms of \textit{score estimation methods (vector fusion)} for the vector-based PRF, we found that, for the retrieval task, $\mathcal{RC}_{\beta}$ and $\mathcal{RC}_{\alpha,\beta}$ achieve the best effectiveness across all datasets. Moreover, R+PRF-R often performs best when using either approximately evenly distributed weights between query/feedback passages, or when the query retains most of the weight.  This is similar for A+PRF-A, although for this method there are cases where the best effectiveness is achieved by giving a higher weight to feedback passages.
When the task of reranking is considered, we found that $\mathcal{RC}_{\alpha}$ and $\mathcal{RC}_{\alpha,\beta}$ record the most improvements. Substantial improvements, however, only occur on TREC DL 2019 and for only MAP, RR, and nDCG@1.

To validate the generalisability of our proposed Vector-based PRF models, we considered five additional dense retrievers (TCT-ColBERT V1, TCT-ColBERT V2 HN+, DistilBERT KD, DistilBERT Balanced, and SBERT) other than ANCE and RepBERT. Experimental results showed that our vector-based PRF approaches consistently improved the effectiveness over deep evaluation metrics.

The \textit{text-based PRF} approaches were only applicable to the reranking task. 
In terms of \textit{depth of PRF}, we found that, in most cases, substantial improvements are achieved when using 3, 5, and 10 passages for PRF. Deeper PRF signals (15 and 20 passages) were found to hurt effectiveness: this is likely because the addition of more feedback passages substantially contributed to query drift.
In terms of \textit{text handling methods}, we found that CA performs best in most cases, but only marginal improvements over the baselines are recorded across datasets and  metrics. 

We also experimented with applying the BERT reranker on top of Vector-based PRF models: this improves the effectiveness for all evaluation metrics, but only significantly for MAP. The significant improvements on R@1000 are purely produced by the Vector-based PRF models: the reranking step does not contribute to these improvements.

Vector-based and text-based PRF approaches could be compared only in the reranking task. In terms of effectiveness, neither of these two types of approaches clearly and consistently outperformed the other across datasets and metrics. However, they greatly differed in terms of \textit{efficiency (query latency)}. When the vector-based PRF was used for retrieval, latency was close to that of the bag-of-words PRF method BM25+RM3, and about double that of BM25. When used for reranking, the latency of the vector-based PRF remained the same. However, to this latency one needs to add that of the preceding part of the retrieval/ranking pipeline: in the case of our experiments that was the latency of BM25+BERT, which was substantial. On the other hand, the latency of text-based PRF was substantially higher. To the latency of the initial retrieval/ranking pipeline, in fact, the text-based PRF added the latency associated with the text handling methods (e.g., Concatenate and Aggregate) -- and this operation involved the execution of many more, costly BERT inferences, rendering the overall latency prohibitive for real-time search engines. The extensive experiments provided in this paper serve as a guide for using state-of-the-art neural retrieval models into low latency environments.

This work opens up a number of avenues for future research. First, we note that the Vector-based PRF method, and especially the Rocchio approach, has a number of limitations that future work could address. For example, Rocchio has parameters ($\alpha$, $\beta$) that need to be properly tuned to achieve the highest performance; we performed this tuning with respect to a validation set. How these parameters could instead be set on a per query basis (e.g., using some form of weight prediction) is still unclear. 

Second, the use-case analysis in Section~\ref{sec:use} revealed that the PRF methods are susceptible to the quality of the PRF signal (i.e. the passages retrieved by the first stage method and then passed as input to PRF). Recent work has made initial inroads to investigate this aspect. Specifically,~\citeauthor{li2022how} have studied the effect of signals of different qualities on the PRF process~\cite{li2022how} -- although they only explored a limited set of possible signals and PRF methods. \citeauthor{zhuang2022implicit} have devised effective methods to de-noise the feedback signal in the case of implicit feedback~\cite{zhuang2022implicit}. The extent to which different PRF methods differ in terms of susceptibility to PRF signals of different qualities is still however unclear, especially beyond the specific methods that we examined in the use-case analysis and the signals and settings explored in the aforementioned recent work.

Third, this paper only considered the passage retrieval and ranking tasks. PRF techniques have, however, long been explored for document retrieval tasks and one wonders how the methods designed here for passages, and that exploit pre-trained language models, generalise when considering documents. We note that adapting these specific passage PRF methods to documents is not straightforward. Documents are sensibly longer in length than passages, and most likely their text exceeds the maximum input length that pre-trained language models like BERT can consider. This is certainly a problem for models like ANCE-PRF and our text-based PRF methods, where the text of the PRF signal needs to be encoded. But, this is also a problem for our Vector-based PRF methods: while the text associated to the PRF signal is not required by the PRF process itself, these documents need still to be encoded so that their dense vector can be used (for the initial round of retrieval, and for the PRF). Tackling this limitation of pre-trained language models for effective document retrieval is not new~\cite{yang2019simple,yilmaz2019applying,yilmaz2019cross,mass2020ad}-- progress so far in this space has not  yet solved this problem (especially if also efficiency considerations are made) and PRF in this context is largely unexplored~\cite{wang2020pseudo,xu2009query,wang2020end}.

Fourth, PRF is typically integrated in wider pipelines, including multiple rankers and a multi-stage ranking architecture. A recent trend with neural rankers is to create hybrid systems that comprise a dense and a sparse rankers~\cite{wang2021bert,arabzadeh2021predicting,lin2021pretrained}. In this paper, we have not considered such architecture type, and how PRF could be integrated. We however refer the interested reader to the recent work of~\citeauthor{li2022interpolate} that has extended our methods to hybrid rankers~\cite{li2022interpolate}.

The implementations of our PRF methods are made publicly available at \url{http://ielab.io/publications/li-tois-2022}, along with the full empirical results.

\begin{acks}
This research is partially funded by the Grain Research and Development Corporation project AgAsk (UOQ2003-009RTX). 
Dr Guido Zuccon is the recipient of an Australian Research Council DECRA Research Fellowship (DE180101579). 
\end{acks}

\bibliographystyle{ACM-Reference-Format}
\bibliography{reference.bib}

\end{document}